\documentclass[11pt,letter]{article}

\usepackage{a4wide}
\usepackage{amsmath,amscd,amssymb}
\usepackage{mathrsfs,fancyhdr,graphics,pifont}
\usepackage{floatflt,subfigure,epsfig,moreverb,multicol,lscape}
\usepackage{supertabular,tabularx,multirow,bar,amssymb}
\usepackage{theorem,psfrag,afterpage,curves,epic,bbm,bbold,color}
\usepackage{wasysym}
\usepackage{pifont}
\usepackage[english]{babel}

\def\statusstring{Submitted to IEEE Transactions on Information Theory \\
                  December 20, 2005}

\newcommand{\ignore}[1]{}

\newcommand{\Int}{\mathbb{Z}}
\newcommand{\Intp}{\mathbb{Z}_{+}}
\newcommand{\Intpp}{\mathbb{Z}_{++}}
\newcommand{\Q}{\mathbb{Q}}
\newcommand{\Qp}{\mathbb{Q}_{+}}
\newcommand{\Qpp}{\mathbb{Q}_{++}}
\newcommand{\R}{\mathbb{R}}
\newcommand{\Rp}{\mathbb{R}_{+}}
\newcommand{\Rpp}{\mathbb{R}_{++}}

\newcommand{\Expec}{\operatorname{E}}
\newcommand{\Var}{\operatorname{Var}}

\newcommand{\supp}{\operatorname{supp}}

\newcommand{\dint}[1]{\,\operatorname{d}{#1}}

\newcommand{\matr}[1]{\mathbf{#1}}
\newcommand{\vect}[1]{\mathbf{#1}}
\newcommand{\code}[1]{\mathcal{#1}}

\newcommand{\set}[1]{\mathcal{#1}}

\newcommand{\graph}[1]{\mathsf{#1}}
\newcommand{\tgraph}[2]{\graph{#1}(\matr{#2})}

\newcommand{\GF}[1]{\mathbb{F}_{#1}}

\newcommand{\defeq}{\triangleq}

\newcommand{\valpha}{\boldsymbol{\alpha}}

\newcommand{\setV}{\set{V}}

\newcommand{\matrunity}{\matr{I}}

\newcommand{\wcol}{w_{\mathrm{col}}}
\newcommand{\wrow}{w_{\mathrm{row}}}

\newcommand{\normaldistlr}[2]{\mathcal{N}\left(#1,#2\right)}

\newcommand{\hM}{\hat M}
\newcommand{\hcT}{\hat{\cover{T}}}
\newcommand{\lift}[2]{{#1}^{\uparrow #2}}

\newcommand{\vLambda}{\boldsymbol{\Lambda}}

\newcommand{\vlambda}{\boldsymbol{\lambda}}

\newcommand{\tvlambda}{\tilde{\boldsymbol{\lambda}}}
\newcommand{\vmu}{\boldsymbol{\mu}}
\newcommand{\vnu}{\boldsymbol{\nu}}

\newcommand{\vomega}{\boldsymbol{\omega}}
\newcommand{\ovomega}{\boldsymbol{\overline{\omega}}}
\newcommand{\hvomega}{\boldsymbol{\hat \omega}}

\newcommand{\hvomegaLPD}{\boldsymbol{\hat \omega}^{\mathrm{LPD}}}
\newcommand{\hvomegaLPDH}[1]{\boldsymbol{\hat \omega}^{\mathrm{LPD}(\matr{#1})}}

\newcommand{\GCDset}[1]{\set{Q}(\matr{#1})}
\newcommand{\cGCDset}[1]{\set{\cover{Q}}(\matr{#1})}
\newcommand{\GCD}[1]{\mathrm{GCD}(\matr{#1})}

\newcommand{\va}{\vect{a}}

\newcommand{\vc}{\vect{c}}

\newcommand{\tvc}{\vect{\tilde c}}

\newcommand{\vv}{\vect{v}}
\newcommand{\cvv}{\vect{\cover{v}}}

\newcommand{\vX}{\vect{X}}

\newcommand{\cvX}{\vect{\cover{X}}}

\newcommand{\vx}{\vect{x}}

\newcommand{\hvx}{\vect{\hat x}}

\newcommand{\hvxMAPD}{\vect{\hat x}^{\mathrm{MAPD}}}
\newcommand{\hvxLPD}{\vect{\hat x}^{\mathrm{LPD}}}

\newcommand{\tvx}{\vect{\tilde x}}
\newcommand{\cvx}{\vect{\cover{x}}}
\newcommand{\hcvx}{\vect{\hat{\cover{x}}}}
\newcommand{\ovx}{\vect{\overline{x}}}
\newcommand{\vY}{\vect{Y}}

\newcommand{\ovY}{\vect{\overline{Y}}}
\newcommand{\cvY}{\vect{\cover{Y}}}

\newcommand{\vy}{\vect{y}}
\newcommand{\ovy}{\vect{\overline{y}}}

\newcommand{\cvy}{\vect{\cover{y}}}

\newcommand{\oX}{\overline{X}}
\newcommand{\ox}{\overline{x}}
\newcommand{\oY}{\overline{Y}}

\newcommand{\oZ}{\overline{Z}}

\newcommand{\wps}{w_{\mathrm{p}}}
\newcommand{\wpsAWGNC}{w_{\mathrm{p}}^{\mathrm{AWGNC}}}

\newcommand{\wpsBSC}{w_{\mathrm{p}}^{\mathrm{BSC}}}
\newcommand{\wpsBEC}{w_{\mathrm{p}}^{\mathrm{BEC}}}
\newcommand{\wpsmin}{w_{\mathrm{p}}^{\mathrm{min}}}
\newcommand{\wpsAWGNCmin}{w_{\mathrm{p}}^{\mathrm{AWGNC,min}}}
\newcommand{\wpsBSCmin}{w_{\mathrm{p}}^{\mathrm{BSC,min}}}
\newcommand{\wpsBECmin}{w_{\mathrm{p}}^{\mathrm{BEC,min}}}
\newcommand{\wpsminc}[1]{\wpsmin(\code{#1})}
\newcommand{\wpsminh}[1]{\wpsmin(\matr{#1})}
\newcommand{\wpsAWGNCminh}[1]{\wpsAWGNCmin(\matr{#1})}
\newcommand{\wpsBSCminh}[1]{\wpsBSCmin(\matr{#1})}
\newcommand{\wpsBECminh}[1]{\wpsBECmin(\matr{#1})}
\newcommand{\wfr}{w_{\mathrm{frac}}}
\newcommand{\wfrminh}[1]{w^{\mathrm{min}}_{\mathrm{frac}}(\matr{#1})}

\newcommand{\wmaxfr}{w_{\mathrm{max-frac}}}
\newcommand{\wmaxfrminh}[1]{w^{\mathrm{min}}_{\mathrm{max-frac}}(\matr{#1})}

\newcommand{\dH}{d_{\mathrm{H}}}
\newcommand{\dHmin}{d_{\mathrm{H}}^{\mathrm{min}}}
\newcommand{\dHminc}[1]{\dHmin(\code{#1})}
\newcommand{\wH}{w_{\mathrm{H}}}
\newcommand{\wHmin}{w_{\mathrm{H}}^{\mathrm{min}}}
\newcommand{\wHminc}[1]{\wHmin(\code{#1})}
\newcommand{\diam}{\delta}
\newcommand{\diamg}[1]{\diam(\graph{#1})}
\newcommand{\girth}{g}
\newcommand{\girthg}[1]{\girth(\graph{#1})}
\newcommand{\Nmax}{N^{\mathrm{max}}}
\newcommand{\fp}[1]{\mathcal{#1}}

\newcommand{\fc}[1]{\mathcal{#1}}

\newcommand{\fph}[2]{\mathcal{#1}(\matr{#2})}

\newcommand{\fch}[2]{\mathcal{#1}(\matr{#2})}

\newcommand{\vs}{\vect{s}}
\newcommand{\tr}{\mathsf{T}}

\newcommand{\PMLD}{P^{\mathrm{MLD}}}
\newcommand{\PGCDLPD}{P^{\mathrm{GCD/LPD}}}

\newcommand{\convhull}{\operatorname{conv}}

\newcommand{\conichull}{\operatorname{conic}}

\newcommand{\relaxation}{\operatorname{relax}}
\newcommand{\relaxationset}[1]{\set{R}(\matr{#1})}
\newcommand{\relaxationsetorder}[2]{\set{R}_{#1}(\matr{#2})}

\newcommand{\SNRb}{\mathrm{SNR}_{\mathrm{b}}}
\newcommand{\SNRc}{\mathrm{SNR}_{\mathrm{c}}}
\newcommand{\Eb}{E_{\mathrm{b}}}
\newcommand{\Ec}{E_{\mathrm{c}}}

\newcommand{\Pii}{\boldsymbol{\Pi}} 

\newcommand{\neighborhood}{\partial}

\newcommand{\cover}[1]{\widetilde{#1}}
\newcommand{\cset}[1]{\cover{\set{#1}}}
\newcommand{\cgraph}[1]{\cover{\graph{#1}}}
\newcommand{\ccode}[1]{\cover{\code{#1}}}
\newcommand{\cmatr}[1]{\cover{\matr{#1}}}
\newcommand{\tgcode}[2]{\code{#1}(\graph{#2})}
\newcommand{\tgmatr}[2]{\matr{#1}(\graph{#2})}
\newcommand{\ctgcode}[2]{\code{#1}(\cover{\graph{#2}})}
\newcommand{\ctgmatr}[2]{\matr{#1}(\cover{\graph{#2}})}

\newcommand{\CheckNode}{B}

\newcommand{\abs}[1]{\lvert #1 \rvert}
\newcommand{\card}[1]{\lvert #1 \rvert}

\newcommand{\onenorm}[1]{\left\lVert #1 \right\rVert_1}
\newcommand{\twonorm}[1]{\left\lVert #1 \right\rVert_2}
\newcommand{\infnorm}[1]{\left\lVert #1 \right\rVert_{\infty}}
\newcommand{\sonenorm}[1]{\onenorm{#1}}

\newcommand{\?}{\mathrm{?}}

\newcommand{\noproof}[1]{}

 \renewcommand{\leq}{\leqslant}
 \renewcommand{\geq}{\geqslant}

\newcommand{\innerprod}[2]{\langle {#1},{#2} \rangle}

\newcommand{\highsup}[1]{\raisebox{0.35ex}{\kern 1pt $\scriptstyle
    {#1} $}}

\newcommand{\Pistar}{{\Pii^{\highsup{\kern-1pt\star}}\kern-1pt}}

\newcommand{\be}[1]{\begin{equation}\label{#1}}
  \newcommand{\ee}{\end{equation}} 

\newcommand{\bc}{\begin{center}} \newcommand{\ec}{\end{center}}

\newtheorem{Lemma}{Lemma}

\newtheorem{Proposition}[Lemma]{Proposition}
\newtheorem{Corollary}[Lemma]{Corollary}

\theoremstyle{plain}

\newtheorem{PreDefinition}[Lemma]{{\textbf{Definition}}}
  \newenvironment{Definition}%
    {\begin{PreDefinition}}{\hfill$\square$\end{PreDefinition}}

\theoremstyle{plain}

\newtheorem{PreRemark}[Lemma]{{\textbf{Remark}}}
    {\begin{PreRemark}\upshape}{\hfill$\square$\end{PreRemark}}

\newtheorem{PreExample}[Lemma]{{\textbf{Example}}}
  \newenvironment{Example}%
    {\begin{PreExample}\upshape}{\hfill$\square$\end{PreExample}}

\newenvironment{Proof}%
  {\noindent \emph{Proof:}}{\hfill$\square$}

\begin{document}


\title{Graph-Cover Decoding and
       Finite-Length Analysis of \\
       Message-Passing Iterative Decoding of LDPC Codes%
  \footnote{The work of P.~O.~Vontobel was supported by NSF Grants CCR 99-84515
    and CCR-0105719 at UIUC and by and by NSF Grants CCR 99-84515, CCR
    01-05719, ATM-0296033, DOE SciDAC, and ONR Grant N00014-00-1-0966 at
    UW-Madison. The work of R.~Koetter was partially supported by NSF Grants
    CCR 99-84515 and CCR-0105719. The material in this paper was presented in
    part at the 3rd International Conference on Turbo Codes and Related
    Topics, Brest, France, September 2003.}  }

\author{Pascal O.~Vontobel%
  \footnote{Was with ECE Department, University of Wisconsin-Madison, 
            1415 Engineering Drive Madison, WI 53706, USA. 
            E-Mail: \texttt{pascal.vontobel@ieee.org}. 
            P.~O.~Vontobel is the corresponding author.}
         \ and Ralf Koetter%
    \footnote{Coordinated Science Laboratory
            and ECE Department,
            University of Illinois at Urbana-Champaign, 1308 West Main Street,
            Urbana, IL 61801, USA.
            E-mail: \texttt{koetter@uiuc.edu}.}
       }

\date{}

\maketitle

\vspace{-7cm}
{
 \begin{flushright}
   \texttt{\statusstring}
 \end{flushright}
}
\vspace{+6cm}


\begin{abstract}

  The goal of the present paper is the derivation of a framework for the
  finite-length analysis of message-passing iterative decoding of low-density
  parity-check codes. To this end we introduce the concept of graph-cover
  decoding. Whereas in maximum-likelihood decoding all codewords in a code are
  competing to be the best explanation of the received vector, under
  graph-cover decoding all codewords in all finite covers of a Tanner graph
  representation of the code are competing to be the best explanation.

  We are interested in graph-cover decoding because it is a theoretical tool
  that can be used to show connections between linear programming decoding and
  message-passing iterative decoding. Namely, on the one hand it turns out
  that graph-cover decoding is essentially equivalent to linear programming
  decoding. On the other hand, because iterative, locally operating decoding
  algorithms like message-passing iterative decoding cannot distinguish the
  underlying Tanner graph from any covering graph, graph-cover decoding can
  serve as a model to explain the behavior of message-passing iterative
  decoding.

  Understanding the behavior of graph-cover decoding is tantamount to
  understanding the so-called fundamental polytope. Therefore, we give some
  characterizations of this polytope and explain its relation to earlier
  concepts that were introduced to understand the behavior of message-passing
  iterative decoding for finite-length codes.

\end{abstract}


\newpage


\section{Introduction}
\label{sec:introduction:1}

Low-density parity-check (LDPC) codes were introduced by
Gallager~\cite{Gallager:62, Gallager:63}. As important as the codes themselves
was also a class of decoding algorithms that he presented. These algorithms
had two common features. Firstly, based on the observed channel output, these
algorithms tried to \emph{iteratively} find the codeword that was sent over
the channel. Secondly, these algorithms operated \emph{locally} in the sense
that they combined partial information that could then be used in other
partial-information combining.

Although revolutionary, these codes and decoding algorithms were forgotten for
a long time. The main reason being that, although these algorithms were
computationally far less demanding than maximum a-posterior decoding (MAPD)
and maximum-likelihood decoding (MLD), they were nevertheless too complex for
that time. Besides some work by Zyablov~\cite{Zyablov:71:1}, Zyablov and
Pinsker~\cite{Zyablov:Pinsker:76:1}, Tanner~\cite{Tanner:81}, and
Margulis~\cite{Margulis:82}, Gallager's ideas lay dormant for about $30$
years. Then, in the mid-1990's, the discovery of turbo codes by Berrou,
Glavieux, and Thitimajshima~\cite{Berrou:Glavieux:Thitimajshima:93}, the
rediscovery of LDPC codes by MacKay and Neal~\cite{MacKay:Neal:96:1,
MacKay:Neal:97:1, MacKay:99:1}, and the work of Wiberg, Loeliger, and
Koetter~\cite{Wiberg:Loeliger:Koetter:95, Wiberg:96} on codes on graphs and
message-passing iterative decoding (MPID) initiated a flurry of research on
iterative decoders and codes amenable to such decoders that continues to these
days. They lead to new and practical approaches not only in communications but
also in signal processing and artifical intelligence. Many of these
developments can be explained nowadays with the help of concepts like the
generalized distributive law as formulated by Aji and
McEliece~\cite{Aji:McEliece:00:1} or factor graphs and the sum-product
algorithm (SPA) by Kschischang, Frey, and
Loeliger~\cite{Kschischang:Frey:Loeliger:01, Loeliger:04:1}.

While MPID has had unparalleled success, it is fair to say that its behavior
for the case of finite-length codes is, at present, not well understood and
many results are based on simulations alone. Before delineating what is known
about the finite-length case, let us however first turn to the infinite-length
case. For LDPC codes with block length going to infinity (where it is assumed
that the length of the smallest cycle in the underlying Tanner graph also goes
to infinity, or where at least the fraction of finite-length cycles vanishes)
it turned out that there is an elegant analysis technique, the so-called
density evolution: this technique was first introduced by Luby et
al.~\cite{Luby:Mitzenmacher:Shokrollahi:Spielman:98} for the binary erasure
channel and then by Richardson, Shokrollahi, and
Urbanke~\cite{Richardson:Urbanke:00:1, Richardson:Shokrollahi:Urbanke:01:1}
for more general channels. These results were very valuable in guiding code
designers how to tweak LDPC codes into well-performing (finite-length)
irregular LDPC codes. There are, however, some drawbacks of these techniques:
firstly, it is not clear, if these results give the best finite-length
irregular codes, and secondly, and more importantly, they do not say if a
specific code exhibits an error floor and if yes, where this error floor is.

Early techniques that tried to tackle the finite-length case focused on
specific families of codes and/or restricted classes of channels. In that
direction, let us mention the analysis of so-called cycle codes\footnote{Cycle
codes are codes with a Tanner graph where all bit nodes have degree two.} by
Wiberg~\cite{Wiberg:96}, tail-biting trellises and graphs with a single cycle
by Anderson and Hladik~\cite{Anderson:Hladik:98:1}, by Aji et
al.~\cite{Aji:Horn:McEliece:98}, and by Forney et
al.~\cite{Forney:Kschischang:Marcus:Tuncel:01:1}. For the binary erasure
channel, influential work was done by Di et
al.~\cite{Di:Proietti:Telatar:Richardson:Urbanke:02:1} utilizing the notion of
{\it stopping sets}. Finally, for more general channels, the idea of
\emph{near-codewords}, \emph{trapping sets}, \emph{extrinsic message degree
(EMD)}, and \emph{instantons} were used by MacKay and
Postol~\cite{MacKay:Postol:03:1}, by Richardson~\cite{Richardson:03:1}, by
Tian et al.~\cite{Tian:Jones:Villasenor:Wesel:04:1, Ramamoorthy:Wesel:04:1},
and by Chernyak et al.~\cite{Chernyak:Chertkov:Stepanov:Vasic:04:1,
Stepanov:Chernyak:Chertkov:Vasic:05:1}, respectively, to empirically
characterize problematic situations for MPID.

A complete understanding of MPID of finite-length codes with finitely many
iterations is essentially given by \emph{computation trees}~\cite{Wiberg:96},
i.e.~by the valid configurations of such computation trees. Some work on
analyzing computation trees was done by Wiberg~\cite{Wiberg:96}, with
subsequent work by Frey et al.~\cite{Frey:Koetter:Vardy:01:1} and Forney et
al.~\cite{Forney:Koetter:Kschischang:Reznik:01:1}. Although this approach is
intuitively very appealing, it seems to be very difficult to get a simple
characterization of the valid configurations on computation trees, a necessary
requirement if one wants to understand MPID. In fact, only extremely simple
codes were analyzed with this technique so far.

Experimental results for codes of reasonable length and rate show that
decision boundaries can be of a rather complex nature, a fact that makes the
above-mentioned problems in trying to analyze the valid configurations on
computation trees not completely unexpected. A complete understanding of MPID
of a given code is probably an illusionary task, therefore we will settle here
for a more modest goal.

In this paper we present an analysis technique for MPID of a given
code. Although the underlying principle of our analysis technique is very
simple, experimentally it seems to give very good predictions of the decoding
behavior; in fact, it gives the correct answers for all the cases where MPID
behavior is understood analytically. The predicted decision boundaries are
hyperplanes in the log-likelihood ratio vector space and it turns out that the
decision boundaries are exactly the same as the ones under so-called linear
programming decoding (LPD) that was recently introduced by Feldman,
Wainwright, and Karger~\cite{Feldman:03:1, Feldman:Wainwright:Karger:05:1}. In
the light of this coincidence one might actually argue that the various MPID
algorithms are nothing else than low-complexity, very efficient, and
aggressive LP solvers that most of the time ``decide'' for the same
(pseudo-)codeword as LPD, but not always.\footnote{When LPD decides for a
pseudo-codeword that is not a codeword, the dynamical behavior of MPID depends
very much on the type of the MPID under consideration.} We have done some work
towards showing the nearness of min-sum algorithm (MSA) decoding and
LPD~\cite{Vontobel:Koetter:04:2} but in this paper we will not discuss this
aspect any further.

The analysis technique that was mentioned in the previous paragraph will be
called graph-cover decoding (GCD): its name stems from the fact that during
GCD all codewords in all finite covers of a given Tanner graph are competing
to be the best explanation of the received vector. Analyzing all the codes in
all the finite covers seems at first to be an infeasible task. However, it
turns out that they can be characterized by the so-called fundamental
polytope. Among other things, we will see in this paper how this fundamental
polytope unifies the notions of stopping sets, pseudo-codewords,
near-codewords, and trapping sets.\footnote{For more references on these
topics, see also~\cite{Pseudocodewords:Website:05:1}.}

The outline of this paper is as follows. In
Sec.~\ref{sec:motivating:example:1} we will discuss the iterative decoding of
a simple code and show the underlying philosophy behind our analysis
technique. After some notational remarks in Sec.~\ref{sec:notation:1}, the
main part of the paper starts in
Sec.~\ref{sec:graph:covers:and:fundamental:polytope:1} which introduces graph
covers and the fundamental polytope. In Sec.~\ref{sec:channels:and:MAPD:LPD:1}
we review MAPD/MLD of codes and by considering relaxations of optimization
problems we make the link to LPD. Then, in
Sec.~\ref{sec:graph:cover:decoding:1} we will show that GCD is essentially
equivalent to LPD and we will see how GCD can be seen as a model for MPID.
Whereas Sec.~\ref{sec:fp:description:1} will discuss various descriptions and
properties of the fundamental polytope and cone,
Sec.~\ref{sec:pseudo:weight:1} will focus on a variety of pseudo-weights and
their properties. A simple upper bound on the AWGNC pseudo-weight will be
presented in Sec.~\ref{sec:pw:upper:bound:1} which implies a sub-linear
asymptotic behavior of the AWGNC pseudo-weight for any family of regular LDPC
codes (under some mild conditions). Finally, in
Sec.~\ref{sec:relationship:fundamental:polytope:other:concepts:1} we explain
the relationship of GCD to other concepts that have been used in the past to
explain the finite-length behavior of MPID, and in
Sec.~\ref{sec:conclusions:1} we offer some conclusions and mention some open
problems.


\subsection{Motivating Example}
\label{sec:motivating:example:1}

Because we are using binary codes, we can without loss of optimality assume
that a decoding algorithm bases its decision on the log-likelihood ratio (LLR)
vector which is given by the observed channel output sequence. The
understanding of a particular decoding algorithm is then tightly related to
the understanding the decision regions in the space of LLR vectors. While the
visualization of decision regions is a very intuitive way of showing how a
decoder works (and of showing differences between different decoders), it is
usually infeasible to show all the aspects of the decision regions since
practical codes have a length of several tens of bits to several ten thousands
of bits which implies that the space of LLR vectors has a dimension of several
tens to several ten thousands.

However, some of the key differences between MAPD/MLD and iterative decoding
can already be seen for very short codes. The aim of this section is to
discuss such a very short code and to introduce an approximate analysis based
on graph covers that explains the main characteristics of the decision regions
of iterative decoding like sum-product algorithm (SPA) and the min-sum
algorithm algorithm (MSA) decoding. (Note that the notation that we will use
in this subsection will be properly introduced in Sec.~\ref{sec:notation:1}
and in later sections.)

\begin{figure}
  \begin{center}
    \epsfig{file=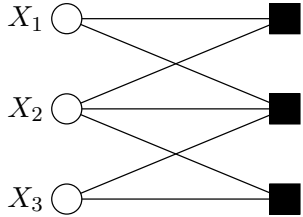}
  \end{center}

  \caption{Tanner graph $\graph{T}$ of the length-$3$ code under
    consideration.}

  \label{fig:trivial:code:tg:1:1}
\end{figure}

\begin{figure}
  \begin{center}
    \epsfig{file=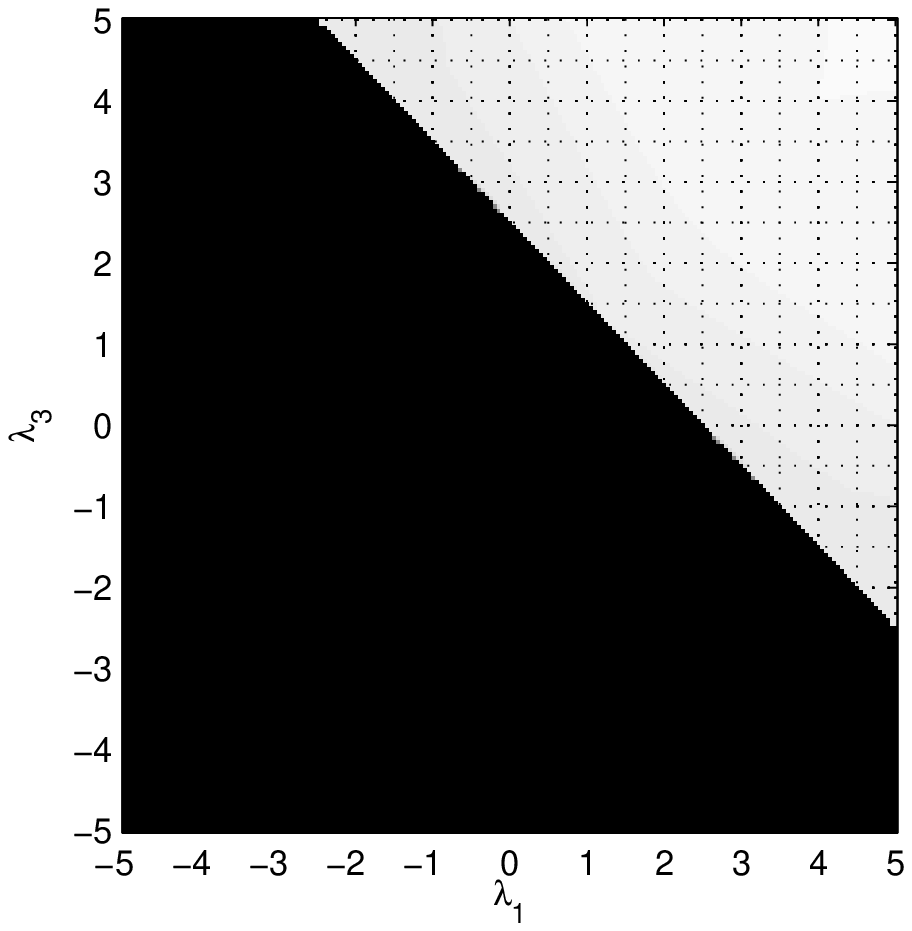,
            width=5cm}
    \epsfig{file=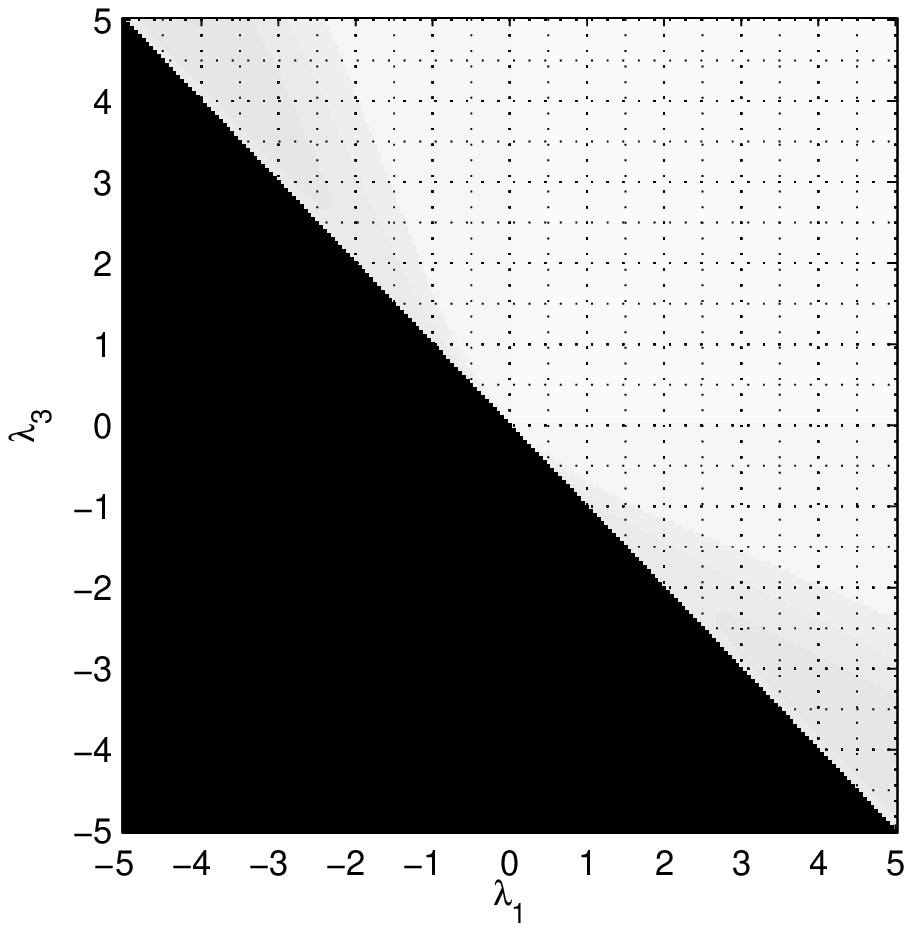,
            width=5cm}
    \epsfig{file=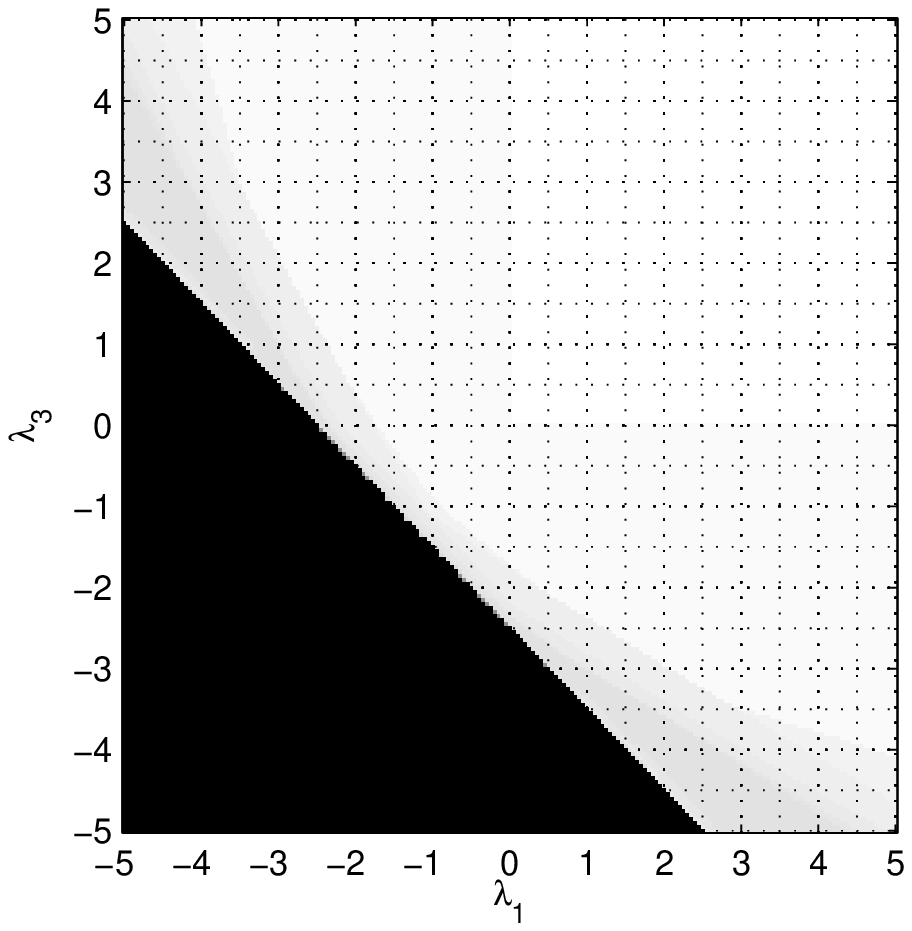,
            width=5cm}
  \end{center}

  \caption{SPA decoding with maximally $60$ iterations of the code $\code{C}$
    that is represented by the Tanner graph $\graph{T}$ in
    Fig.~\ref{fig:trivial:code:tg:1:1}. The gray-scale indicates after how
    many iterations the algorithm converged to the all-zeros codeword with the
    implication that in the black region the decoder did not converge (see
    text for more details). From left to right: $(\lambda_1, \lambda_3)$-plane
    for $\lambda_2 = -2.5, \, 0, \, +2.5$.}

  \label{fig:trivial:code:conv:time:1:1}
\end{figure}

We consider a code $\code{C}$ of length $n = 3$ defined by the parity-check
matrix
\begin{align}
  \matr{H}
    &\defeq
       \begin{pmatrix}
         1 & 1 & 0 \\
	 1 & 1 & 1 \\
	 0 & 1 & 1 \\
       \end{pmatrix},
         \label{eq:trivial:code:pcm:1}
\end{align}
whose Tanner graph $\graph{T} \defeq \tgraph{T}{H}$ is depicted in
Fig.~\ref{fig:trivial:code:tg:1:1}. Because $\matr{H}$ has rank $3$, the
dimension of the code is $0$ and therefore $\code{C}$ contains only one
codeword:
\begin{align*}
  \code{C}
    &\defeq
       \left\{
         (x_1,x_2,x_3) \in \GF{2}^3
         \ 
         \left|
         \
           (x_1,x_2,x_3) \cdot \matr{H}^\tr = \vect{0}
         \right.
       \right\}
     = \big\{
         (0, 0, 0)
       \big\}.
\end{align*}
While it, at first, may seem strange to consider a zero-rate code, it is
indeed an ideal candidate to investigate problematic behaviors of iterative
decoding. Assume that we are using the code for data transmission over an
additive white Gaussian noise channel (AWGNC) and that the LLR vector is
$\vlambda = (\lambda_1, \lambda_2, \lambda_3)$.

Consider first block-wise MAPD (which is equivalent to block-wise MLD since we
assume that all codewords are transmitted equally likely). It is immediately
apparent that for such a decoder there is only one decision region: we decide
$\hvx = (0,0,0)$ independently of $\vlambda$.\footnote{Note that using a
\emph{symbol-wise} maximum a-posteriori decoder has also only one decision
region: we decide for $\hat x_1 = 0$, $\hat x_2 = 0$, $\hat x_3 = 0$
independently of $\vlambda$.}

We now turn to MPID, more precisely decoding based on the SPA and
MSA~\cite{Kschischang:Frey:Loeliger:01} where one iteration consists in
updating the messages at all variable nodes and then updating the messages at
all check nodes. The SPA decoding convergence behavior as a function of
$(\lambda_1, \lambda_2, \lambda_3)$ is depicted in
Fig.~\ref{fig:trivial:code:conv:time:1:1}: the gray-scale indicates after how
many iterations the SPA converged to the all-zeros codeword.

In practical applications, the SPA and the MSA are performed for a certain
pre-defined number of iterations. The binary vector that is obtained at the
end of these iterations is then considered to be the decision on the
transmitted codeword. Very often, the following termination rule is used
additionally: the algorithm terminates if the binary vector found by the
algorithm is a codeword, i.e.~the syndrome is the all-zeros vector.

However, for our investigations of the code $\code{C}$ we did not adopt this
latter termination rule: the reason is that there are only eight binary
vectors of length $3$ and therefore it is not unlikely that at some point the
algorithm obtains the all-zeros vector even if the internal state of the
iterative process has not converged to a stable point.\footnote{For reasonably
long codes this is hardly an issue. E.g.~for a rate-$1/2$ code of length
$200$, the probability that the algorithm accidentally finds a codeword is
$2^{100}/2^{200} = 2^{-100}$.} So, for obtaining the plots in
Fig.~\ref{fig:trivial:code:conv:time:1:1} we did the following: for each
$(\lambda_1, \lambda_2, \lambda_3)$ point we performed $60$ iterations of the
SPA and we considered the algorithm to have converged once the decision vector
remained the all-zeros codeword over subsequent
iterations. Fig.~\ref{fig:trivial:code:conv:time:1:1} shows then the decision
regions and the convergence times under SPA decoding after performing $60$
iterations. It is evident that these decision regions are clearly
\emph{different} from the decision regions for block-wise MAPD/MLD! Indeed,
the plots in Fig.~\ref{fig:trivial:code:conv:time:1:1} suggest that there is a
decision boundary described by the equation $\lambda_1 + \lambda_2 + \lambda_3
= 0$: for $\lambda_1 + \lambda_2 + \lambda_3 > 0$ the SPA does converge and
for $\lambda_1 + \lambda_2 + \lambda_3 < 0$ the SPA does not converge to the
all-zeros codeword.

How can these differences in the decision regions between MAPD/MLD on the one
hand and MPID on the other hand be explained? In this paper we argue that the
key difference between the block-wise MAPD/MLD (or symbol-wise MAPD/MLD) and
any MPID algorithm is the following: whereas the former algorithms use
\emph{global} information and constraints to find the optimal solution, the
latter algorithms base their decisions on information that was gathered by
processing information \emph{locally}. This locality, which on one hand leads
to huge savings in terms of the number of computations needed, is on the other
hand also the main weakness of any MPID algorithm.

\begin{figure}
  \begin{center}
    \begin{minipage}[c]{0.3\linewidth}
      \begin{center}
        \epsfig{file=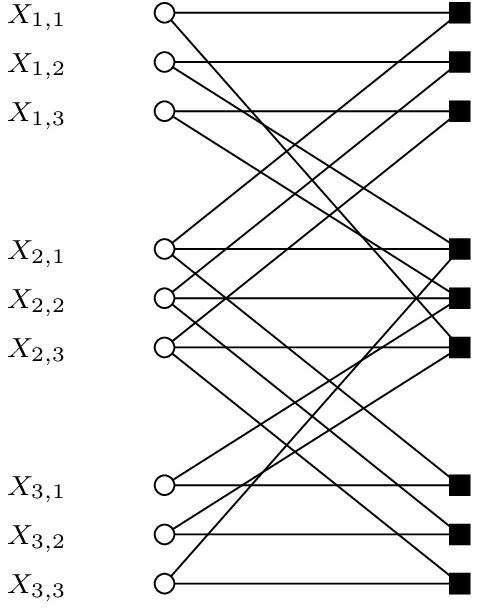, height=4cm}
      \end{center}
    \end{minipage}
    \quad\quad\quad
    \begin{minipage}[c]{0.3\linewidth}
      \begin{center}
        \epsfig{file=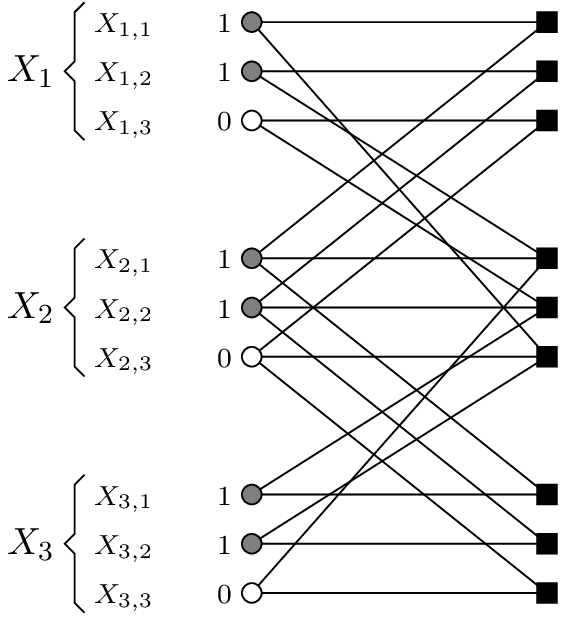, height=4.2cm}
      \end{center}
    \end{minipage}
  \end{center}

  \caption{Left: a possible triple cover $\cgraph{T}$ of the Tanner graph
    $\graph{T}$. Right: a non-zero codeword of the code defined by
    $\cgraph{T}$.}
  
  \label{fig:trivial:code:1:2}
\end{figure}

\begin{figure}
  \begin{center}
    \epsfig{file=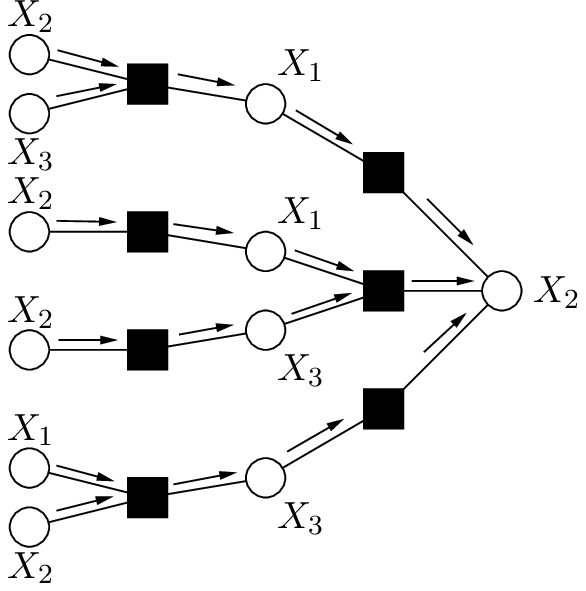, width=4cm}
    \quad\quad
    \epsfig{file=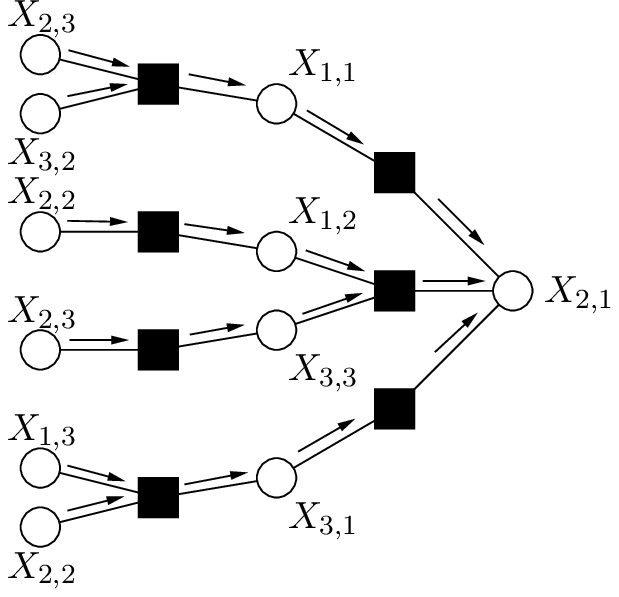, width=4cm}
  \end{center}
  \caption{Left: computation tree with root $X_2$ after two iterations 
           when decoding code $\code{C}$. Right: computation tree with root
           $X_{2,1}$ after two iterations when decoding code $\ccode{C}$.}
  \label{fig:trivial:code:1:3}
\end{figure}

Let us briefly outline how we will use this global-vs-local perspective to
obtain an unterstanding of the differences between MAPD/MLD and MPID.  Consider
the code $\ccode{C}$ of length $9$ that is defined by the Tanner graph
$\cgraph{T}$ in Fig.~\ref{fig:trivial:code:1:2} (left). Assume that we use
this code for data transmission over an AWGNC and assume that at the receiver
the hypothetical LLR vector is
\begin{align*}
  \tvlambda
    &\defeq (\lambda_{1,1}{:}\lambda_{1,2}{:}\lambda_{1,3},\
             \lambda_{2,1}{:}\lambda_{2,2}{:}\lambda_{2,3},\
             \lambda_{3,1}{:}\lambda_{3,2}{:}\lambda_{3,3}).
\end{align*}
In the same way that we used the SPA for decoding the code $\code{C}$ whose
Tanner graph is shown in Fig.~\ref{fig:trivial:code:tg:1:1}, we can use the
analogous message-passing-based decoding algorithm for decoding the code
$\ccode{C}$.

For both cases we can draw the computation trees~\cite{Wiberg:96}:
Fig.~\ref{fig:trivial:code:1:3} (left) shows the computation tree with root
$X_2$ after two iterations when decoding code $\code{C}$ whereas
Fig.~\ref{fig:trivial:code:1:3} (right) shows the computation tree with root
$X_{2,1}$ after two iterations when decoding code $\ccode{C}$. The topological
equivalence with the computation tree in Fig.~\ref{fig:trivial:code:1:3}
(left) might at first appear as a coincidence. However, this is not a
coincidence. The reason is that the Tanner graph $\cgraph{T}$ has a special
relationship with respect to the Tanner graph $\graph{T}$; in fact,
$\cgraph{T}$ is a so-called $3$-cover of $\graph{T}$. This means that
$\cgraph{T}$ has three times more nodes but locally it is indistinguishable
from $\graph{T}$.

Moreover, if we assume that
\begin{align*}
  \tvlambda
    &= (\lambda_1{:}\lambda_1{:}\lambda_1,\
        \lambda_2{:}\lambda_2{:}\lambda_2,\ 
        \lambda_3{:}\lambda_3{:}\lambda_3)
\end{align*}
then not only are the computation trees topologically equivalent, but also the
messages are identical! Therefore, for this special choice of $\tvlambda$ (in
relation to a given $\vlambda$), the message-passing-based decoding algorithm
cannot distinguish if it is decoding code $\code{C}$ or $\ccode{C}$. In fact,
it cannot distinguish if it is decoding code $\code{C}$ or any code defined by
any graph cover of $\graph{T}$. The harmful effect of the codes that are given
by the graph covers is that they contain codewords that {\em cannot} be
explained as liftings of codewords in $\code{C}$. E.g.~code $\ccode{C}$
contains the codeword $(0{:}0{:}0,\ 0{:}0{:}0,\ 0{:}0{:}0)$ which is a lifting
of the codeword $(0,0,0)$ in $\code{C}$. However, code $\ccode{C}$ contains
also the codeword $(1{:}1{:}0,\ 1{:}1{:}0,\ 1{:}1{:}0)$,
cf.~Fig.~\ref{fig:trivial:code:1:2} (right), which is {\em not} a lifting of a
codeword in $\code{C}$.\footnote{In total, $\ccode{C}$ contains four
codewords, three of them are not liftings of any codeword of $\code{C}$.}

We emphasize two crucial observations:
\begin{itemize}

  \item In principle, locally operating decoding algorithms {\em cannot
    distinguish} if they are operating on a Tanner graph $\graph{T}$ or any
    finite cover of this graph as, for example, the cubic cover depicted in
    Fig.~\ref{fig:trivial:code:1:2} (left).

  \item In general, the binary codes defined by finite covers of a Tanner
    graph support codewords that are not liftings of codewords in the original
    Tanner graph. Such a codeword is indicated in
    Fig.~\ref{fig:trivial:code:1:2} (right) for the cubic cover in
    Fig.~\ref{fig:trivial:code:1:2} (left).
\end{itemize}

It is clear, that any locally operating MPID will automatically take into
account all possible codewords in all finite graph covers of the original
graph. In other words, whereas in MAPD/MLD decoding all the codewords are
competing to be the best explanation of the received vector, under MPID all
codewords in all finite graph covers compete to be the best explanation of the
received vector. In the case of our example code, the existence of non-zero
codewords in finite covers of the original graph explains to large extents the
observed behavior of SPA- and MSA-based decoding: indeed, for the specific
code at hand it can be shown that \emph{any} non-zero codeword in a finite
cover of $\graph{T}$ (like the codeword in the triple cover shown in
Fig.~\ref{fig:trivial:code:1:2} (right)) has the same effect as a virtually
present, all-one codeword.

At first glance it seems to be a formidable task to characterize all possible
codewords being introduced by the union of finite covers of any degree. (The
number of finite covers of a graph grows faster than exponential with the
covering degree). However, it turns out that this becomes an object that
itself is elegantly described and compactly represented in the original Tanner
graph.

Let us emphasize that this paper uses graph covers as an \emph{analysis}
technique. In the past, there have been various researchers who have used
graph covers (sometimes also called graph liftings) but they used them for
\emph{constructing} LDPC codes that have some desirable symmetries, see
e.g.~Tanner et al.~\cite{Tanner:99:2, Tanner:Sridhara:Fuja:01:1}.

\begin{figure}
  \begin{center}
    \epsfig{file=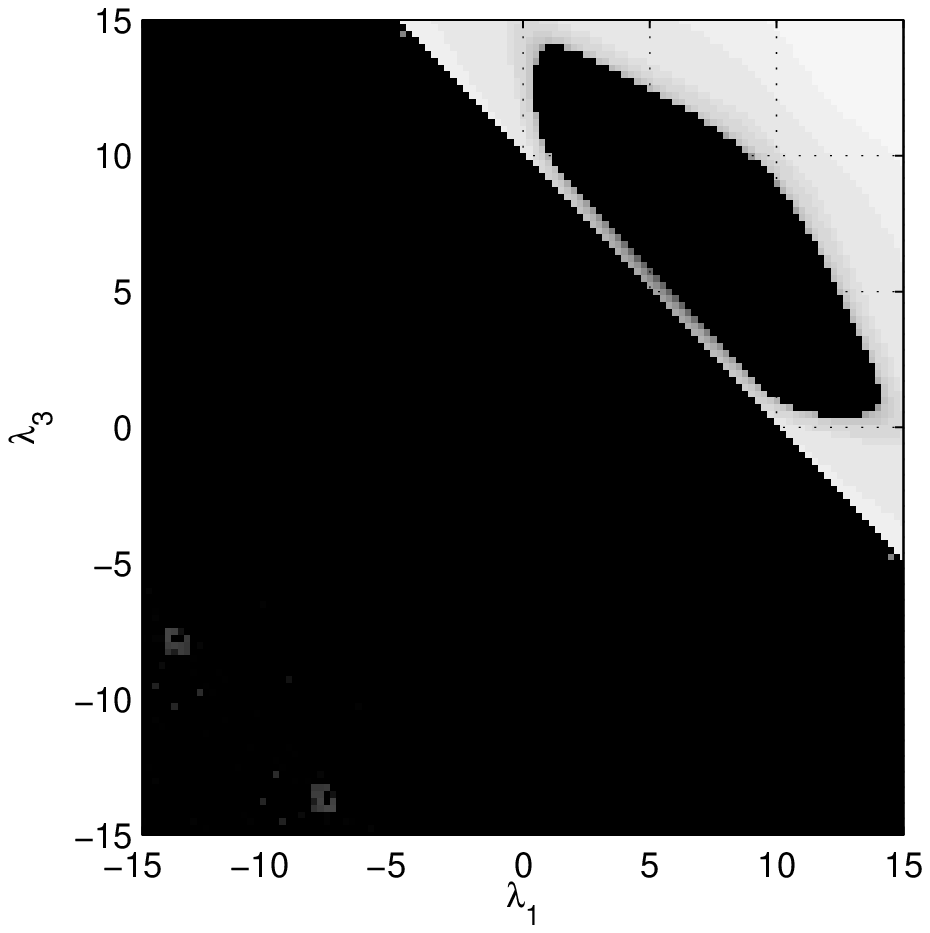,
            width=5cm}
    \
    \epsfig{file=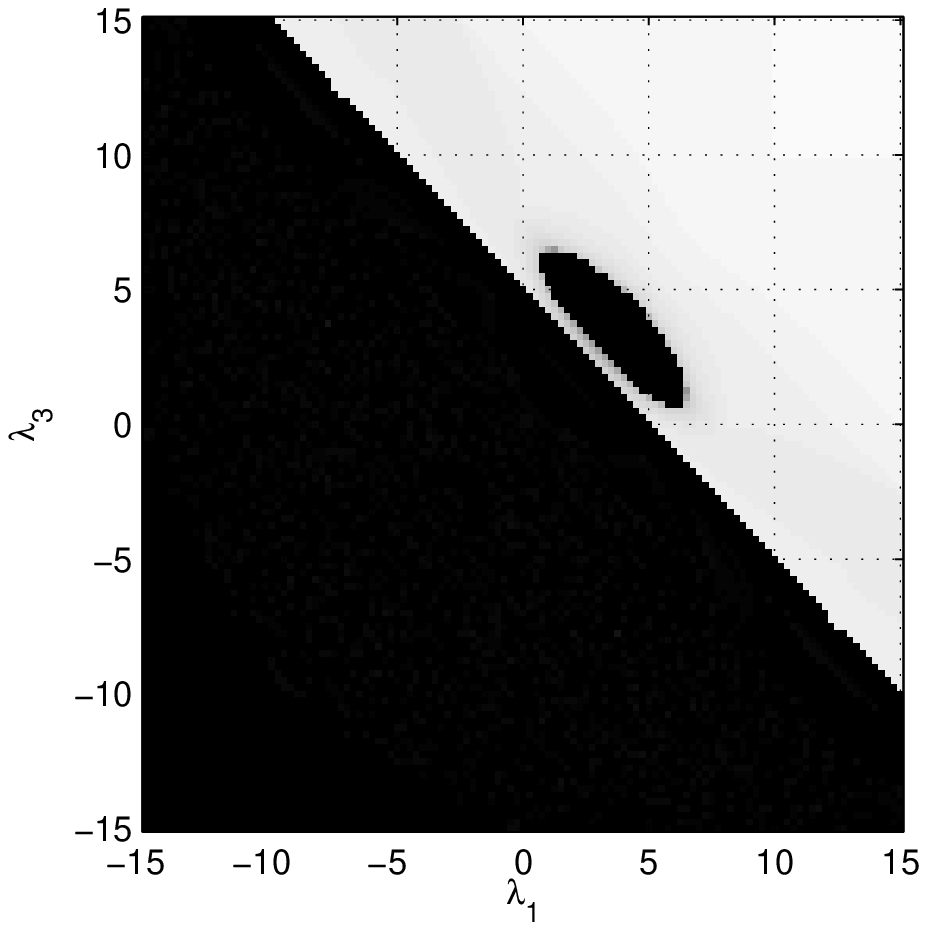,
            width=5cm}
    \
    \epsfig{file=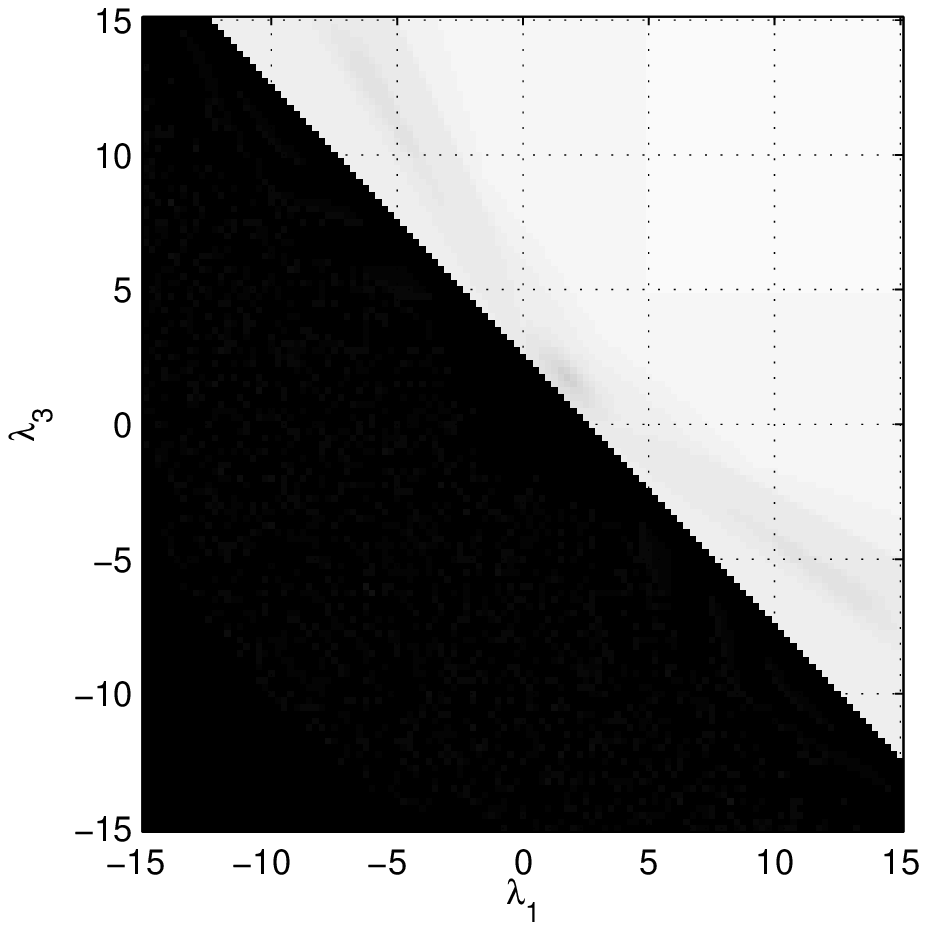,
            width=5cm} \\
    \epsfig{file=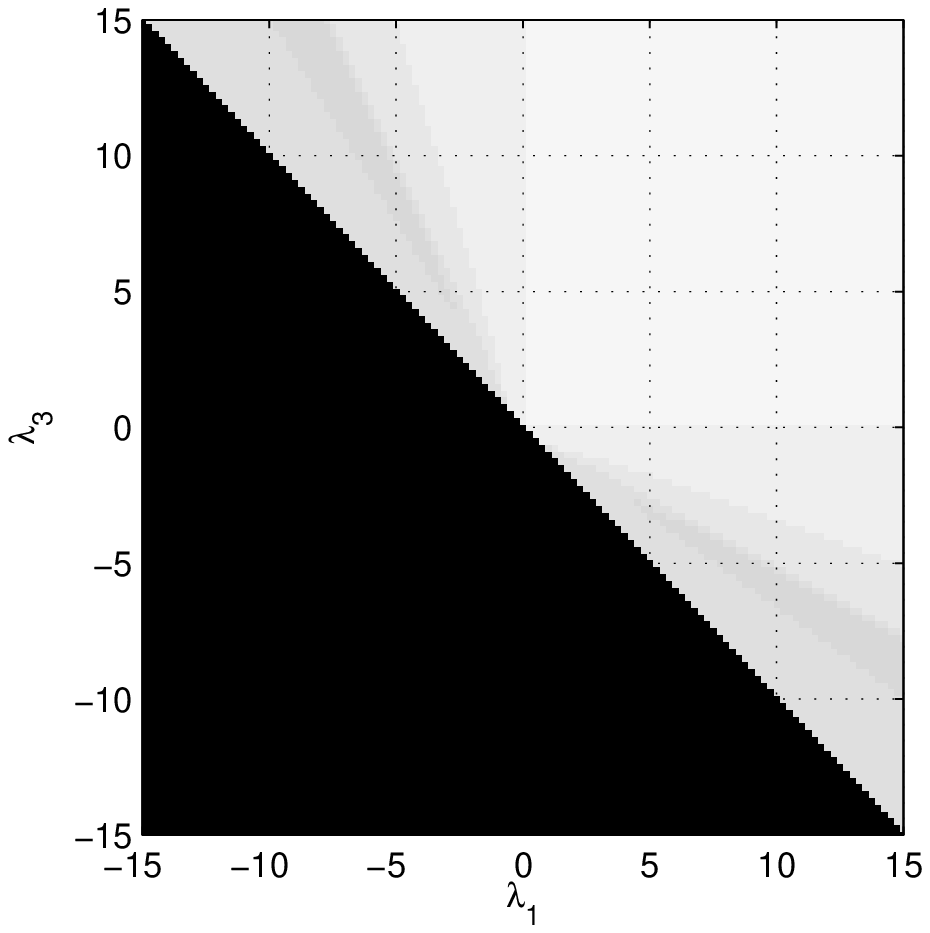,
            width=5cm} \\
    \epsfig{file=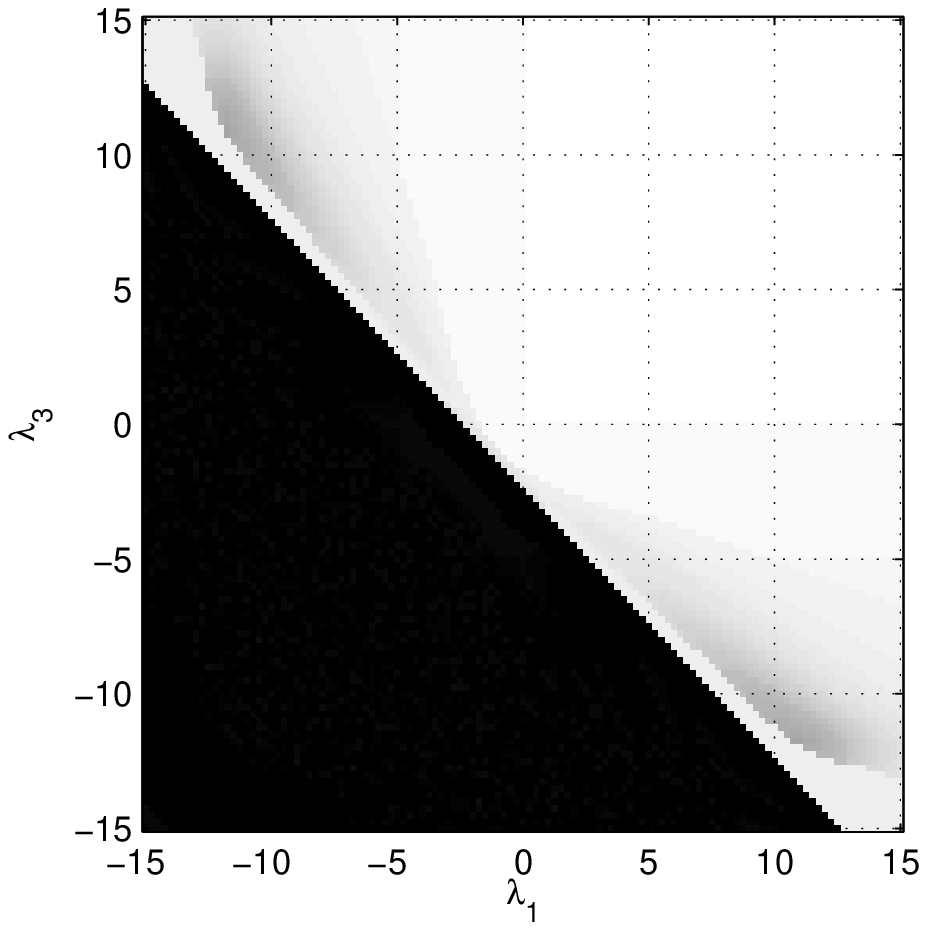,
            width=5cm}
    \
    \epsfig{file=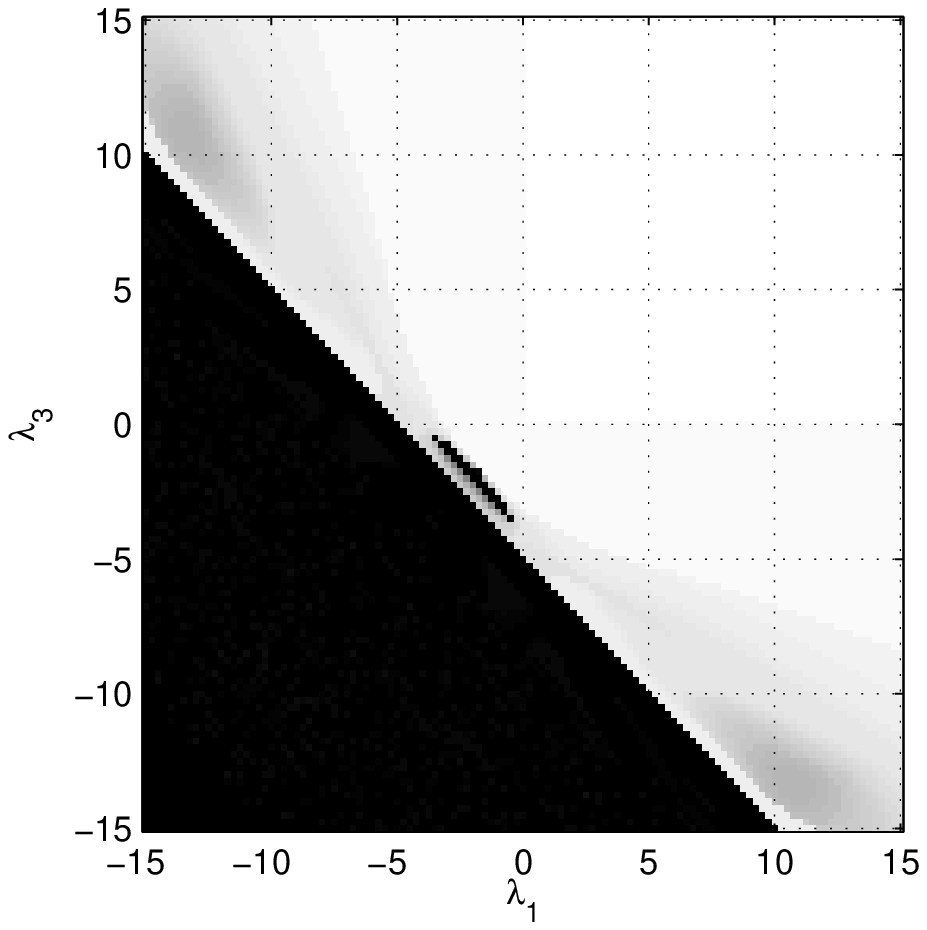,
            width=5cm}
    \
    \epsfig{file=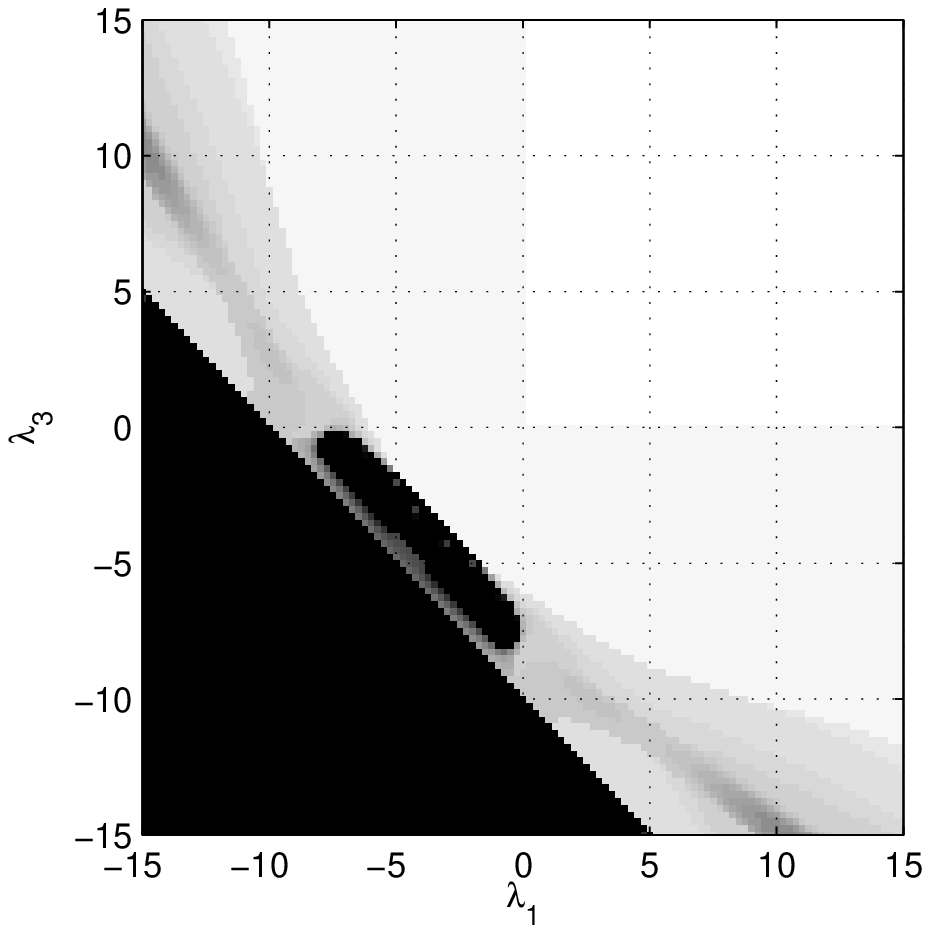,
            width=5cm}
  \end{center}

  \caption{SPA decoding with maximally $60$ iterations of the code $\code{C}$
    that is represented by the Tanner graph $\graph{T}$ in
    Fig.~\ref{fig:trivial:code:tg:1:1}. The gray-scale indicates after how
    many iterations the algorithm converged to the all-zeros codeword with the
    implication that in the black region the decoder did not converge (see
    text for more details). From top-left to bottom-right: $(\lambda_1,
    \lambda_3)$-plane for $\lambda_2 = -10, \, -5, \, -2.5, \, 0, \, +2.5, \,
    +5, \, +10$.}

  \label{fig:trivial:code:conv:time:1:2}
\end{figure}

\begin{figure}
  \begin{center}
    \epsfig{file=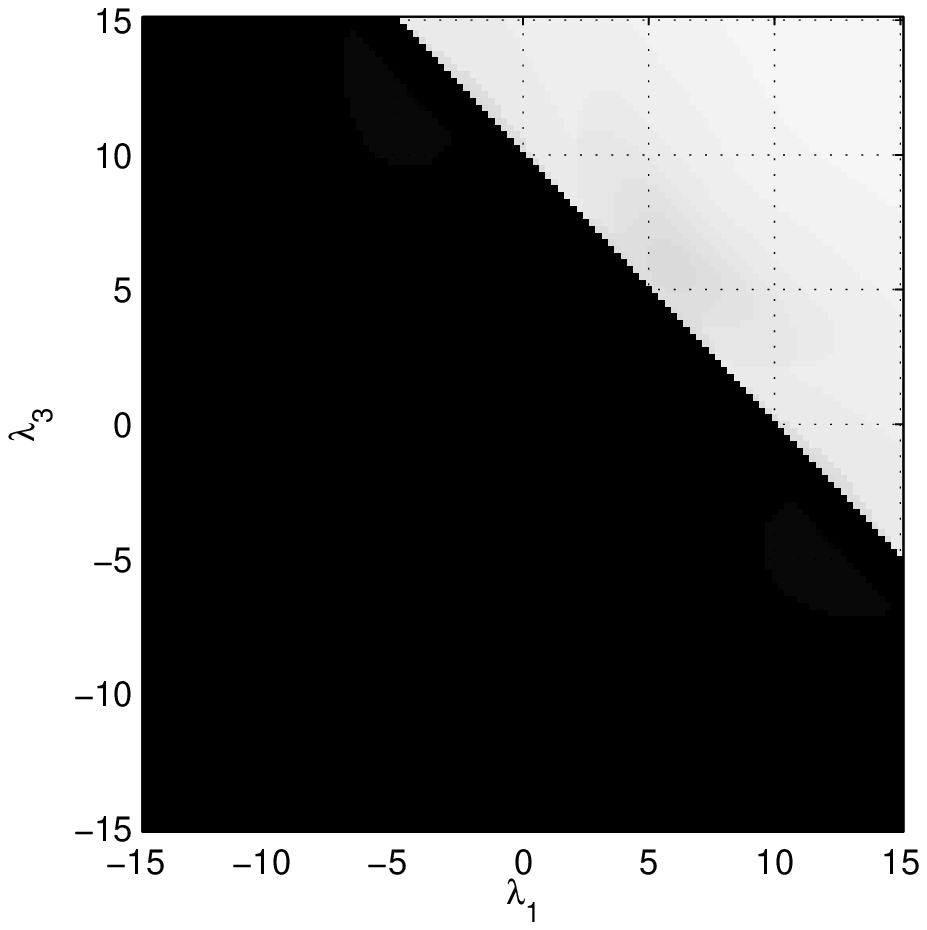,
            width=5cm}
    \
    \epsfig{file=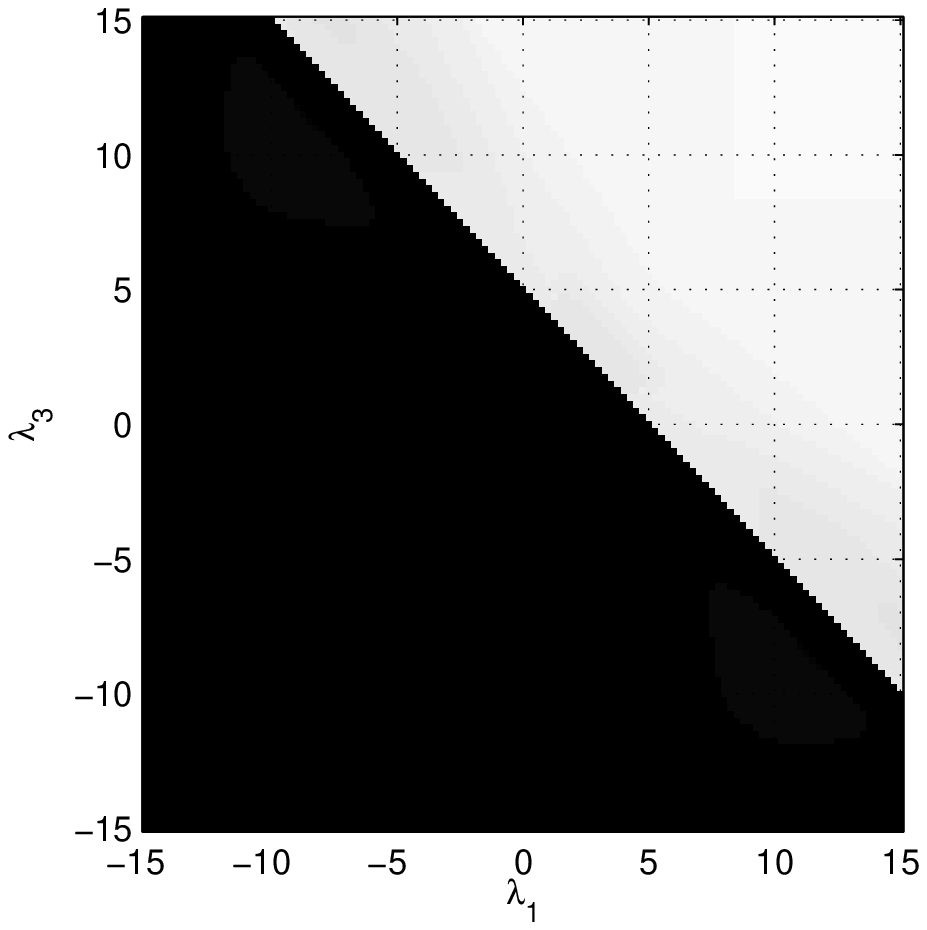,
            width=5cm}
    \
    \epsfig{file=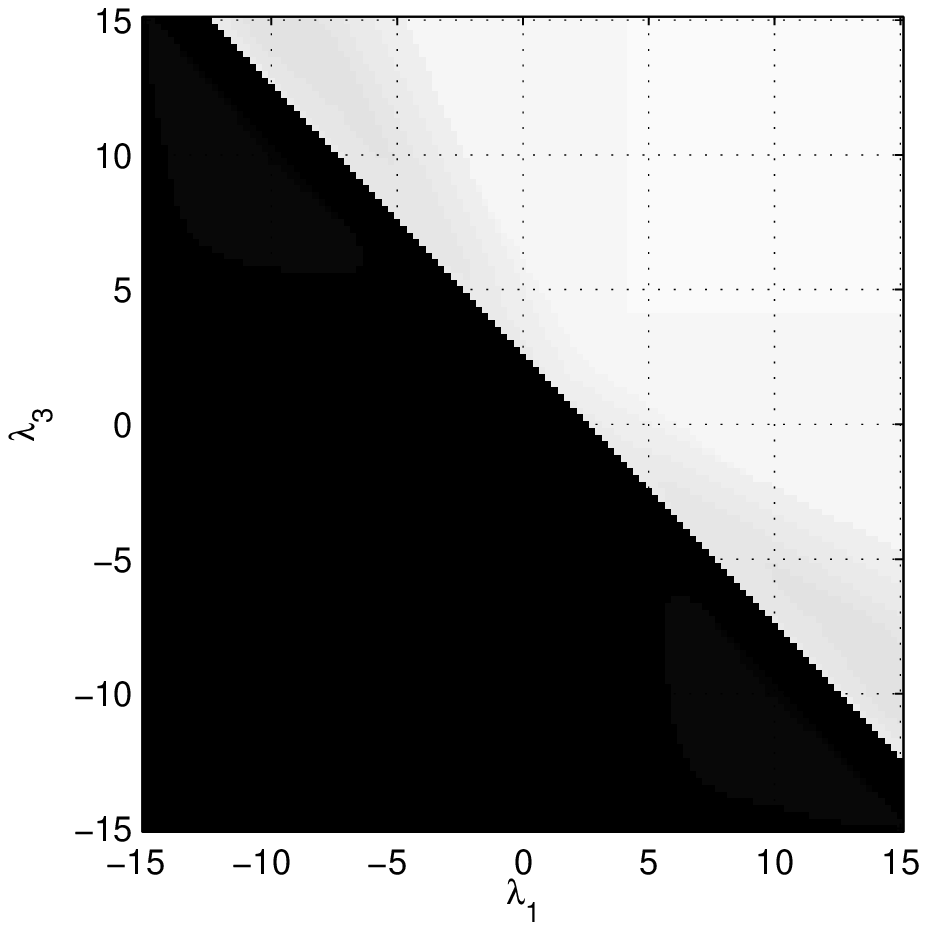,
            width=5cm} \\
    \epsfig{file=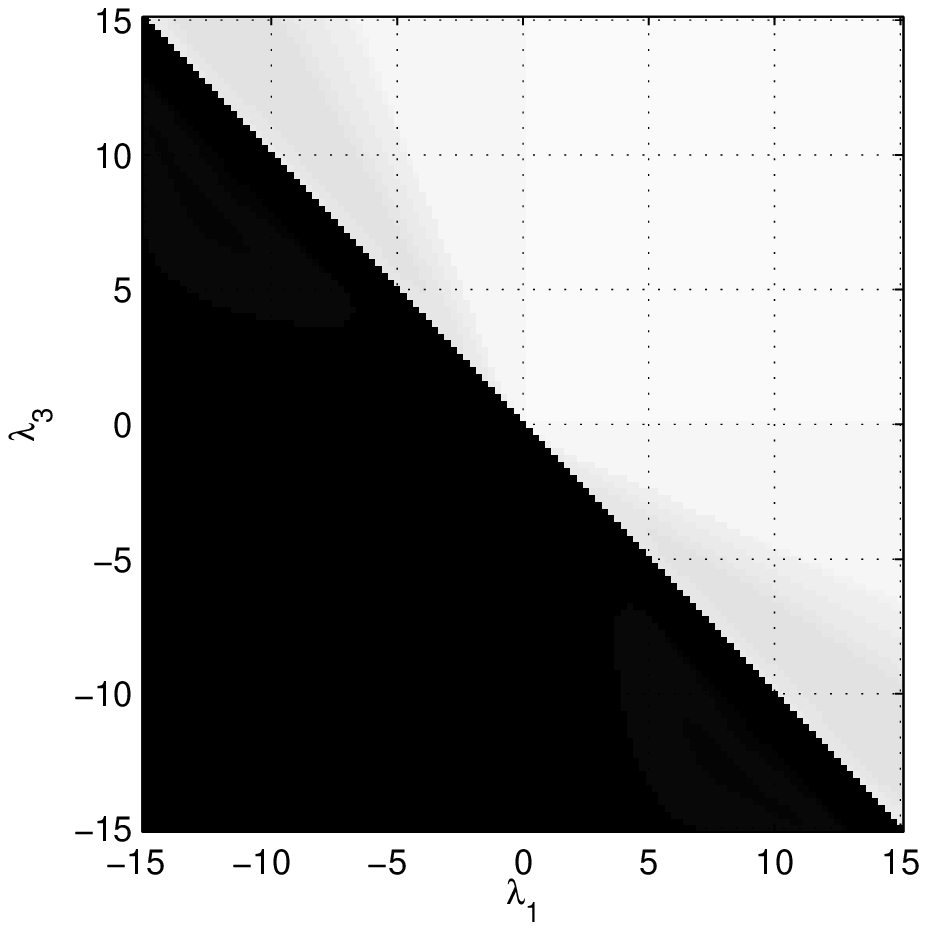,
            width=5cm} \\
    \epsfig{file=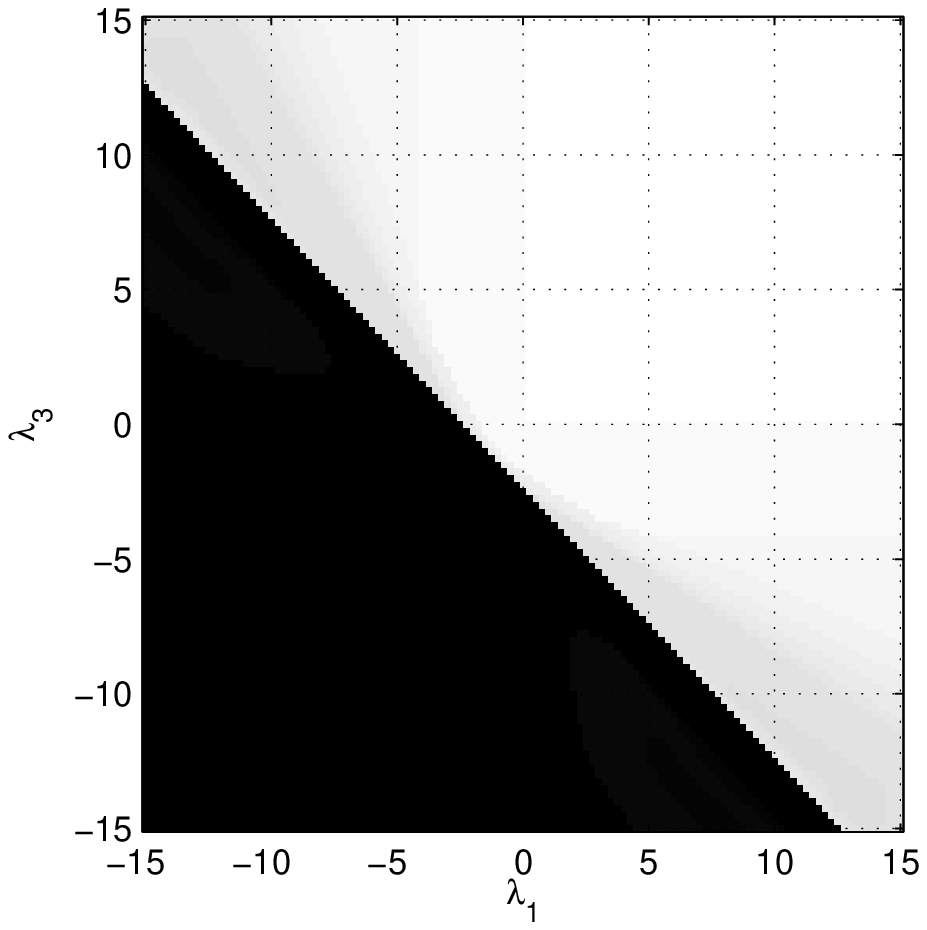,
            width=5cm}
    \
    \epsfig{file=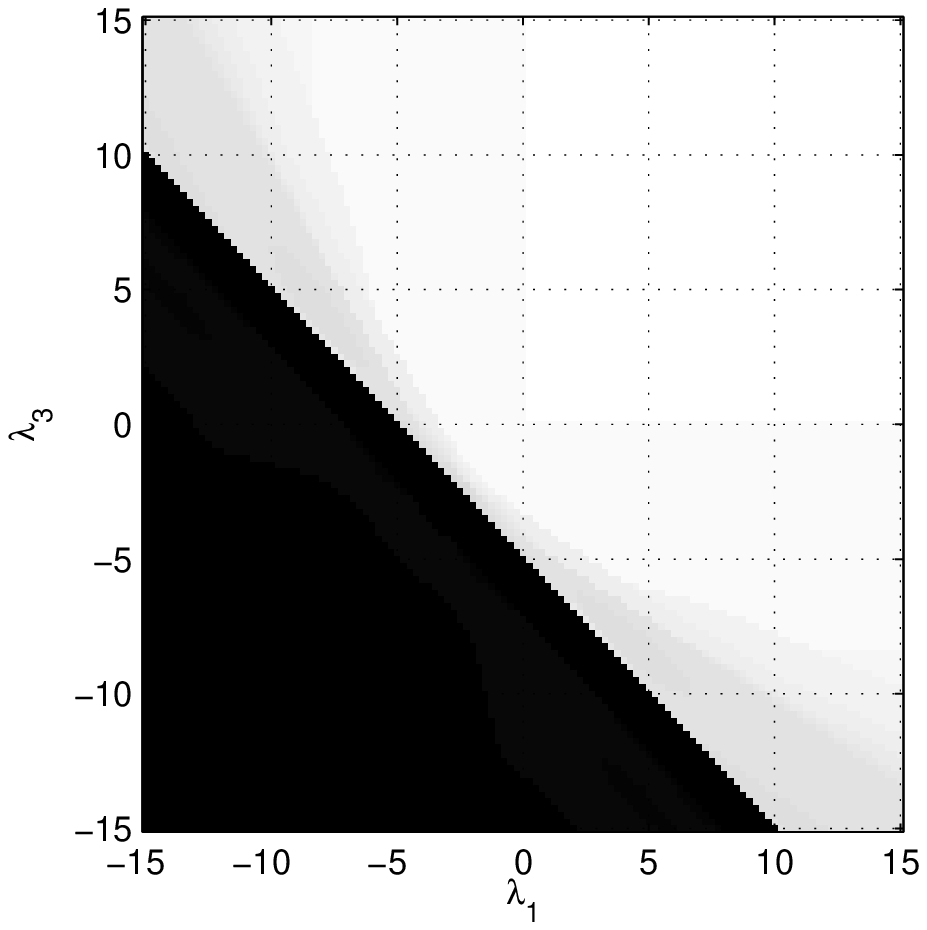,
            width=5cm}
    \
    \epsfig{file=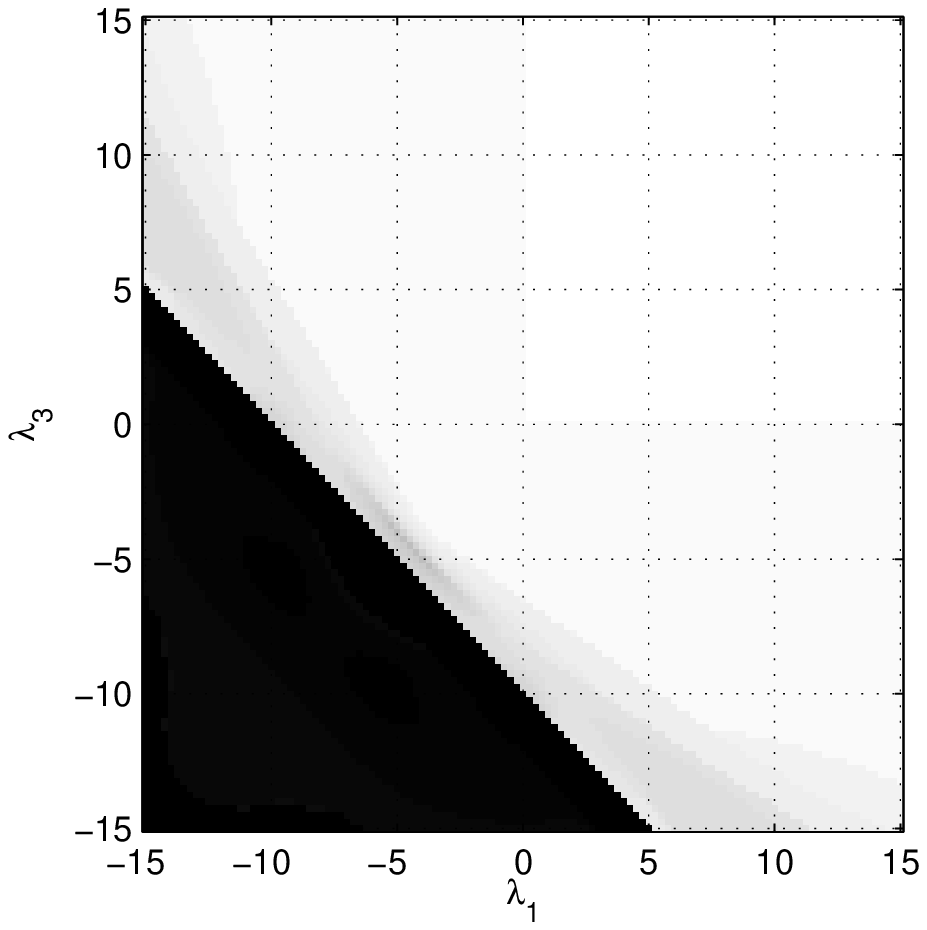,
            width=5cm}
  \end{center}

  \caption{Modified SPA decoding ($\alpha = 0.85$) with maximally $60$
    iterations of the code $\code{C}$ that is represented by the Tanner graph
    $\graph{T}$ in Fig.~\ref{fig:trivial:code:tg:1:1}. The gray-scale
    indicates after how many iterations the algorithm converged to the
    all-zeros codeword with the implication that in the black region the
    decoder did not converge (see text for more details). From top-left to
    bottom-right: $(\lambda_1, \lambda_3)$-plane for $\lambda_2 = -10, \, -5,
    \, -2.5, \, 0, \, +2.5, \, +5, \, +10$.}

  \label{fig:trivial:code:conv:time:1:3}
\end{figure}

Before concluding this motivating example let us mention some unexplained
behaviour of SPA decoding for larger LLR values, see
Fig.~\ref{fig:trivial:code:conv:time:1:2}. Besides the decision boundary
$\lambda_1 + \lambda_2 + \lambda_3 = 0$ that we have already discussed above,
there appears an oval-shaped region where the SPA seems to have a problem in
converging to the all-zeros codeword. Upon applying a slight modification to
the SPA decoder, these oval-shaped regions disappear however, see
Fig.~\ref{fig:trivial:code:conv:time:1:3}. The modification that we applied
was the following. Letting $\vmu^{(t)}$ and $\tilde \vmu^{(t)}$ be the LLR
messages at iteration $t$ from the bit nodes to the check nodes and from check
to bit nodes, respectively, the usual SPA message updates can be written as
$\vmu^{(t)} := f(\tilde \vmu^{(t-1)},\vlambda)$, \ $\tilde \vmu^{(t)} :=
\tilde f(\vmu^{(t)})$ for some suitably chosen functions $f$ and $\tilde
f$. The modified SPA message update rules are then $\vmu^{(t)} := \alpha
\cdot f(\tilde \vmu^{(t-1)},\vlambda) + (1-\alpha) \cdot \vmu^{(t-1)}$, \
$\tilde \vmu^{(t)} := \tilde f(\vmu^{(t)})$ for some $\alpha$ where $0 \leq
\alpha \leq 1$.\footnote{Let us mention that while disussing trapping sets and 
their influence, Laendner and Milenkovic~\cite{Laendner:Milenkovic:05:1}
observed a similar slight change in behavior upon modifying the SPA
slightly. However, whereas they are ``averaging'' the probability messages, we
are ``averaging'' the LLR messages.} Note that this modified SPA still
operates locally and so it cannot distinguish if it is decoding the code
described by the base Tanner graph or any of the codes described by the finite
covers of the base Tanner graph.

\subsection{Notation}
\label{sec:notation:1}

This section discusses the various notations that we will use in this
paper. We start with some sets. We let $\Int$, $\Intp$, $\Intpp$, $\Q$, $\Qp$,
$\Qpp$, $\R$, $\Rp$, and $\Rpp$ be the set of integers, the set of
non-negative integers, the set of positive integers, the set of quotients, the
set of non-negative quotients, the set of positive quotients, the set of real
numbers, the set of non-negative real numbers, and the set of positive real
numbers, respectively. We let $\GF{2} \defeq \{ 0, 1 \}$ be the Galois field
with two elements; as a set, $\GF{2}$ will be considered as a subset of
$\R$. The size of a set $\set{S}$ is denoted by $\card{\set{S}}$.

In the following, all scalars, entries of vectors, and entries of matrices
will be considered to be in $\R$, unless noted otherwise. So, if an addition
or a multiplication is not in the real field, we will indicate this, e.g.~by
writing $a + b \text{ (in $\GF{2}$)}$ or $\vect{a} + \vect{b} \text{ (in
$\GF{2}$)}$. Moreover, when $\set{T} \in \Int^N$ and $\set{S} \subseteq
\GF{2}^N$ then an expression like $\set{T} \subseteq \set{S} \text{ (in
$\GF{2}$)}$ means that $\vect{t} \ (\operatorname{mod} 2)$ lies in $\set{S}$
for all $\vect{t} \in \set{T}$. As usually done in coding theory, we use only
row vectors. An inequality of the form $\vect{a} \geq \vect{b}$ involving two
vectors of length $N$ is to be understood component-wise, i.e.~$a_i \geq b_i$
for all $1 \leq i \leq N$. We let $\vect{1}_N$ be the row-vector of length $N$
and the matrix $\matrunity_{N}$ be the identity matrix of size $N \times N$;
when the length (size) of this vector (matrix) are obvious from the context,
we will omit the index. The support $\supp(\vx)$ of a vector will be the set
of indices where $\vx$ is nonzero.

Square brackets will be used in different ways: if $L$ is some positive
integer then $[L]$ will denote the set $\{ 1, 2, \ldots, L \}$. If $\matr{A}$
is some matrix then $[\matr{A}]_{k,\ell}$ will denote the element in the
$k$-th row and $\ell$-th column of $\matr{A}$. If $S$ is a statement (for
example $\vx \in \code{C}$) then $[S] = 1$ if $S$ is true and $[S] = 0$
otherwise.

By $\langle \vx, \vy \rangle \defeq \sum_{i} x_i y_i$ we will denote the
standard inner product of two vectors having the same length. The
$\ell_1$-norm of a vector $\vx$ is $\onenorm{\vx} \defeq \sum_{i} \abs{x_i}$,
the $\ell_2$-norm of a vector $\vx$ is $\twonorm{\vx} \defeq \sqrt{\sum_{i}
|x_i|^2}$, and the $\ell_{\infty}$-norm (also called the max-norm) of a vector
$\vx$ is $\infnorm{\vx} \defeq \max_{i} |x_i|$. Note that $\sonenorm{\vx} =
\innerprod{\vx}{\vect{1}}$ if and only if $\vx \geq \vect{0}$. Let $\vx, \vy
\in \GF{2}^N$ be two vectors of length $N$. The Hamming weight $\wH(\vx)$ of
$\vx$ is the number of non-zero positions of $\vx$, and the Hamming distance
$\dH(\vx,\vy)$ between $\vx$ and $\vy$ is the number of positions where $\vx$
and $\vy$ disagree.

Unless stated otherwise, the code $\code{C}$ will be a binary linear code of
length $n$ and will be defined by some $m \times n$ parity-check matrix
$\matr{H}$, i.e.~$\code{C} \defeq \{ \vx \in \GF{2}^n \ | \ \vx \matr{H}^\tr
{=} \vect{0} \}$.\footnote{Note the following convention: a row index of
$\matr{H}$ will be denoted by $j$ and a column index of $\matr{H}$ will be
denoted by $i$.} We let $\set{I} \defeq \set{I}(\matr{H}) \defeq \{ 1, \ldots,
n \}$ be the set of codeword indices, $\set{J} \defeq \set{J}(\matr{H}) \defeq
\{ 1, \ldots, m \}$ be the set of check indices, $\set{J}_i \defeq
\set{J}_i(\matr{H}) \defeq \{ j \in \set{J} \ | \ [\matr{H}]_{j,i} {=} 1 \}$
be the set of check indices that involve the $i$-th codeword position, and
$\set{I}_j \defeq \set{I}_j(\matr{H}) \defeq \{ i \in \set{I} \ | \
[\matr{H}]_{j,i} {=} 1 \}$ be the set of codeword positions that are involved
in the $j$-th check.

If $\vx \in \GF{2}^n$ and $\set{S} \subseteq \set{I}$, we let $\vx_{\set{S}}$
be the sub-vector of those positions of $\vx$ whose indices are elements of
$\set{S}$, i.e.~the projection of $\vx$ onto $\set{S}$. Similarly,
$\code{C}_{\set{S}} \defeq \{ \vx_{\set{S}} \ | \ \vx \in \code{C} \} $ will
be the projection of $\code{C}$ onto the index set $\set{S}$.\footnote{In
coding language, this is often called puncturing the code $\code{C}$ at
positions $\set{I} \setminus \set{S}$~\cite{MacWilliams:Sloane:98}.} A
$(\wcol,\wrow)$-regular binary LDPC code is a code that has a parity-check
matrix $\matr{H}$ where all columns have weight $\wcol$ and all rows have
weight $\wrow$. The dimension of a code $\code{C}$ is the logarithm (to the
base $2$) of the number of codeword and the rate is the ratio of the dimension
divided by the length. Note that the dimension of $\code{C}$ is at least $1 -
\card{\set{J}}/n$, with equality if and only if $\matr{H}$ has full
rank.

If $\code{C}$ is a code then the minimum Hamming weight $\wHminc{C}$ is the
minimum Hamming weight of all nonzero codewords of $\code{C}$, and the minimum
Hamming distance $\dHminc{C}$ is the minimum Hamming distance between any two
distinct codewords of $\code{C}$. It is well known that for linear codes
$\wHminc{C} = \dHminc{C}$. A code of length $n$, dimension $k$, and minimum
distance $d$ will be called an $[n,k,d]$ code.

Let us introduce some notions from convex geometry (see
e.g.~\cite{Boyd:Vandenberghe:04:1}). Let $\vx^{(1)}, \ldots, \vx^{(k)}$ be $k$
points in $\R^N$. A point of the form $\theta_1 \vx^{(1)} + \cdots + \theta_k
\vx^{(k)}$ with $\theta_1 + \cdots + \theta_k = 1$ and $\theta_i \geq 0$, $i
\in [k]$ is called a convex combination of $\vx^{(1)}, \ldots, \vx^{(k)}$. A
set $\set{S} \subseteq \R^N$ is called convex if every possible convex
combination of two points of $\set{S}$ is in $\set{S}$. By
$\convhull(\set{S})$ we denote the convex hull of the set $\set{S}$, i.e.~the
set that consists of all possible convex combinations of all the points in
$\set{S}$; equivalently, $\convhull(\set{S})$ is the smallest convex set that
contains $\set{S}$.

Again, let $\vx^{(1)}, \ldots, \vx^{(k)}$ be $k$ points in $\R^N$.  A point of
the form $\theta_1 \vx^{(1)} + \cdots + \theta_k \vx^{(k)}$ with $\theta_i
\geq 0$, $i \in [k]$, is called a conic combination of $\vx^{(1)}, \ldots,
\vx^{(k)}$. A set $\set{K} \subseteq \R^N$ is called a cone if every possible
conic combination of two points of $\set{K}$ is in $\set{K}$. A cone $\set{K}$
is called a proper cone if it satisfies the following conditions: $\set{K}$ is
convex, $\set{K}$ is closed, $\set{K}$ is solid (i.e.~it has nonempty
interior), and $\set{K}$ is pointed (i.e., it contains no line or,
equivalently, if $\vx \in \set{K}$ and $-\vx \in K$, then $\vx =
\vect{0}$). By $\conichull(\set{S})$ we denote the conic hull of the set
$\set{S}$, i.e.~the set that consists of all possible conic combinations of
all the points in $\set{S}$; equivalently, $\conichull(\set{S})$ is the
smallest conic set that contains $\set{S}$.

Let us now introduce polytopes and polyhedra. On the one hand, a polytope in
$\R^N$ is defined to be the convex hull of a finite set of points in
$\R^N$. On the other hand, a polyhedron $\set{P}$ in $\R^N$ is defined as the
solution set of a finite number of linear equalities and inequalities:
\begin{align*}
  \set{P}
    &\defeq
       \left\{
         \vx \in \R^N
           \ \left| \
         \innerprod{\va^{(j)}}{\vx} \leq b_j, j \in [m],\ 
         \innerprod{\vc^{(j)}}{\vx}  =    d_j, j \in [p]
           \right. 
       \right\},
\end{align*}
where $\va^{(j)}$, $j \in [m]$, and $\vc^{(j)}$, $j \in [p]$, are vectors of
the same length as $\vx$ and $b_j$, $j \in [m]$, and $d_j$, $j \in [p]$, are
scalars. From this definition we see that a polyhedron is the intersection of
a finite number of half-spaces and hyperplanes and it is also easy to see that
a polyhedron is a convex set. By the Weyl-Minkowski Theorem,
cf.~e.g.~\cite[p.~55]{Barvinok:02:1}, a bounded polyhedron is a polytope.

An \emph{undirected} graph $\graph{G} = \graph{G}\big( \set{V}(\graph{G}),
\set{E}(\graph{G}) \big)$ consists of a vertex-set $\set{V}(\graph{G})$ and an
edge-set $\set{E}(\graph{G})$ whereby the elements of $\set{E}(\graph{G})$ are
$2$-subsets of $\set{V}(\graph{G})$. By a graph (without further
qualifications) we will always mean an undirected graph without loops and
multiple edges. The smallest length of any cycle will be called the girth
$\girthg{G}$ and the largest graph distance between any to vertices will be
called the diameter $\diamg{G}$. If the graph has more than one component then
$\diamg{G} = \infty$. The neighborhood $\neighborhood(v)$ of a vertex $v \in
\graph{G}$ is the set of vertices of $\graph{G}$ that are adjacent to $v$. It
follows that $|\neighborhood(v)|$ is the degree of the vertex $v$.


\section{Graph Covers and the Fundamental Polytope}

\label{sec:graph:covers:and:fundamental:polytope:1}

After recalling the definitions of finite graph covers and Tanner graphs, we
will introduce the fundamental polytope, a notion that will turn out to be the
crucial definition for the rest of the present paper.

\begin{Definition}
  [Graph cover, see e.g.~\cite{Massey:77:1, Stark:Terras:96:1}] An {\em
  unramified, finite cover}, or, simply, a {\em cover} of a (base) graph
  $\graph{G}$ is a graph $\cgraph{G}$ along with a surjective map $\phi:
  \cgraph{G} \to \graph{G}$ which is a graph homomorphism, i.e., which takes
  adjacent vertices of $\cgraph{G}$ to adjacent vertices of $\graph{G}$, such
  that for each vertex $v \in \set{V}(\graph{G})$ and each $\cover{v} \in
  \phi^{-1}(v)$, the neighborhood $\neighborhood(\cover{v})$ of $\cover{v}$ is
  mapped bijectively to $\neighborhood(v)$. For a positive integer $M$, an
  {\em $M$-cover} of $\graph{G}$ is an unramified finite cover $\phi:
  \cgraph{G} \to \graph{G}$ such that for each vertex $v \in
  \set{V}(\graph{G})$ of $\graph{G}$, $\phi^{-1}(v)$ contains exactly $M$
  vertices of $\cgraph{G}$. An $M$-cover of $\graph{G}$ is sometimes also
  called an $M$-sheeted covering of $\graph{G}$ or a cover of $\graph{G}$ of
  degree $M$.\footnote{It is important not to confuse the degree of a covering
  and the degree of a vertex.}
\end{Definition}

A consequence of this definition is that if $\cgraph{G}$ is an $M$-cover of
$\graph{G}$ then we can choose $\set{V}(\cgraph{G})$ to be
$\set{V}(\cgraph{G}) \defeq \set{V}(\graph{G}) \times [M]$: if $(v,m) \in
\set{V}(\cgraph{G})$ then $\phi\big( (v,m) \big) = v$ and if $\big( (v_1,m_1),
(v_2,m_2) \big) \in \set{E}(\cgraph{G})$ then $\phi\big( \big\{ (v_1,m_1),
(v_2,m_2) \big\} \big) = \{ v_1, v_2 \}$. Another consequence is that any
$M_2$-cover of any $M_1$-cover of the base graph is an $(M_1 \cdot M_2)$-cover
of the base graph.

\begin{figure}
  \begin{center}
    \epsfig{file=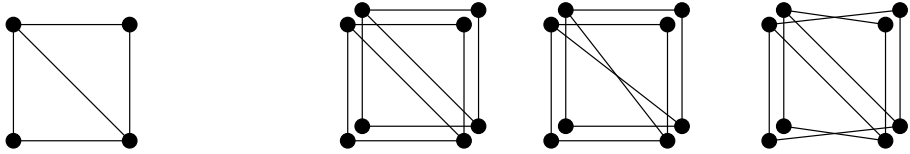, width=12cm}
  \end{center}
  \caption{Left: base graph $\graph{G}$. Right: sample of possible $2$-covers
           of $\graph{G}$.}
  \label{fig:graph:cover:samples:1}
\end{figure}

\begin{figure}
  \begin{center}
    \epsfig{file=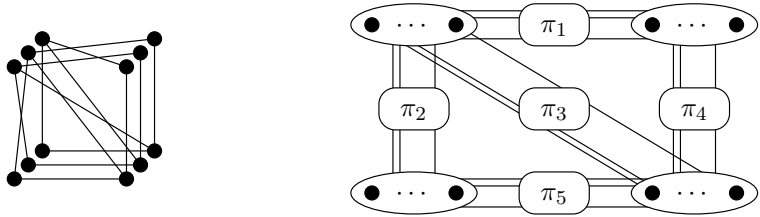, width=10cm}
  \end{center}
  \caption{Left: a possible $3$-cover of $\graph{G}$. Right: a possible
           $M$-cover of $\graph{G}$.}
  \label{fig:graph:cover:samples:2}
\end{figure}

\begin{Example}
  Let $\graph{G}$ be a (base) graph with $4$ vertices and $5$ edges as shown
  in Fig.~\ref{fig:graph:cover:samples:1}
  (left). Figs.~\ref{fig:graph:cover:samples:1} (right)
  and~\ref{fig:graph:cover:samples:2} (left), show possible $2$- and
  $3$-covers of $\graph{G}$, respectively. Any $M$-cover of $\graph{G}$ is
  entirely specified by $\card{\set{E}(\graph{G})}$ permutations: this is
  represented by Fig.~\ref{fig:graph:cover:samples:2} (right). Note that any
  $2$-cover of $\graph{G}$ must have $8 = 2 \cdot 4$ vertices and $10 = 2
  \cdot 5$ edges and any $3$-cover of $\graph{G}$ must have $12 = 3 \cdot 4$
  vertices and $15 = 3 \cdot 5$ edges.
\end{Example}

In general, a graph $\graph{G}$ has $(M!)^{|\set{E}(\graph{G})|}$ possible
$M$-covers, some of them might be isomorphic. Moreover, an $M$-cover of
$\graph{G}$ may consist of several components also if $\graph{G}$ consists of
only one component. Before we can consider graph covers of Tanner graphs, we
briefly recall the definition of Tanner graphs.

\begin{Definition}
  [Tanner graph~\cite{Tanner:81, Wiberg:Loeliger:Koetter:95,
  Kschischang:Frey:Loeliger:01}] To a binary parity-check matrix $\matr{H}$
  that defines the code $\code{C}$ we can associate a bipartite graph
  $\tgraph{T}{H}$, the so-called Tanner graph of $\matr{H}$. This graph has
  vertex set $\set{V} \defeq \{ X_i \ | \ i \in \set{I} \} \cup \{
  \CheckNode_j \ | \ j \in \set{J} \}$ and edge set $\set{E} \defeq \{
  (X_i,\CheckNode_j) \ | \ i \in \set{I}, j \in \set{J}_i \} = \{
  (X_i,\CheckNode_j) \ | \ j \in \set{J}, i \in \set{I}_j \}$. On the other
  hand, given a Tanner graph $\graph{T}$ we can associate to $\graph{T}$ a
  code $\tgcode{C}{\graph{T}}$ with parity-check matrix
  $\tgmatr{H}{\graph{T}}$ in the obvious manner.
\end{Definition}

We will use some language from behavioral theory~\cite{Polderman:Willems:98}:
an assignment of $\GF{2}$-values to the variable nodes will be called a
configuration, and a configuration that fulfills all the checks will be called
valid. In that sense, a codeword corresponds to a valid configuration and a
code corresponds to the set of all valid configurations.

From the above definition of a Tanner graph it follows that
$\neighborhood(X_i) = \{ \CheckNode_j \ | \ j \in \set{J}_i \}$ for all $i \in
\set{I}(\matr{H})$ and $\neighborhood(\CheckNode_j) = \{ X_i \ | \ i \in
\set{I}_j \}$ for all $j \in \set{J}(\matr{H})$. Moreover, the degree
$|\neighborhood(X_i)|$ of the node $X_i$ is equal to the Hamming weight of the
$i$-th column of $\matr{H}$ and the degree
$\card{\neighborhood(\CheckNode_j)}$ of the node $\CheckNode_j$ is equal to
the Hamming weight of the $j$-th row of $\matr{H}$. Therefore, Tanner graphs
of LDPC codes are sparse because of the sparseness of the parity-check matrix
of LDPC codes.

\begin{figure}
  \begin{center}
    \epsfig{figure=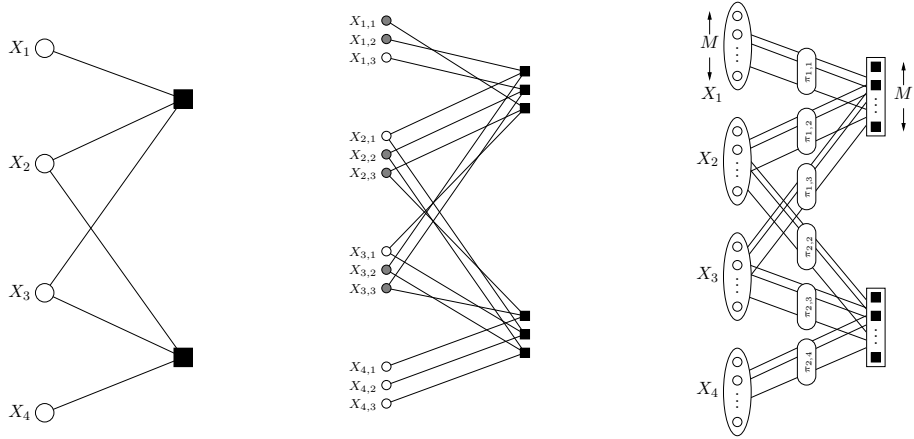,width=12cm}
  \end{center}
  \caption{Left: Tanner graph $\tgraph{T}{H}$ of the simple binary linear code
           in Ex.~\ref{ex:simple:code:1}. Middle: Possible $3$-cover of
           $\tgraph{T}{H}$. The shading of the symbol nodes indicates the
           codeword $\cvx$ found in Ex.~\ref{ex:simple:code:2}. Right:
           Possible $M$-cover of $\tgraph{T}{H}$.}
  \label{fig:simple:code:1}
\end{figure}

\begin{Example}
  \label{ex:simple:code:1}

  Let $\code{C}$ be a binary $[4,2]$ code with parity-check
  matrix\footnote{Note that this is the same parity-check matrix as in the
  Example after Th.~2 in~\cite{Feldman:Karger:Wainwright:03:2}.}
  \begin{align*}
    \matr{H}
      &\defeq
         \begin{pmatrix}
           1 & 1 & 1 & 0 \\
           0 & 1 & 1 & 1
         \end{pmatrix}.
  \end{align*}
  Obviously, $\code{C} = \big\{ (0,0,0,0), (0,1,1,0), (1,0,1,1), (1,1,0,1)
  \big\}$, $\set{J} = \{ 1, 2 \}$, $\set{J}_1 = \{ 1 \}$, $\set{J}_2 = \{ 1, 2
  \}$, $\set{J}_3 = \{ 1, 2 \}$, $\set{J}_4 = \{ 2 \}$, $\set{I} = \{ 1, 2, 3,
  4 \}$, $\set{I}_1 = \{ 1, 2, 3 \}$, and $\set{I}_2 = \{ 2, 3, 4 \}$. The
  Tanner graph $\graph{T}(\matr{H})$ that is associated to $\matr{H}$ is shown
  in Fig.~\ref{fig:simple:code:1} (left). 

  An $M$-fold cover $\graph{\widetilde T}$ (as shown in
  Fig.~\ref{fig:simple:code:1} (right)) of $\graph{T}$ is specified by
  defining the permutations $\pi_{1,1}$, $\pi_{1,2}$, $\pi_{1,3}$
  (corresponding to the first row of $\matr{H}$) and the permutations
  $\pi_{2,2}$, $\pi_{2,3}$, $\pi_{2,4}$ (corresponding to the second row of
  $\matr{H}$).
\end{Example}

  Let $\code{C}$ be a binary code with parity-check matrix $\matr{H}$ and
  Tanner graph $\graph{T} \defeq \tgraph{T}{H}$. For a positive integer $M$,
  let $\cgraph{T}$ be an arbitrary $M$-fold cover of $\graph{T}$, let
  $\ccode{C} \defeq \ctgcode{C}{T}$ be the binary code described by
  $\cgraph{T}$, and let the codeword positions of $\ccode{C}$ be indexed by
  $\cset{I} \defeq \set{I} \times [M]$ and the check equations by $\cset{J}
  \defeq \set{J} \times [M]$.

  Knowing the graph $\graph{T}$, the graph $\cgraph{T}$ is completely
  specified by defining for all $j \in \set{J}$, $i \in \set{I}_j$ the
  permutations $\pi_{j,i}$ that map $[M]$ onto itself. The meaning of
  $\pi_{j,i}(m)$, $m \in [M]$, is the following: the $m^{\text{th}}$ copy of
  check node $j$ is connected to the $\pi_{j,i}(m)^{\text{th}}$ copy of
  codeword symbol $X_i$, i.e.~check node $\cover{\CheckNode}_{j,m}$ is
  connected to codeword symbol $\cover{X}_{i,\pi_{j,i}(m)}$. It follows that
  $\cvx \in \code{\tilde C}$ if and only if
  \begin{align}
    \sum_{i \in \set{I}_j}
      \cover{x}_{i,\pi_{j,i}(m)}
      &= 0 \quad \text{(in $\GF{2}$)}
             \label{eq:remark:parity:check:equality:cover:code:1}
  \end{align}
  for all $(j,m) \in \cset{J}$. The parity check matrix $\matr{\tilde H}$ that
  expresses this fact can be defined as follows. Let the entries of
  $\matr{\tilde H}$ be indexed by $(j,m) \in \cset{J}$ and $(i,m') \in
  \cset{I}$. Then
  \begin{align}
    [\matr{\tilde H}]_{(j,m),(i,m')}
      &\defeq
         \begin{cases}
           1 & \text{if $i \in \set{I}_j$ and $m' = \pi_{j,i}(m)$} \\
           0 & \text{otherwise}.
         \end{cases}
  \end{align}

\begin{Example}
  \label{ex:simple:code:2}

  We continue Ex.~\ref{ex:simple:code:1}. The parity-check matrix $\cmatr{H}
  \defeq \ctgmatr{H}{T}$ associated to a possible $3$-fold cover Tanner graph
  $\cgraph{T}$ as shown in Fig.~\ref{fig:simple:code:1} (middle) looks like
  \begin{align*}
    \cmatr{H}
      &\defeq
         \left(
           \begin{array}{ccc|ccc|ccc|ccc}
             0 & 1 & 0  &  1 & 0 & 0  &  0 & 1 & 0  &  0 & 0 & 0 \\
             0 & 0 & 1  &  0 & 1 & 0  &  0 & 0 & 1  &  0 & 0 & 0 \\
             1 & 0 & 0  &  0 & 0 & 1  &  1 & 0 & 0  &  0 & 0 & 0 \\
             \hline
             0 & 0 & 0  &  0 & 0 & 1  &  0 & 0 & 1  &  1 & 0 & 0 \\
             0 & 0 & 0  &  1 & 0 & 0  &  1 & 0 & 0  &  0 & 1 & 0 \\
             0 & 0 & 0  &  0 & 1 & 0  &  0 & 1 & 0  &  0 & 0 & 1
           \end{array}
         \right).
  \end{align*}
  This parity-check matrix defines a code $\ccode{C} = \ctgcode{C}{T}$:
  e.g.~the configuration $\cvx = (1{:}1{:}0, \ 0{:}1{:}1, \ $ $0{:}1{:}1, \
  0{:}0{:}0)$ that is highlighted in Fig.~\ref{fig:simple:code:1} (middle) is
  a codeword in this code. Note also that $\ccode{C}$ contains the liftings of
  all codewords to $\cgraph{T}$, namely if $(x_1, x_2, x_3, x_4) \in \code{C}$
  then $(x_1{:}x_1{:}x_1, \ x_2{:}x_2{:}x_2, \ $ $x_3{:}x_3{:}x_3, \
  x_4{:}x_4{:}x_4) \in \ccode{C}$. The last statement follows from the
  following argument: since $\graph{T}$ and $\cgraph{T}$ look locally the
  same, the fact that a codeword $\vx$ in $\code{C}$ fulfills the checks
  imposed by $\graph{T}$ implies that the lifting of $\vx$ to $\cgraph{T}$
  fulfills all the checks imposed by $\cgraph{T}$, i.e.~that it is a codeword
  in $\ccode{C}$.
\end{Example}

\begin{Definition}
  \label{def:pseudo:codeword:1}
  
  Let $\code{C}$ be a binary linear (base) code with parity-check matrix
  $\matr{H}$ and let $\graph{T} \defeq \tgraph{T}{H}$ be the corresponding
  Tanner graph. For any positive integer $M$, let $\cgraph{T}$ be an $M$-fold
  cover of $\graph{T}$ and let $\ccode{C} \defeq \ctgcode{C}{T}$. The {\em
  (scaled) pseudo-codeword} associated to $\cvx$ is the rational vector
  $\vomega(\cvx) \defeq \big( \omega_1(\cvx), \omega_2(\cvx), \ldots,
  \omega_n(\cvx) \big)$ with
  \begin{align}
    \omega_i(\cvx)
      &\defeq
        \frac{1}{M}
           \sum_{m \in [M]}
             \widetilde x_{i,m},
               \label{eq:def:pseudo:codeword:1:1}
  \end{align}
  where the sum is taken in $\mathbb{R}$ (not in $\GF{2}$). In fact, any
  multiple (by a positive scalar) of $\vomega(\cvx)$ will be called a
  pseudo-codeword associated with $\cvx$. Because of its importance, we give a
  special name to the vector $M \cdot \vomega(\cvx)$, namely we will call it
  the \emph{unscaled} pseudo-codeword associated to $\cvx$. Additionally, we
  define $\vomega(\ccode{C})$ to be the set
  \begin{align*}
    \vomega(\ccode{C})
      &\defeq
         \left\{
           \vomega(\cvx)
             \ \left| \
           \cvx \in \ccode{C}
               \right.
         \right\}.
  \end{align*}
  Obviously, $\vomega(\ccode{C}) \subseteq [0,1]^n \cap \Q^n$.
\end{Definition}

Note that whereas a pseudo-codeword as defined in
Def.~\ref{def:pseudo:codeword:1} has length $\card{\set{I}(\matr{H})}$,
i.e.~equal to the length of the code $\code{C}$, a codeword like $\tvc \in
\code{\widetilde{C}}$ has length $M \cdot \card{\set{I}(\matr{H})}$ where $M$
is the degree of the corresponding cover Tanner graph. Because $\graph{T}$ is
a $1$-cover of a Tanner graph $\graph{T}$ we see that any codeword is also a
pseudo-codeword.

\begin{Example}
  \label{ex:simple:code:1:pseudo:codeword:1}

  We continue Ex.~\ref{ex:simple:code:1}. We saw that $\tvc = (1{:}1{:}0, \
  0{:}1{:}1, \ 0{:}1{:}1, \ 0{:}0{:}0)$ was a codeword of the code
  $\code{\widetilde{C}}$. Applying Def.~\ref{def:pseudo:codeword:1} we see
  that the corresponding pseudo-codeword is $\vomega(\tvc) = (\frac{2}{3},
  \frac{2}{3}, \frac{2}{3}, 0)$. (Note that this pseudo-codeword cannot be
  written as a convex combination of the codewords in $\code{C}$.) The
  corresponding unscaled pseudo-codeword is $3 \cdot \vomega(\tvc) =
  (2,2,2,0)$ and comparing this vector with Fig.~\ref{fig:simple:code:1}
  (middle), we see the intuitive meaning of its components: the first
  component corresponds to the number of shaded variable nodes $X_{1,m}$, $m
  \in [M]$, the second component corresponds to the number of shaded variable
  nodes $X_{2,m}$, $m \in [M]$, etc.
\end{Example}

We would like to investigate the question if it is possible to characterize
the union of the set of all (scaled) pseudo-codewords obtained by all finite
covers of the Tanner graph of a binary linear code, i.e.~we would like to
understand the set
\begin{align}
  \cGCDset{H}
    &\defeq
       \bigcup_{
         \stackrel{M \in \Intpp}
                   {\stackrel{\cgraph{T}:\ \cgraph{T}
                              \text{ is an $M$-fold cover of }
                              \tgraph{T}{H}}
                             {\cvx \in \ctgcode{C}{T}}
                   }
               }
         \left\{
           \left(
             M, \cgraph{T}, \cvx
           \right)
         \right\}
           \label{eq:cgcs:def:1}
\end{align}
and its ``projection''\footnote{We could have defined
\begin{align*}
  \GCDset{H}
    &\defeq
       \bigcup_{(M,\cgraph{T},\cvx) \in \cGCDset{H}}
         \left\{
           \left(
             M, \cgraph{T}, \vomega(\cvx)
           \right)
         \right\},
\end{align*}
but the definition of $\GCDset{H}$ in~\eqref{eq:gcs:def:1} contains enough
information for our purposes.}
\begin{align}
  \GCDset{H}
    &\defeq
       \bigcup_{(M, \cgraph{T}, \cvx) \in \cGCDset{H}}
         \big\{ \vomega(\cvx) \big\}.
           \label{eq:gcs:def:1}
\end{align}
From the properties of $\vomega(\ccode{C})$ it follows that $\GCDset{H}
\subseteq [0,1]^n \cap \Q^n$. Observe that
\begin{align}
  \GCDset{H}
    &= \bigcup_{\cgraph{T}: \  
         \cgraph{T} \text{ is a finite-cover graph of } \tgraph{T}{H}}
       \left\{
         \vomega
           \big( \ctgcode{C}{T} \big)
       \right\}.
           \label{eq:gcs:def:2}
\end{align}
This set has a surprisingly simple characterization. It will turn out that
$\GCDset{H}$ is essentially given by the fundamental polytope introduced in the
next definition. Before we turn to that definition, let us observe that the
code $\code{C}$ can be written as the intersection
\begin{align*}
  \code{C}
    &= \bigcap_{j \in \set{J}}
         \code{C}_j
\end{align*}
of the codes
\begin{alignat}{2}
  \code{C}_j
    &\defeq
       \code{C}_j(\matr{H})
    &\defeq
       \big\{
          \vx \in \GF{2}^n
          \ \big| \
          \langle \vx, \vect{h}_j \rangle = 0 \text{ (in $\GF{2}$)}
       \big\}
    &= \big\{
         \vx \in \GF{2}^n
         \ \big| \
         \langle \vx_{\set{I}_j}, \vect{1} \rangle = 0 \text{ (in $\GF{2}$)}
       \big\},
         \label{eq:code:C:j:1}
\end{alignat}
where for each $j \in \set{J}$ we let $\matr{h}_j$ be the $j$-th row of
$\matr{H}$. For $j \in \set{J}$, we will also use the codes
\begin{alignat}{2}
  \code{C}'_j
    &\defeq
       \code{C}'_j(\matr{H})
    &\defeq
       \left\{
          \vx' \in \GF{2}^{\card{\set{I}_j}}
          \ \big| \
          \left\langle \vx', (\vect{h}_j)_{\set{I}_j} \right\rangle = 0 
            \text{ (in $\GF{2}$)}
       \right\}
    &= \left\{
         \vx' \in \GF{2}^{\card{\set{I}_j}}
         \ \big| \
         \left\langle \vx', \vect{1} \right\rangle = 0 \text{ (in $\GF{2}$)}
       \right\}.
         \label{eq:code:C:prime:j:1}
\end{alignat}
The codes $\code{C}_j$ and $\code{C}'_j$ are related as follows. First,
$\code{C}'_j$ is the projection of $\code{C}_j$ onto $\set{I}_j$,
i.e.~$\code{C}'_j = (\code{C}_j)_{\set{I}_j}$. Secondly, the convex hulls of
$\code{C}_j$ and of $\code{C}'_j$ fulfill
\begin{align}
  \convhull
    \big(
      \code{C}_j
    \big)
    &= \big\{
          \vomega \in \R^n
          \ \big| \
          \vect{0} \leq \vomega \leq \vect{1},\ 
          \vomega_{\set{I}_j} \in 
            \convhull(\code{C}'_j)
       \big\}.
         \label{eq:code:conv:hull:C:j:1}
\end{align}
We are now ready for the main definition of this paper.

\begin{Definition}
  \label{def:fundamental:polytope:1}
  The fundamental polytope $\fp{P} \defeq \fph{P}{H}$ of $\matr{H}$ is defined
  to be the set
  \begin{align}
    \fp{P}
      &\defeq
         \bigcap_{j \in \set{J}}
           \convhull(\code{C}_j)
             \label{eq:fp:def:1} \\
      &= \bigcap_{j \in \set{J}}
           \big\{
            \vomega \in \R^n
            \ \big| \
            \vect{0} \leq \vomega \leq \vect{1},\ 
            \vomega_{\set{I}_j} \in 
              \convhull(\code{C}'_j)
         \big\}
             \label{eq:fp:def:2} \\
      &= [0,1]^n
         \cap
         \bigcap_{j \in \set{J}}
           \big\{
            \vomega \in \R^n
            \ \big| \
            \vomega_{\set{I}_j} \in 
              \convhull(\code{C}'_j)
         \big\}.
           \label{eq:fp:def:3}
  \end{align}
\end{Definition}

As can be seen from the notation $\fph{P}{H}$, the fundamental polytope is a
function of the parity-check matrix $\matr{H}$ that describes the code
$\code{C}$. This means that different parity-check matrices for the same code
can (and usually do) yield different fundamental polytopes. 

In the same way as all codewords of a code described by a parity-check matrix
$\matr{H}$ are all the valid configurations in a Tanner graph $\tgraph{T}{H}$,
we see that~\eqref{eq:fp:def:3} yields a similar description for all
pseudo-codewords, i.e.~for all the vectors that lie in the fundamental
polytope $\fph{P}{H}$. Indeed, we redefine the Tanner graph as follows: each
bit node $X_i$ is now labeled $\Omega_i$ and can take on values in the
interval $[0,1]$ and each check node $B_j$ is replaced by the indicator
function of the convex hull of $\code{C}'_j$. (We can use the results of
Lemmas~\ref{lemma:properties:fp:fc:1} and~\ref{lemma:properties:fp:fc:2} in
Sec.~\ref{sec:fp:description:1} to formulate these indicator functions.)

\begin{figure}
  \begin{center}
    \epsfig{file=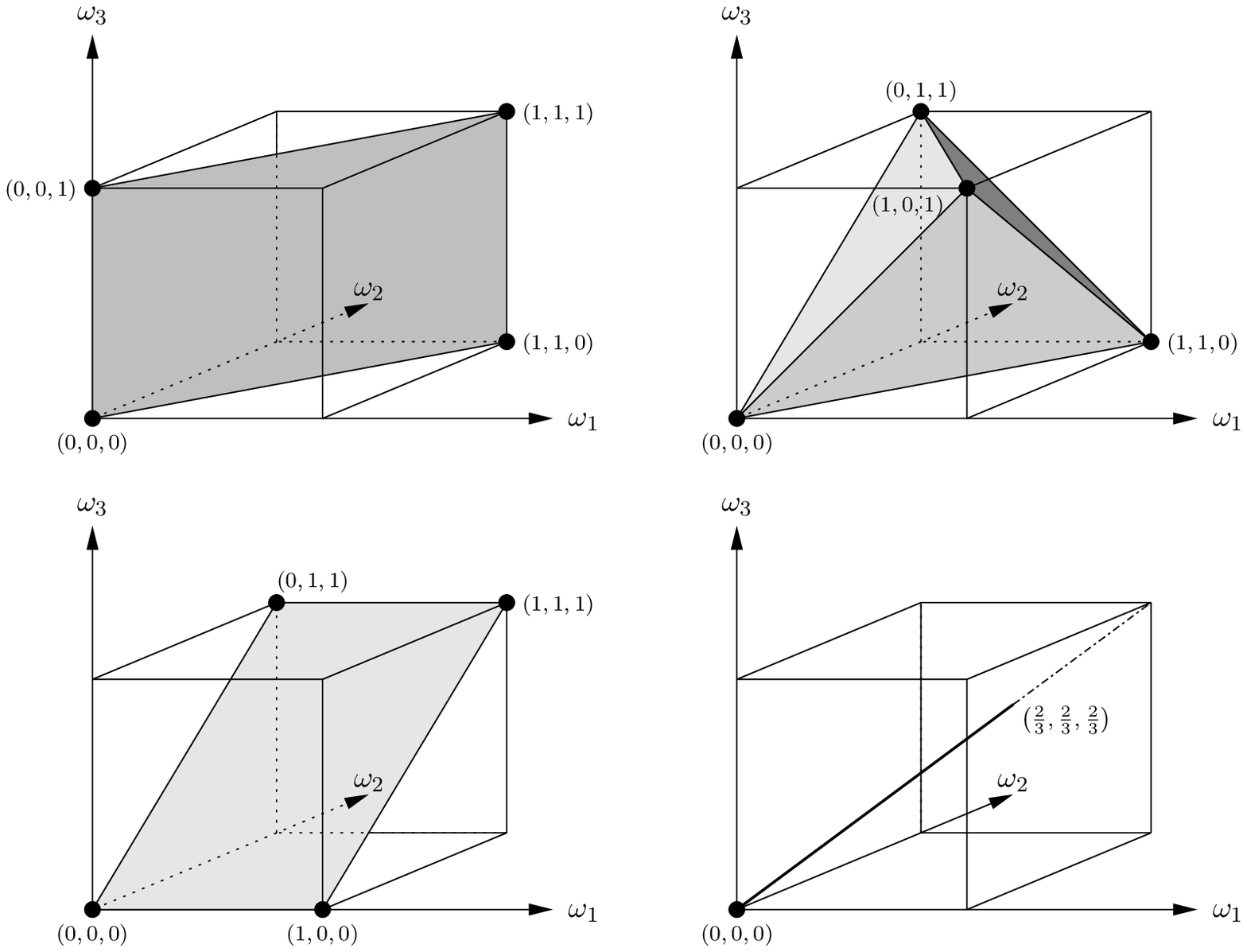, width=10cm}
  \end{center}
  \caption{$\code{C}_j(\matr{H})$, $j \in \set{J}$ and $\fph{P}{H}$ for the
    parity-check matrix $\matr{H}$ in~\eqref{eq:trivial:code:pcm:1}. Top left:
    $\convhull(\code{C}_1(\matr{H}))$. Top right:
    $\convhull(\code{C}_2(\matr{H}))$. Bottom left:
    $\convhull(\code{C}_3(\matr{H}))$. Bottom right: $\fph{P}{H} = \cap_{j \in
    \set{J}} \convhull(\code{C}_j(\matr{H}))$.}
  \label{fig:trivial:code:fp:1:1}
\end{figure}

\begin{Example}
  \label{ex:trivial:code:fundamental:polytope:1}

  We continue discussing the code that was introduced in
  Sec.~\ref{sec:motivating:example:1} whose parity-check matrix is shown
  in~\eqref{eq:trivial:code:pcm:1}. For this parity-check matrix the codes
  $\code{C}_j$, $j \in \set{J}$ turn out to be
  \begin{alignat*}{2}
    \code{C}_1
      &&= \{ (0,0), (1,1) \} \times \{ (0), (1) \}
       &= \{ (0,0,0), (0,0,1), (1,1,0), (1,1,1) \} , \\
    \code{C}_2
      &&
       &= \{ (0,0,0), (0,1,1), (1,0,1), (1,1,0) \} , \\
    \code{C}_3
      &&= \{ (0), (1) \} \times \{ (0,0), (1,1) \}
       &= \{ (0,0,0), (1,0,0), (0,1,1), (1,1,1) \}.
  \end{alignat*}
  We can easily check that $\code{C} = \cap_{j \in \set{J}} \code{C}_j = \{
  (0,0,0) \}$. Fig.~\ref{fig:trivial:code:fp:1:1} visualizes these codes,
  their convex hulls, and the fundamental polytope $\fph{P}{H} = \cap_{j \in
  \set{J}} \convhull(\code{C}_j) = \big\{ (\omega, \omega, \omega) \ | \ 0
  \leq \omega \leq \frac{2}{3} \big\}$. Note that here the fundamental
  polytope has only two vertices: $(0,0,0)$ and
  $(\frac{2}{3},\frac{2}{3},\frac{2}{3})$ where the former is the
  pseudo-codeword corresponding to the all-zeros assignment in any finite
  cover and where the latter is e.g.~the pseudo-codeword corresponding to the
  configuration in the triple cover shown Fig.~\ref{fig:trivial:code:1:2}
  (right).

  Moreover, using Prop.~\ref{prop:image:in:fp:2} below, it can be shown that
  $\GCDset{H}$ equals the set of all the rational points of
  $\fph{P}{H}$. Accepting this fact, we can also verify the statement made in
  Prop.~\ref{prop:image:in:fp:2} that all vertices of $\fph{P}{H}$ are in
  $\GCDset{H}$.
\end{Example}

Note that usually the effective dimension of the fundamental polytope equals
the length $n$ of the code. In cases where the parity-check matrix has checks
that involve only one or two codeword symbols, there is a reduction in
effective dimensionality. The above example is a witness of this fact.

After having seen the definition of the fundamental polytope we are in a
position to formulate the main theorem of this paper which relates the set
$\GCDset{H}$ with the fundamental polytope $\fph{P}{H}$.

\begin{Proposition}
  \label{prop:image:in:fp:2}

  Let $\code{C}$ be an arbitrary binary linear code and let $\matr{H}$ be its
  parity-check matrix. It holds that
  \begin{align}
    \GCDset{H}
      &= \fph{P}{H}
         \cap
         \Q^n, 
           \label{eq:sets:s:and:fp:1} \\
    \fph{P}{H}
      &= \overline{\GCDset{H}},
           \label{eq:sets:s:and:fp:2}
  \end{align}
  where the over-bar denotes the closure of the corresponding set under the
  usual topology of $\R^n$. Moreover, all vertices of $\fph{P}{H}$ are in
  $\GCDset{H}$.
\end{Proposition}

\begin{Proof}
  See Sec.~\ref{sec:proof:prop:image:in:fp:2}.
\end{Proof}

Before finishing this section let us mention that the fundamental polytope and
related concepts can not only be defined for a code whose Tanner graph
consists only of single parity-check codes but also for codes described by a
Tanner graph where some or all of the check nodes represent more complicated
subcodes or for codes described by a factor graph that represents a
tail-biting trellis. The generalization is relatively straightforward and will
not be discussed any further in this paper.


\section{Channels, MAP Decoding, and LP Decoding}
\label{sec:channels:and:MAPD:LPD:1}

We consider the problem of data communication over a memoryless channel with
input alphabet $\set{X}$, output alphabet $\set{Y}$, and with channel law
$P_{Y|X}(y|x)$. In this paper we only consider channels with binary input,
i.e.~with $\set{X} = \{ 0, 1 \}$. In order to achieve reliable communication
over such a channel, we will use a binary code $\code{C} \subseteq \GF{2}^n$
of length $n$ and rate $R$ that is defined by some parity-check matrix
$\matr{H}$. We assume that every codeword $\vx \in \code{C}$ is transmitted
with equal probability, i.e.~$P_{\vX}(\vx) = 2^{-nR}$ if $\vx \in \code{C}$
and $P_{\vX}(\vx) = 0$ otherwise, where $R$ is the rate of the code.

Upon observing the output $\vect{Y} = \vect{y}$, block-wise maximum
a-posteriori decoding (MAPD) can be formulated as the following optimization
problem:\footnote{Note that the resulting decision rule equals also the
maximum-likelihood decision rule because all possible codewords $\vx$ occur
with the same probability.}
\begin{align}
  \hvxMAPD(\vy)
    &= \arg \max_{\vx \in \GF{2}^n} \
         P_{\vX,\vY}(\vx,\vy)
     = \arg \max_{\vx \in \GF{2}^n} \
         P_{\vX}(\vx) \cdot P_{\vY|\vX}(\vy|\vx)
           \nonumber \\
    &= \arg \max_{\vx \in \code{C}} \
         P_{\vY|\vX}(\vy|\vx)
     = \arg \min_{\vx \in \code{C}} \
            -
            \log P_{\vY|\vX}(\vy|\vx),
              \label{eq:blockwise:MAPD:1}
\end{align}
where $P_{\vX,\vY}(\vx,\vy) = P_{\vX}(\vx) \cdot P_{\vY|\vX}(\vy|\vx)$ is
the joint pmf/pdf of the the coded (but un-modulated) channel input $\vX$ and
the channel output $\vY$. Ties are resolved in a systematic way.

In the following we will use the fact that $P_{\vY | \vX}(\vy | \vx) =
\prod_{i \in \set{I}} P_{Y_i | X_i}(y_i | x_i) = \prod_{i \in \set{I}} P_{Y |
X}(y_i | x_i)$ holds for memoryless channels (that are used without
feedback). The random variable
\begin{align}
  \Lambda_i
    &\defeq \Lambda_i(Y_i)
     \defeq
       \log
         \frac{P_{Y|X}(Y_i|0)}
              {P_{Y|X}(Y_i|1)}
           \label{eq:log:likelihood:definition:1}
\end{align}
with realization $\vlambda$ will be be called the channel log-likelihood ratio
for the $i$-th codeword symbol.\footnote{Because of the memoryless property of
the channel it also follows that $p_{\vLambda|\vX}(\vlambda|\vx) = \prod_{i
\in \set{I}} p_{\Lambda_i | X_i}(\lambda_i | x_i) = \prod_{i \in \set{I}}
p_{\Lambda| X}(\lambda_i | x_i)$. \label{footnote:lambda:vect:pdf:1}}
Block-wise MAPD can therefore be rewritten to read
\begin{align}
  \hvxMAPD(\vy)
    &= \arg \min_{\vx \in \code{C}}
         \log
           \frac{P_{\vY|\vX}(\vy|\vect{0})}
                {P_{\vY|\vX}(\vy|\vx)}
     = \arg \min_{\vx \in \code{C}}
         \sum_{i \in \set{I}}
           x_i \lambda_i
     = \arg \min_{\vx \in \code{C}}
         \langle
           \vx, \vlambda
         \rangle,
             \label{eq:blockwise:MAPD:2}
\end{align}
where ties are resolved in a systematic manner.

From this expression it is not far anymore to linear programming decoding
(LPD)~\cite{Feldman:03:1, Feldman:Wainwright:Karger:05:1}. In a first step,
let us reformulate~\eqref{eq:blockwise:MAPD:2} as
\begin{align}
  \hvxMAPD(\vy)
    &= \arg \min_{\vx \in \convhull(\code{C})}
         \langle
           \vx, \vlambda
         \rangle,
             \label{eq:blockwise:MAPD:3}
\end{align}
where ties are resolved in a systematic manner. This expression follows from
two facts: all codewords in $\code{C}$ are vertices of $\convhull(\code{C})$
and because the cost function is linear, the set of optimal solutions must
always include at least one vertex of $\convhull(\code{C})$.\footnote{In case
a whole face of of $\convhull(\code{C})$ is optimal we decide in favor of one
of the vertices in it.} The resulting optimization problem on the right-hand
side of~\eqref{eq:blockwise:MAPD:3} is a linear program (LP). Although it is of
course desirable to solve such a problem, for arbitrary codes this problem
turns out to be hard, a reason being that the number of inequalities needed to
describe $\convhull(\code{C})$ usually grows exponentially in the block
length. A standard way in optimization theory to circumvent such complexity
issues is to solve a closely related problem: instead of minimizing over
$\convhull(\code{C})$ we will minimize over a relaxation polytope
$\relaxation(\convhull(\code{C}))$ of this polytope, i.e.~over a larger
polytope:
\begin{align}
  \hvomegaLPD(\vy)
    &= \arg \min_{\vomega \in \relaxation(\convhull(\code{C}))}
         \langle
           \vomega, \vlambda
         \rangle,
             \label{eq:LPD:1}
\end{align}
Of course, this new polytope should have a low description complexity, yet be
a good approximation of $\convhull(\code{C})$ so that it is highly likely that
$\hvxLPD(\vy) = \hvxMAPD(\vy)$. In particular, all codewords in $\code{C}$
should be vertices of $\relaxation(\convhull(\code{C}))$.

Probably one of the easiest ways of obtaining a reasonable relaxation is the
following. Observe that
\begin{align*}
  \code{C}
    &= \bigcap_{j \in \set{J}(\matr{H})}
         \code{C}_j(\matr{H}),
\end{align*}
where $\code{C}_j(\matr{H})$ was defined in~\eqref{eq:code:C:j:1}. Consider
now the set
\begin{align}
  \relaxationset{H}
    &\defeq
       \bigcap_{j \in \set{J}(\matr{H})}
         \convhull
           \big(
             \code{C}_j(\matr{H})
           \big).
             \label{eq:canonical:relaxation:def:1}
\end{align}
The fact that the set $\relaxationset{H}$ is a relaxation of
$\convhull(\code{C})$ can be seen from the following chain of reasoning:
firstly, the set $\relaxationset{H}$ is the intersection of convex sets and is
therefore convex itself; secondly, the set $\relaxationset{H}$ contains all
codewords in $\code{C}$; thirdly, $\convhull(\code{C})$ is the smallest convex
set that contains $\code{C}$; combining these three observations leads to the
conclusion that $\convhull(\code{C}) \subseteq \relaxationset{H}$. Note that
$\convhull(\code{C}) = \relaxationset{H}$ is possible though strict inclusion
turns out to be what happens usually. Of course, the set $\relaxationset{H}$
in~\eqref{eq:canonical:relaxation:def:1} equals the set $\fph{P}{H}$ defined
in Def.~\ref{def:fundamental:polytope:1}: the solution of the LP decoder when
choosing $\relaxation(\convhull(\code{C})) \defeq \relaxationset{H} =
\fph{P}{H}$ will henceforth be called $\hvomegaLPDH{H}(\vy)$.

The next definition introduces another class of relaxations.

\begin{Definition}
  \label{def:relaxation:of:order:r:1}

  Let $\matr{H}$ be an arbitrary parity-check matrix that defines a code
  $\code{C}$. For some $r \geq 1$, let
  \begin{align*}
    \relaxationsetorder{r}{H}
      &\defeq
         \bigcap_{\vect{h}}
           \convhull
             \big(
               \code{C}(\vect{h})
             \big),
  \end{align*}
  where the intersection is over all vectors $\vect{h} \in \GF{2}^n$ that can
  be written as the modulo-$2$ sum of at most $r$ rows of $\matr{H}$.  We call
  $\relaxationsetorder{r}{H}$ the $r$-th relaxation of $\convhull(\code{C})$
  with respect to $\matr{H}$. Note that $\relaxationsetorder{r}{H} =
  \fp{P}(\matr{H}')$ where $\matr{H}'$ is the parity-check matrix consisting
  of all rows of $\matr{H}$, the modulo-$2$ sums of all pairs of rows of
  $\matr{H}$, \ldots, the modulo-$2$ sum of all $r$-tuples of rows of
  $\matr{H}$.
\end{Definition}

Some of the consequences of this definition will be explored in
Sec.~\ref{sec:why:four:cycles:potentiall:bad:1}.

Let us define three channels that will be of prime interest in this paper: the
binary-input additive white Gaussian noise channel (BI-AWGNC or simply AWGNC),
the binary symmetric channel (BSC), and the binary erasure channel (BEC).

\begin{Example}
  \label{ex:awgnc:def:1}

  The binary input additive white Gaussian noise channel (BI-AWGNC) with input
  energy per channel symbol $\Ec$ and noise power $\sigma^2$ has output
  alphabet $\set{Y} = \R$ and channel law\footnote{In the case
  of the AWGNC we will denote the output symbols by $\oY_i$ and not by $Y_i$
  so that all (random) variables that can be represented in a signal space
  have an over-bar.}
  \begin{align}
    P_{\overline{Y}|X}(\overline{y}|x)
      &= \begin{cases}
           \frac{1}{\sqrt{2\pi} \sigma}
             \exp
               \left(
                 -
                 \frac{(\overline{y} - \sqrt{\Ec})^2}{2\sigma^2}
               \right)
             & \text{(if $x = 0$)} \\
           \frac{1}{\sqrt{2\pi} \sigma}
             \exp
               \left(
                 -
                 \frac{(\overline{y} + \sqrt{\Ec})^2}{2\sigma^2}
               \right)
             & \text{(if $x = 1$)}
         \end{cases}.
           \label{eq:awgnc:channel:law:1}
  \end{align}
  Defining the input energy per information symbol to be $\Eb$, this quantity
  is related to $\Ec$ through $\Ec = R \cdot \Eb$. Introducing $N_0 \defeq
  2\sigma^2$, two different signal-to-noise ratios can be defined, namely
  $\SNRb \defeq \Eb/N_0$ and $\SNRc \defeq \Ec/N_0$, which are related through
  $\SNRc = R \cdot \SNRb$. Defining $\ox(x) \defeq \sqrt{\Ec} \cdot (1-2x)$
  for $x \in \GF{2} \subset \R$ we can write \eqref{eq:awgnc:channel:law:1} as
  \begin{align*}
    P_{Y|X}(y|x)
      &= \frac{1}{\sqrt{2\pi} \sigma}
           \exp
             \left(
               -
               \frac{\big( \overline{y} - \ox(x) \big)^2}{2\sigma^2}
             \right).
  \end{align*}
  If $\vx \in \GF{2}^n \subset \R^n$ is the codeword to be transmitted, then
  the modulated word is $\ovx \defeq \ovx(\vx) \defeq \sqrt{\Ec} \cdot
  (\vect{1} - 2\vx)$. So, upon sending $\ox_i$ we receive $\oY_i = \ox_i +
  \oZ_i$ where $\oZ_i$ is normally distributed with mean zero and variance
  $\sigma^2$. Therefore, $\ovY_i$ given $X_i = 0$ is normally distributed with
  mean $+\sqrt{\Ec}$ and variance $\sigma^2$, whereas $\ovY_i$ given $X_i = 1$
  is normally distributed with mean $-\sqrt{\Ec}$ and variance $\sigma^2$. For
  the BI-AWGNC we have a simple relationship between $\ovY$ and $\vLambda$,
  namely by simplifying the definition of LLR for the $i$-th symbol we see
  that
  \begin{align*}
    \Lambda_i
      &\defeq \Lambda_i(\oY_i)
       \defeq
         \log
           \frac{P_{\oY_i|X_i}(\oY_i|0)}
                {P_{\oY_i|X_i}(\oY_i|1)}
       = \log
           \frac{P_{\oY_i|\oX_i}(\oY_i|+\sqrt{\Ec})}
                {P_{\oY_i|\oX_i}(\oY_i|-\sqrt{\Ec})}
       = 4
           \cdot
           \frac{\sqrt{R \Eb}}
                {N_0}
           \cdot
           \oY_i,
  \end{align*}
  i.e.~$\vLambda$ is just a scaled version of $\ovY$. From this, it can easily
  be calculated that $\Lambda_i$ given $X_i = 0$ is normally distributed with
  mean $4 R \cdot \SNRb$ and variance $8 R \cdot \SNRb$, whereas $\Lambda_i$
  given $X_i = 1$ is normally distributed with mean $- 4 R \cdot \SNRb$ and
  variance $8 R \cdot \SNRb$.

  Finally, let us note that block-wise MAPD can not only be written as in
  (\ref{eq:blockwise:MAPD:1}) and (\ref{eq:blockwise:MAPD:2}) but also as
  \begin{align}
    \hvxMAPD(\ovy)
      &= \arg \max_{\vx \in \code{C}}
           \sum_{i \in \set{I}}
             \ox_i(x_i) \lambda_i
       = \arg \max_{\vx \in \code{C}}
           \left\langle
             \ovx(\vx), \vlambda
           \right\rangle,
             \label{eq:blockwise:MAPD:BIAWGNC:1}
  \end{align}
  i.e.~decoding can be written as finding the $\ovx(\vx)$, $\vx \in \code{C}$,
  with the largest standard inner product with $\vlambda$. The decoding rule
  in (\ref{eq:blockwise:MAPD:BIAWGNC:1}) is also known as the correlation
  decoding rule.
\end{Example}

\begin{Example}
  \label{ex:bsc:def:1}

  The binary symmetric channel (BSC) with cross-over probability $0 \leq
  \varepsilon \leq \frac{1}{2}$ has output alphabet $\set{Y} = \{ 0, 1 \}$ and
  channel law $P_{Y|X}(y|x) = 1 - \varepsilon$ if $y = x$ and $P_{Y|X}(y|x) =
  \varepsilon$ otherwise. The log-likelihood ratio for the $i$-th bit is the
  random variable
  \begin{align}
  \Lambda_i
    &\defeq \Lambda_i(Y_i)
     \defeq
       \log
         \frac{P_{Y_i|X_i}(Y_i|0)}
              {P_{Y_i|X_i}(Y_i|1)}
     = \begin{cases}
         +\log
            \frac{1-\varepsilon}{\varepsilon}
              & \text{(if $Y_i = 0$)} \\
         -\log
            \frac{1-\varepsilon}{\varepsilon}
              & \text{(if $Y_i = 1$)}
       \end{cases}.
  \end{align}
  Note that $\log \frac{1-\varepsilon}{\varepsilon} \geq 0$. Upon sending $X_i
  = 0$, $\Lambda_i(Y_i)$ takes on the value $+\log
  \frac{1-\varepsilon}{\varepsilon}$ with probability $1 - \varepsilon$ and
  the value $-\log \frac{1-\varepsilon}{\varepsilon}$ with probability
  $\varepsilon$.
\end{Example}

\begin{Example}
  \label{ex:bec:def:1}

  The binary erasure channel (BEC) with erasure probability $0 \leq \epsilon
  \leq 1$ has output alphabet $\set{Y} = \{ 0, 1, \? \}$ and channel law
  $P_{Y|X}(y|x) = 1 - \epsilon$ if $y = x$, $P_{Y|X}(y|x) = \epsilon$ if $y =
  \?$, and $P_{Y|X}(y|x) = 0$ otherwise. The log-likelihood ratio for the
  $i$-th bit is the random variable
  \begin{align}
  \Lambda_i
    &\defeq \Lambda_i(Y_i)
     \defeq
       \log
         \frac{P_{Y_i|X_i}(Y_i|0)}
              {P_{Y_i|X_i}(Y_i|1)}
     = \begin{cases}
         +\infty
              & \text{(if $Y_i = 0$)} \\
         -\infty
              & \text{(if $Y_i = 1$)} \\
         0
              & \text{(if $Y_i = \?$)} \\
       \end{cases}.
  \end{align}
  Upon sending $X_i = 0$, $\Lambda_i(Y_i)$ takes on the value $+\infty$ with
  probability $1 - \epsilon$ and the value $0$ with probability $\epsilon$.
\end{Example}

\begin{Definition}
  \label{def:output:symmetric:channel:1}

  A binary-input memoryless channel $(\set{X} \defeq \{ 0, 1 \}, \set{Y},
  P_{Y|X})$ is called output-symmetric if there is a involution\footnote{An
  involution is a mapping of order two, i.e.~$\sigma(\sigma(y)) = y$ for all
  $y \in \set{Y}$.} $\sigma:\ \set{Y} \to \set{Y}$ and two (possibly
  overlapping) sets $\set{Y}'$ and $\set{Y}''$ such that:
  \begin{itemize}
  
    \item $\set{Y}'' = \sigma(\set{Y}')$, $\set{Y}' = \sigma(\set{Y}'')$,
      $\set{Y}' \cup \set{Y}'' = \set{Y}$.

    \item For every $y' \in \set{Y}'$ we have $P_{Y|X}(y'|0) = P_{Y|X}(y''|1)$
      and $P_{Y|X}(y'|1) = P_{Y|X}(y''|0)$ where $y'' \defeq \sigma(y')$.
  
  \end{itemize}
\end{Definition}

It is easy to see that the three previously discussed channels are
output-symmetric. For the AWGNC one can e.g.~choose $\set{Y}' = \Rp$ and
$\sigma(y') = -y'$, for the BSC one can e.g.~choose $\set{Y}' = \{ 0 \}$ and
$\sigma(y') = 1-y'$, and for the BEC one can e.g.~choose $\set{Y}' = \{ 0, \?
\}$, $\sigma(0) = 1$, $\sigma(1) = 0$, and $\sigma(\?) = \?$.

In the rest of this paper we will focus on a specific class of codes,
channels, and decoders:
\begin{itemize}

  \item The codes are assumed to be binary and linear. (Note that a binary
    code that is defined by a parity-check matrix is automatically binary and
    linear.)

  \item The channels are assumed to be binary-input output-symmetric
    memoryless channels.

  \item The decoders are symmetric with respect to codewords.

\end{itemize}
For this scenario it turns out that the conditional decoding error probability
is independent of the codeword that was sent. Therefore, for understanding
decoders it is sufficient to analyze the case where the all-zeros codeword was
transmitted.

\begin{figure}
  \begin{center}
    \epsfig{file=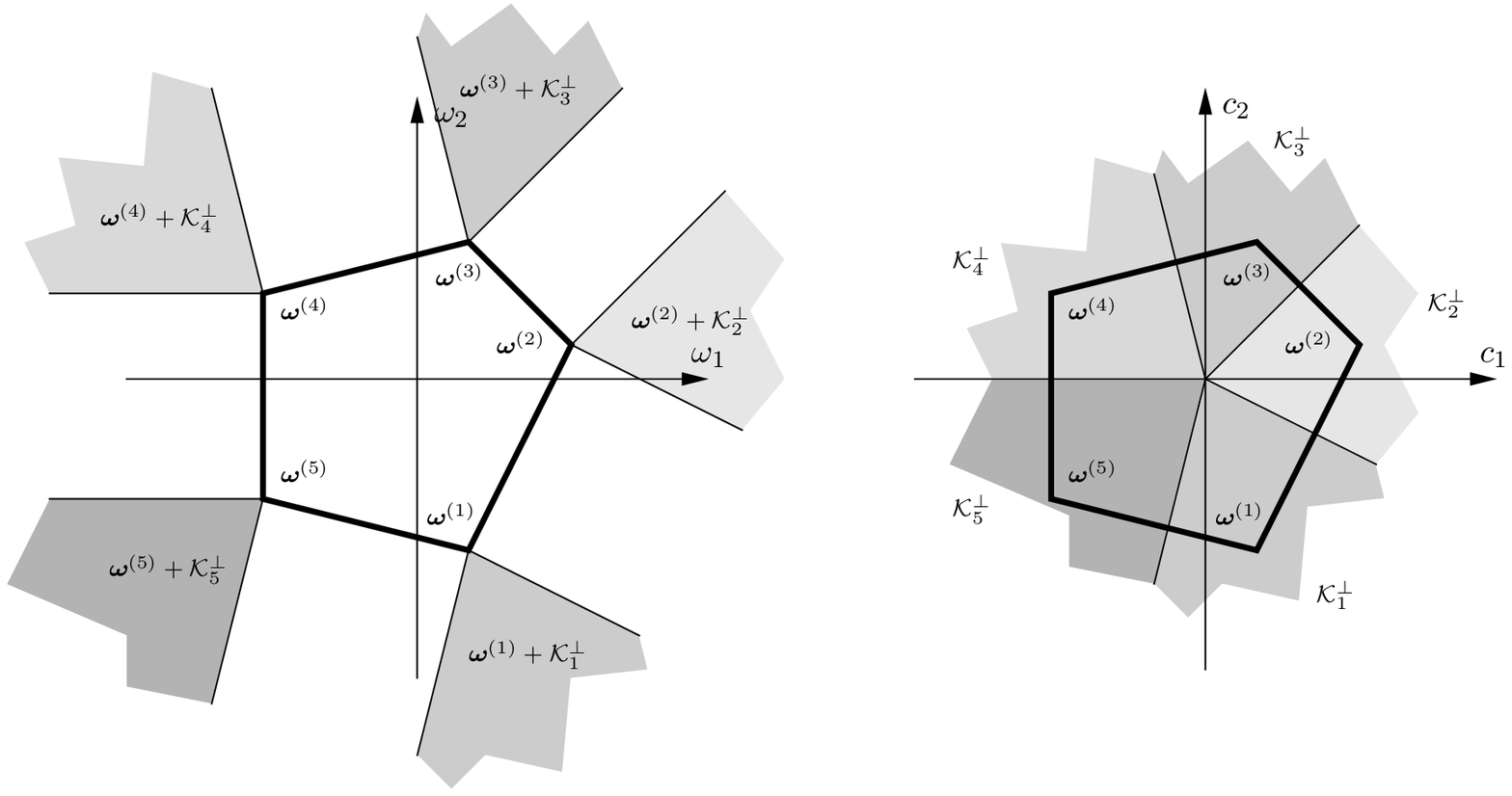, width=10cm, height=6cm}
  \end{center}
  \caption{The set $\set{A} = \convhull\big( \big\{ \vomega^{(1)},
    \vomega^{(2)}, \vomega^{(3)}, \vomega^{(4)}, \vomega^{(5)} \big\} \big)$
    used in Ex.~\ref{example:linprog:1:1}. When the cost vector lies in
    $\set{K}_i^{\perp}$ then the linear program decides in favor vertex
    $\vomega^{(i)}$. Note that the half-rays that constitute the boundaries
    between the decision regions are perpendicular to the corresponding edge
    of the polytope. (In $n$-dimensional space the half-rays that span a
    decision cone are perpendicular to the corresponding facets of the
    polytope.)}
  \label{fig:linprog:1:1}
\end{figure}

\begin{figure}
  \begin{center}
    \epsfig{file=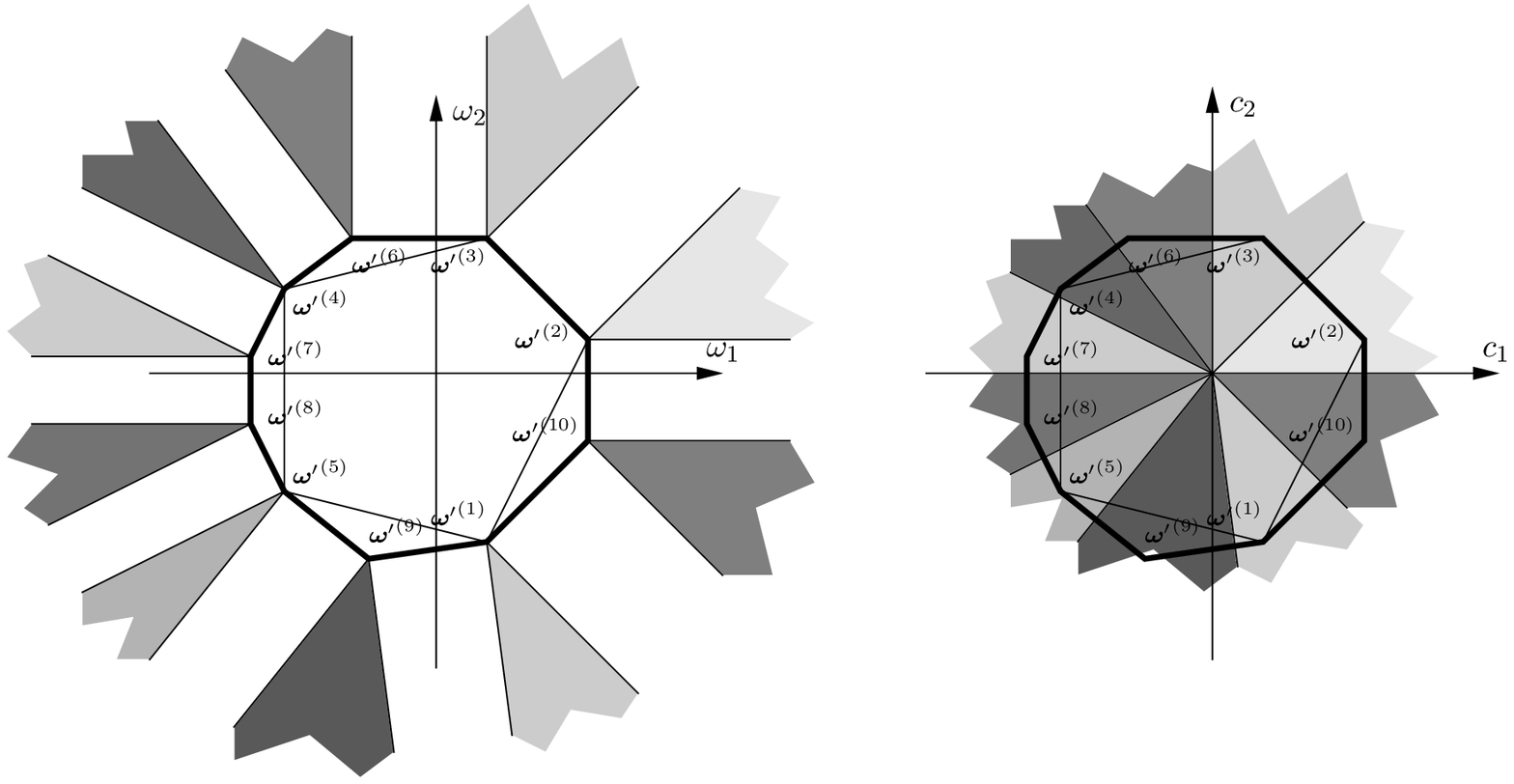, width=10cm, height=6cm}
  \end{center}
  \caption{The set $\set{A} = \convhull\big( \big\{ \vomega^{(1)}, \ldots,
    \vomega^{(10)} \big\} \big)$ used in Ex.~\ref{example:linprog:1:3}.}
  \label{fig:linprog:relaxed:1:1}
\end{figure}

\begin{figure}
  \begin{center}
    \epsfig{file=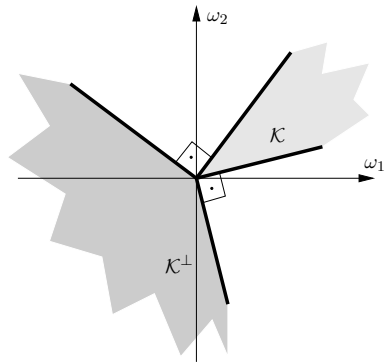, width=5cm}
  \end{center}
  \caption{A cone $\set{K}$ in $\R^2$ and its dual cone
    $\set{K}^{\perp}$. Because $\set{K}$ is a proper cone it holds that
    $\set{K}^{\perp\perp} = \set{K}$.}
  \label{fig:cone:dualcone:1:1}
\end{figure}

The rest of this section will be devoted to recalling some facts from linear
programming that will help to better understand the LPD. Let $n$ be some
positive integer. Consider the following optimization problem
\begin{align}
  \max_{\vomega \in \set{A}} \ 
    \langle \vomega, \vect{c} \rangle
      \label{eq:linprog:1:1}
\end{align}
where $\set{A}$ is a polyhedron in $\R^n$ and cost vector $\vect{c} \in
\R^n$. Such an optimization problem is called a linear program (LP) and the
set of all $\vomega$ that achieve the maximum for a give $\vect{c}$ is called
the optimum set. Because the polyhedra that we are interested in are bounded
we can actually assume that $\set{A}$ is a polytope.\footnote{Here are some
commonly used terms when talking about polytopes: the intersection of an
$n$-dimensional polytope with a tangent hyperplane is called a face,
zero-dimensional faces are known as vertices, one-dimensional faces as edges,
$(n-2)$-dimensional faces as ridges, and $(n-1)$-dimensional faces as
facets. Note that edges and facets of two-dimensional polytopes are both
one-dimensional objects; therefore one must be careful when generalizing a
certain setup to a higher-dimensional space.}

\begin{Example}
  \label{example:linprog:1:1}

  Fig.~\ref{fig:linprog:1:1} (left) shows a possible polytope $\set{A}$ in $n
  = 2$ dimensions with vertices $\vomega^{(i)}$, $i \in [5]$. One way to
  describe the set $\set{A}$ is as the convex combination of the set of
  vertices: $\set{A} = \convhull\big( \big\{ \vomega^{(1)}, \vomega^{(2)},
  \vomega^{(3)}, \vomega^{(4)}, \vomega^{(5)} \big\} \big)$. Another way is to
  describe the set $\set{A}$ as the intersection of half-spaces where each of
  the half-spaces is described by a single linear (affine) inequality.
\end{Example}

A special feature of an LP as in~\eqref{eq:linprog:1:1} is that for any given
$\vect{c}$ there is always a vertex that is optimal.\footnote{For a generic
vector $\vect{c}$ the set of optimal points will contain exactly one vertex of
the polytope. However, for any face of the polytope there is at least one
cost vector $\vect{c}$ such that this face is the optimal set.} Let
$\vomega^{(*)}$ be a vertex of $\set{A}$. An interesting question to ask is
for which vectors $\vect{c}$ the vertex $\vomega^{(*)}$ will be in the optimal
set. To answer this question it is useful to introduce so-called dual cones.

\begin{Definition}
  \label{def:dual:cone:1}

  Let $\set{K}$ be a cone in $\R^n$. The dual cone $\set{K}^{\perp}$ is then
  defined to be set\footnote{The dual cone can be defined by
  $\innerprod{\vx}{\vy} \leq 0$ or by $\innerprod{\vx}{\vy} \geq 0$, here we
  have chosen the first possibility.}
  \begin{align}
    \set{K}^{\perp}
      \defeq \big\{
               \vomega' \in \R^n
               \ | \
               \langle \vomega', \vomega \rangle \leq 0 \ 
               \forall \vomega \in \set{K}
             \big\}.
               \label{eq:def:dual:cone:1}
  \end{align}
\end{Definition}

If $\set{K}$ is a proper cone (cf.~Sec.~\ref{sec:notation:1}) it turns out
that $\set{K}^{\perp}$ is also proper and that $\set{K}^{\perp\perp} =
\set{K}$. Fig.~\ref{fig:cone:dualcone:1:1} shows a possible cone in two
dimensions along with its dual cone. Cones can either be described as the
conic hull of a set of vectors, as the intersection of half-spaces, or a
combination of both. When a cone is described as the conic hull of a set of
vectors then this yields immediately the representation of the dual cone as
the intersection of certain half-spaces. On the other hand, when a cone is
described as the intersection of half-spaces then this yields immediately the
representation of the dual cone as the conic hull of a certain set of vectors.

\begin{Example}
  \label{example:linprog:1:2}

  Consider the same setup as in Ex.~\ref{example:linprog:1:1} and fix some $i
  \in [5]$. It turns out that the set of vectors $\vect{c}$ where
  $\vomega^{(i)}$ is in the optimal set is the set $\set{K}_i^{\perp}$ where
  $\set{K}_i \defeq \conichull\big( \set{A} - \vomega^{(1)} \big)$. The set
  $\set{K}_i^{\perp}$ is shown in Fig.~\ref{fig:linprog:1:1} (right). It is
  also instructive to plot the translated set $\vomega^{(i)} +
  \set{K}_i^{\perp}$ in Fig.~\ref{fig:linprog:1:1} (left). (Note that when the
  maximum operator in~\eqref{eq:linprog:1:1} is replaced by a minimum operator
  then the optimal set is $-\set{K}_i^{\perp}$ where $\set{K}_i \defeq
  \conichull\big( \set{A} - \vomega^{(1)} \big)$ as above.)
\end{Example}

Often it turns out that the linear program in~\eqref{eq:linprog:1:1} is too
complicated to be solved. A possibility is then to solve a tightly related
problem and then to try to infer the solution of the original problem from the
related problem. A popular way of obtaining a related problem is to relax the
set $\set{A}$ to the set $\set{A}'$ and to solve
\begin{align}
  \max_{\vomega \in \set{A}'} \ 
    \langle \vomega, \vect{c} \rangle
      \label{eq:linprog:relaxed:1:1}
\end{align}
Of course, the set $\set{A}'$ should have some desirable properties:
$\set{A}'$ should not be much larger than $\set{A}$ and all vertices of
$\set{A}$ should be vertices of $\set{A}'$.

\begin{Example}
  \label{example:linprog:1:3}

  Consider the same setup as in Ex.~\ref{example:linprog:1:1}. Instead of
  solving~\eqref{eq:linprog:1:1} for the set $\set{A}$ as in
  Fig.~\ref{fig:linprog:1:1} (left) we can solve the relaxed linear
  program~\eqref{eq:linprog:relaxed:1:1} with the set $\set{A}' = \convhull(\{
  {\vomega'}^{(1)}, \ldots, {\vomega'}^{(10)})$ as in
  Fig.~\ref{fig:linprog:relaxed:1:1} (left). We see that $\set{A}'$ fulfills
  the desirable properties that were listed above: $\set{A}'$ is not much
  larger than $\set{A}$ and ${\vomega'}^{(i)} = \vomega^{(i)}$, $i \in
  [5]$. Fig.~\ref{fig:linprog:relaxed:1:1} (right) shows for which $\vect{c}$
  we decide for which vertex. Of course, the regions fulfill
  ${\set{K}'_i}^{\perp} \subseteq \set{K}_i^{\perp}$ for $i \in
  [5]$. Moreover, the fact that $\set{A}'$ tightly resembles $\set{A}$ can
  also be seen from the fact that ${\set{K}'_i}^{\perp}$ is nearly as large as
  $\set{K}_i^{\perp}$ for $i \in [5]$.
\end{Example}

Contemplating Figs.~\ref{fig:linprog:1:1} and~\ref{fig:linprog:relaxed:1:1},
it does not look as if this relaxation really bought us anything. In fact, the
optimization has to be carried out over a more complex region. However, for
higher-dimensional problems the relaxation approach can work very
nicely. E.g.~the fundamental polytope $\fph{P}{H}$ is a relaxation of the set
$\convhull(\code{C})$~\cite{Feldman:03:1, Feldman:Wainwright:Karger:05:1}
which seems to be quite tight especially in the case of LDPC codes. Whereas
$\convhull(\code{C})$ is usually very difficult to describe\footnote{An
exception are e.g.~convolutional codes with not too many states.}, we will see
that the fundamental polytope has a relatively simple description.

We conclude this section with a warning to the uninitiated reader: whereas
two-dimensional pictures of polytopes and cones are very useful to get an
initial understanding of the various definitions, higher dimensional polytopes
and cones can behave quite differently. Note that in the channel coding case
the high-dimensional spaces are unavoidable since it is well known from
information theory that well-performing codes need to have a certain length.


\section{Graph-Cover Decoding}
\label{sec:graph:cover:decoding:1}

This section introduces graph-cover decoding (GCD) which is the theoretical
tool that will help to link LPD and MPID. On the one hand, GCD will be shown
to be essentially equivalent to LPD. On the other hand, we will discuss how
GCD can serve as a model of what is going on in MPID. Sometimes it is an exact
model but usually it is just a very good approximation. The findings in this
section will be corroborated by some simulation results that will be presented
at the end of Sec.~\ref{sec:fp:description:1}.

In the following we assume that we consider data transmission over a channel
as discussed in Sec.~\ref{sec:channels:and:MAPD:LPD:1}.

\begin{Definition}[Lifting]
  Let $\cgraph{T}$ be an arbitrary $M$-cover of $\tgraph{T}{H}$. The
  $M$-lifting of a length-$n$ vector $\vv$ is the vector $\cvv \defeq
  \lift{\vv}{M}$ with entries $\cover{v}_{i,m} \defeq v_i$ for all $(i,m) \in
  \set{I}(\matr{H}) \times [M]$, i.e.~$\cvv$ is a vector of length $Mn$ where
  each entry is repeated $M$ times.
\end{Definition}

We remind the reader of the MAPD/MLD decision rule formulation
in~\eqref{eq:blockwise:MAPD:1} and~\eqref{eq:blockwise:MAPD:2}. That rule aims
to find the codeword that gives the largest log-likelihood ratio given that
$\vy$ was received. GCD extends this idea in the following way: instead of
trying to find the codeword that gives the largest log-likelihood ratio that
$\vy$ was received we want to find the codeword in any finite graph cover that
gives the largest log-likelihood ratio that $\vy$ was received. In order to
obtain a fair comparison we will rescale the log-likelihood ratios by the
order of the cover degree.

However, before formulating GCD more precisely we have to extend the
definition of the channel law. Let $P_{Y|X}(y|x)$ be the channel law of a
memoryless channel. We define the extended joint conditional pmf/pdf of
receiving a vector $\cvy$ of length $Mn$ upon sending a vector $\cvx$ of
length $Mn$ to be
\begin{align}
  P_{\cvY|\cvX}(\cvy|\cvx)
    &= \prod_{i \in [n]}
         \prod_{m \in [M]}
           P_{Y|X}(y_{i,m}|x_{i,m})
\end{align}

\begin{Definition}
  \label{def:graph:cover:decoder:1}

  We define graph-cover decoding (GCD) to be the following decision rule:
  \begin{align}
    (\hM, \hcT, \hcvx)^{\GCD{H}}(\vy) 
      &= \arg \max_{(M, \cgraph{T}, \cvx) \in \cGCDset{H}}
           \frac{1}{M}
             \log
               P_{\cvY|\cvX}(\lift{\vy}{M}|\cvx),
                 \label{eq:graph:cover:decoder:1}
  \end{align}
  where ties are resolved in a systematic or arbitrary way. Moreover, let
  $\hvomega^{\GCD{H}}(\vy) \defeq \vomega\big( \hcvx^{\GCD{H}}(\vy) \big)$.
\end{Definition}

The $1/M$ factor on the right-hand side of~\eqref{eq:graph:cover:decoder:1} is
the promised rescaling factor that makes a fair comparison of the
log-likelihood ratios. Note that the expression
in~\eqref{eq:graph:cover:decoder:1} is also well-defined in the following
sense: let $\vx$ be a codeword in $\code{C}$. Then, for any $M$-cover graph
$\cgraph{T}$ of $\tgraph{T}{H}$ the vector $\cvx \defeq \lift{\vx}{M}$ is a
codeword in $\ctgcode{C}{T}$ with the property that
\begin{align}
  \log
    P_{\vY|\vX}(\vy|\vx)
    &= \frac{1}{M}
       \log
         P_{\cvY|\cvX}(\lift{\vy}{M}|\cvx).
\end{align}
(A similar statement can be made about the relationship of a codeword in some
finite cover to its liftings in finite covers of that finite cover.)

The next proposition shows that GCD and the LPD are essentially equivalent.

\begin{Proposition}
  \label{prop:relationship:gc:dec:lp:dec:1}

  For a given received vector $\vy$, let $\hvomega^{\GCD{H}}(\vy)$ be the GCD
  decision as defined as in Def.~\ref{def:graph:cover:decoder:1} and let
  $\hvomegaLPDH{H}(\vy)$ be the LPD decision of as given in \eqref{eq:LPD:1}
  with $\relaxation(\convhull(\code{C})) = \fph{P}{H}$. Then
  \begin{align}
    \hvomega^{\GCD{H}}(\vy)
      &= \hvomegaLPDH{H}(\vy).
  \end{align}
  (For this statement we assume that if ties appear in either decoder that
  they are resolved in the same way.)
\end{Proposition}

\begin{Proof}
  See Sec.~\ref{sec:proof:prop:relationship:gc:dec:lp:dec:1}.
\end{Proof}

Let us now turn our attention to the connection between GCD and MPID. Recall
our discussion about MPID for the trivial code in
Sec.~\ref{sec:motivating:example:1}. On the one hand, we considered MPID of
the received vector $\vy$ on the base Tanner graph $\graph{T}$ shown in
Fig.~\ref{fig:trivial:code:tg:1:1} and on the other hand, we considered MPID
of $\cvy$ on the triple cover $\cgraph{T}$ shown in
Fig.~\ref{fig:trivial:code:1:2} (left). Because $\graph{T}$ and $\cgraph{T}$
look locally the same, the computation tree for variable node $X_i$ after $t$
iterations will be identical to the computation tree for variable node
$X_{i,m}$ after $t$ iterations, where $m \in [3]$ is arbitrary. This is shown
in Fig.~\ref{fig:trivial:code:1:3} for the variable node $X_2$ and after $t =
2$ iterations. Moreover, under the assumption that $\cvy = \lift{\vy}{3}$ it
can readily be verified that the messages on the two computation trees are the
same. In that way we see that because MPID is operating locally on Tanner
graphs, MPID cannot distinguish if it is decoding the code defined by the base
Tanner $\graph{T}$ graph or any of the codes defined by the finite covers of
$\graph{T}$. If the decoding of these codes is done in a MAPD/MLD fashion,
then MPID is essentially equivalent to GCD, otherwise GCD is just a (usually
very good) approximation to MPID.

There are cases were GCD is the right model for MPID. The list includes Tanner
graphs that are trees (i.e.~have no cycle), codes represented by trellises,
codes represented by tail-biting trellises, and cycle codes (i.e.~codes where
all bit nodes have degree two). Additionally, when we transmit over the BEC
then GCD is also the right model, independently of the Tanner graph of the
code.

In conclusion, we see that the locality, which makes MPID a low-complexity
algorithm, is also the main weakness of MPID.


\section{Properties of Fundamental Polytopes and Cones}
\label{sec:fp:description:1}

The fundamental polytope was introduced in
Def.~\ref{def:fundamental:polytope:1}. In the meantime we have seen that it is
one of the objects of central interest in this paper, namely it turns up when
considering GCD and LPD and because of the closeness of MPID and GCD it seems
to be also important for MPID. It is therefore natural to try to better
understand this object. To that end, this section will look at different ways
of describing the fundamental polytope and will discuss various properties of
it. Actually, we will mostly look at the fundamental cone which is the
fundamental polytope around the vertex $\vect{0}$ and blown up to infinity, in
other words, the conic hull of the fundamental polytope. Understanding the
fundamental cone is sufficient because we restrict ourself to using
binary-input output-symmetric memoryless channels, as was outlined in
Sec.~\ref{sec:channels:and:MAPD:LPD:1}.

\begin{Definition}
  \label{def:fundamental:cone:1}

  The fundamental cone $\fch{K}{H}$ is defined to be the conic hull of the
  fundamental polytope $\fph{P}{H}$, i.e.
  \begin{align*}
    \fch{K}{H}
      &\defeq
         \conichull(\fph{P}{H}).
  \end{align*}
\end{Definition}

From this definition it follows easily that $\fph{P}{H} \subset \fch{K}{H}$
and that for any $\vomega \in \fch{K}{H}$ there is an $\alpha \in \Rpp$ (in
fact, a whole interval of $\alpha$'s) such that $\alpha \cdot \vomega \in
\fph{P}{H}$.

In Ex.~\ref{example:linprog:1:2} we saw that the set of cost vectors where
$\vomega^{(i)}$ is in the optimal set is given by the set $\big(\conichull(
\set{A} - \vomega^{(1)})\big)^{\perp}$. In the case of LPD and GCD, we see
that $\vect{0}$ is in the optimal set when $\vlambda$ lies in
$-\big(\conichull(\fph{P}{H} - \vect{0})\big)^{\perp}$, which equals
$-\fch{K}{H}^{\perp}$.\footnote{Note that LPD/GCD is formulated as a
minimization and not as a maximization problem, therefore the minuses in front
of the dual cones.} This observation emphasize the fact that the fundamental
cone contains all the relevant information and it is sufficient to study the
fundamental cone (instead of the fundamental polytope). For that reason, all
vectors in $\fch{K}{H}$ will be called pseudo-codewords. Moreover, if $\vomega
\in \fch{K}{H}$ and $\{ \alpha \cdot \vomega \, | \, \alpha \in \Rp \}$ is an
edge of the fundamental cone then we call $\vomega$ a \emph{minimal}
pseudo-codeword. This generalizes the notion of minimal
codewords~\cite{Hwang:79:1, Agrell:96:1, Ashikhmin:Barg:98:1,
Borissov:Manev:Nikova:01:1}\footnote{A side remark: interestingly, Decoding
Algorithm~$1$ in~\cite{Hwang:79:1} can be seen as a simplex-type algorithm on
$\convhull(\code{C})$ to solve the LP in~\eqref{eq:blockwise:MAPD:3}.} which
are the edges of $\conichull(\code{C})$.\footnote{For a further discussion of
minimal pseudo-codewords and minimal pseudo-codeword enumerators,
see~\cite{Vontobel:Smarandache:Kiyavash:Teutsch:Vukobratovic:05:1,
Vontobel:Smarandache:05:1, Smarandache:Wauer:05:1:subm}.} Note that although
all codewords are vertices of the fundamental polytope~\cite{Feldman:03:1,
Feldman:Wainwright:Karger:05:1}, a minimal codeword need \emph{not}
necessarily be a minimal pseudo-codeword!  (Given a minimal codeword there are
simple conditions to check if it is a minimal pseudo-codeword; however, we are
not aware of a general result that says when a minimal codeword is also a
minimal pseudo-codeword. Having e.g.~a Tanner graph with girth six is neither
sufficient nor necessary to have all minimal codewords being minimal
pseudo-codewords.)

In Sec.~\ref{sec:graph:covers:and:fundamental:polytope:1} we have seen that
$\GCDset{H}$ and $\fph{P}{H}$ are tightly related. Not surprisingly, there is a
connection between $\cGCDset{H}$ and $\fch{K}{H}$, a connection that is
explored in the following lemma.

\begin{Lemma}
  \label{lemma:integrality:1}

  Remember that if $\cvx$ is a codeword in some $M$-cover $\cgraph{T}$ of
  $\tgraph{T}{H}$, then $M \vomega(\cvx) \in \Intp^n$ is called the unscaled
  pseudo-codeword corresponding to $\cvx$. Let
  \begin{align}
    \set{Z}(\matr{H})
      &\defeq
         \bigcup_{(M,\cgraph{T},\cvx) \in \cGCDset{H}}
           \{ M \vomega(\cvx) \}
  \end{align}
  be the set of all these unscaled pseudo-codewords. This set fulfills
  $\set{Z}(\matr{H}) = \fch{K}{H} \cap \Int^n$ and $\set{Z}(\matr{H}) =
  \code{C} \text{ (in $\GF{2}$)}$. Moreover, for every minimal pseudo-codeword
  $\vomega$ there is an $\alpha \in \Rpp$ (in fact, a whole set of $\alpha$'s)
  such that $\alpha \vomega \in \set{Z}(\matr{H})$.
\end{Lemma}

\begin{Proof}
  See Sec.~\ref{sec:proof:lemma:integrality:1}.
\end{Proof}

\mbox{}

The following lemmas discuss different representations of the fundamental
polytope and cone.

\begin{Lemma}
  \label{lemma:properties:fp:fc:1}

  Let ${\matr{P}'}^{(j)}$ be a $2^{\card{\set{I}_j}-1} \times \set{I}_j$
  matrix containing all the binary vectors of length $|\set{I}_j|$ with even
  Hamming weight, i.e.~the codewords of $\code{C}'_j$, i.e.~the codewords of a
  single-parity-check code of length $\card{\set{I}_j}$.  Let
  ${\matr{P}''}^{(j)}$ be a ${\card{\set{I}_j} \choose 2} \times \set{I}_j$
  matrix containing all the binary vectors of length $|\set{I}_j|$ with
  Hamming weight two. The fundamental polytope $\fp{P} \defeq \fph{P}{H}$ and
  the fundamental cone $\fc{K} \defeq \fch{K}{H}$ can be described by the
  following sets of linear inequalities, respectively:
  \begin{align}
    \fp{P}
      &= \left\{
           \vomega \in \R^n
         \ \left| \
           \begin{array}{ll}
             \forall i \in \set{I}:
               & 0 \leq \omega_i \leq 1 \\
             \forall j \in \set{J}:
               & \vomega_{\set{I}_j} = \valpha^{(j)} {\matr{P}'}^{(j)},\ 
                 \valpha^{(j)} \in \R^{2^{\card{\set{I}_j}-1}},\ 
                 \vect{0} \leq \valpha^{(j)},\  
                 \langle \valpha^{(j)}, \vect{1} \rangle = 1
           \end{array}
           \right.
         \right\},
           \label{eq:properties:fp:1:1} \\
    \fp{K}
      &= \left\{
           \vomega \in \R^n
         \ \left| \
           \begin{array}{ll}
             \forall i \in \set{I}:
               & 0 \leq \omega_i \\
             \forall j \in \set{J}:
               & \vomega_{\set{I}_j} = \valpha^{(j)} {\matr{P}''}^{(j)},\ 
                 \valpha^{(j)} \in \R^{\card{\set{I}_j} \choose 2},\ 
                 \vect{0} \leq \valpha^{(j)}
           \end{array}
           \right.
         \right\}.
           \label{eq:properties:fc:1:1}
      \end{align}
\end{Lemma}

\begin{Proof}
  The expression for $\fp{P}$ is a direct consequence of the definition given
  in~\eqref{eq:fp:def:2} and the expression for $\fc{K}$ is obtained by taking
  the conic hull of $\fp{P}$. Note that because all binary vectors of even
  Hamming weight with Hamming weight larger than two can be written as the
  (integer) sum of several binary vectors of Hamming weight two, we were able
  to replace the matrices $\{ {\matr{P}'}^{(j)} \}$ by the matrices $\{
  {\matr{P}''}^{(j)} \}$ in the expression for $\fc{K}$.
\end{Proof}

\begin{Lemma}
  \label{lemma:properties:fp:fc:2}

  The fundamental polytope $\fp{P} \defeq \fph{P}{H}$ and the fundamental cone
  $\fc{K} \defeq \fch{K}{H}$ can be described by the following sets of linear
  inequalities, respectively:\\[-0.9cm]

  {\small
  \begin{align*}
    \fp{P}
      &= \left\{
           \vomega \in \R^n
         \ \left| \
           \begin{array}{ll}
             \forall i \in \set{I}:
               & 0 \leq \omega_i \leq 1 \\
             \forall j \in \set{J}(\matr{H}),\ 
             \forall \set{I}'_j \subseteq \set{I}_j,\ 
             \card{\set{I}'_j} \text{ odd}:
               & \sum_{i \in \set{I}'_j}
                   \omega_i
                 +
                 \sum_{i \in (\set{I}_j \setminus \set{I}'_j)}
                   (1-\omega_i)
                   \leq \card{\set{I}_j} - 1
           \end{array}
           \right.
         \right\} \\
    \fp{K}
      &= \left\{
           \vomega \in \R^n
         \ \left| \
           \begin{array}{ll}
             \forall i \in \set{I}: 
               & 0 \leq \omega_i \\
             \forall j \in \set{J}(\matr{H}),\ 
             \forall i' \in \set{I}_j:
               & \omega_{i'}
                 -
                 \sum_{i \in (\set{I}_j \setminus \{ i' \})}
                   \omega_i
                   \leq 0
           \end{array}
           \right.
         \right\}
  \end{align*}
      }%
\end{Lemma}

\begin{Proof}
  We do not go into the details of deriving these inequalities. For a
  discussion, see e.g.~\cite{Feldman:Wainwright:Karger:05:1,
  Koetter:Li:Vontobel:Walker:05:1:subm}. Note that the inequalities that
  describe $\fch{K}{H}$ are exactly those inequalities describing $\fph{P}{H}$
  which are homogenous, i.e.~that define half-spaces that go through the
  origin.
\end{Proof}

\begin{table}
  \begin{center}
    \begin{tabular}{|c|c|c|}
      \hline
        Object & Number of variables & Number of (in)equalities \\
      \hline
      \hline
        $\fp{P}$ in Lemma~\ref{lemma:properties:fp:fc:1}
          & $n + \card{\set{J}} 2^{\wrow-1}$ 
          & $2n + \card{\set{J}} (\wrow + 2^{\wrow-1} + 1)$ \\
        $\fc{K}$ in Lemma~\ref{lemma:properties:fp:fc:1}
          & $n + \card{\set{J}} {\wrow \choose 2}$
          & $n + \card{\set{J}} \left(\wrow + {\wrow \choose 2} \right)$ \\
        $\fp{P}$ in Lemma~\ref{lemma:properties:fp:fc:2}
          & $n$
          & $2n + \card{\set{J}} 2^{\wrow-1}$ \\
        $\fc{K}$ in Lemma~\ref{lemma:properties:fp:fc:2}
          & $n$
          & $n + \card{\set{J}} \wrow$ \\
      \hline
    \end{tabular}
  \end{center}
  \caption{The description complexity of the fundamental polytope and cone for
    a $(\wcol,\wrow)$-regular LDPC code of length $n$. Here, $\card{\set{J}} =
    n \wcol / \wrow$.}
  \label{table:fp:fc:description:complexity:1}
\end{table}

Let us consider the description complexities of the various characterizations
of the fundamental polytope and cone in Lemmas~\ref{lemma:properties:fp:fc:1}
and~\ref{lemma:properties:fp:fc:2}.  For reasons of simplicity we consider a
$(\wcol,\wrow)$-regular binary LDPC code, but similar expressions can be
obtained for irregular binary LDPC codes. The number of variables and
(in)equalities that are needed are listed in
Tab.~\ref{table:fp:fc:description:complexity:1}. For the fundamental polytope
we observe a linear behavior in the block length $n$ but an exponential
behavior in the row weight $\wrow$. For binary LDPC codes, where $\wrow$ is a
small number this is usually not a problem because $2^{\wrow}$ is of
reasonable magnitude. But for codes where $\wrow$ is on the order of the block
length $n$ the description complexity obviously grows exponentially in
$n$. Interestingly, as shown
in~\cite[Appendix~II]{Feldman:Wainwright:Karger:05:1}, there is a way to
obtain a description of the fundamental polytope where the number of variables
and the number of (in)equalities grow only polynomially and not exponentially
in $\wrow$. Indeed, the description complexity for that representation turns
out to be on the order of $O(n \card{\set{J}} + \card{\set{J}} \wrow^2 + n
\wcol \wrow)$. While this representation is obviously favorable for $\wrow$'s
on the order of $n$, it is clearly inferior for codes with small
$\wrow$.

Because understanding GCD and LPD is tightly related to understanding the
fundamental cone, the following lemma lists some reformulations on the
(in)equalities that describe the fundamental cone.

\begin{Lemma}
  \label{lemma:reformulations:fc:1}

  For a vector $\vomega \in \R^n$, $\vomega \geq \vect{0}$, the following
  conditions are equivalent
  \begin{itemize}

    \item $\vomega \in \fch{K}{H}$.

    \item For each $j \in \set{J}$ we have
      \begin{align*}
        -\omega'_1 + \omega'_2 + \omega'_3 + \cdots +
         \omega'_{|\set{I}_j|}
          &\geq 0, \\
        +\omega'_1 - \omega'_2 + \omega'_3 + \cdots +
         \omega'_{|\set{I}_j|}
          &\geq 0, \\
        +\omega'_1 + \omega'_2 - \omega'_3 + \cdots +
         \omega'_{|\set{I}_j|}
          &\geq 0, \\
        & \ \,\vdots \\
        +\omega'_1 + \omega'_2 + \omega'_3 + \cdots -
         \omega'_{|\set{I}_j|}
          &\geq 0,
      \end{align*}
      where $\vomega' \defeq \vomega_{\set{I}_j}$.

    \item For each $j \in \set{J}$ we have $\big( \matr{1}_{|\set{I}_j| \times
    |\set{I}_j|} - 2 \cdot \matrunity_{|\set{I}_j| \times |\set{I}_j|} \big)
    \cdot \vomega_{\set{I}_j}^\tr \geq \vect{0}^\tr$, where
    $\matr{1}_{|\set{I}_j| \times |\set{I}_j|}$ is the all-ones matrix of size
    $|\set{I}_j| \times |\set{I}_j|$ and where $\matrunity_{|\set{I}_j| \times
    |\set{I}_j|}$ is the identity matrix of size $|\set{I}_j| \times
    |\set{I}_j|$.

    \item For each $j \in \set{J}$ we have for each $i' \in \set{I}_j$:
      $\sum_{i \in \set{I}_j \setminus \{ i' \}} \omega_i \geq \omega_{i'}$,
      or, equivalently, $\sum_{i \in \set{I}_j} \omega_i \geq 2 \omega_{i'}$.

    \item For each $j \in \set{J}$ we have: $\sum_{i \in \set{I}_j} \omega_i
      \geq 2 \cdot \left( \max_{i \in \set{I}_j} \omega_i \right)$, which can
      also be written as $\sonenorm{\vomega_{\set{I}_j}} \geq 2 \cdot
      \infnorm{\vomega_{\set{I}_j}}$.

  \end{itemize}

\end{Lemma}

\begin{Lemma}
  \label{lemma:fp:tree:1}

  Assume that the Tanner graph $\tgraph{T}{H}$ of a code with parity-check
  matrix $\matr{H}$ is a forest, i.e.~it has no cycles. Then $\fph{P}{H} =
  \convhull(\code{C})$, i.e.~$\fph{P}{H}$ is the convex hull of all the
  codewords.
\end{Lemma}

\begin{Proof}
  See Sec.~\ref{sec:proof:lemma:fp:tree:1}.
\end{Proof}

\mbox{}

One of the consequences of Lemma~\ref{lemma:fp:tree:1} is that GCD and LPD
equal MAPD/MLD for codes that are described by cycle-free Tanner
graphs. Moreover, as is well-known from graphical models, the max-product
algorithm is also equal to the MAPD/MLD in the cycle-free Tanner
case. Unfortunately, as was shown in~\cite{Etzion:Trachtenberg:Vardy:99},
cycle-free Tanner graphs of binary codes, where all constraint nodes are
simple parity-checks, support only weak codes.

\begin{figure}[t]
  \begin{center}
    \epsfig{file=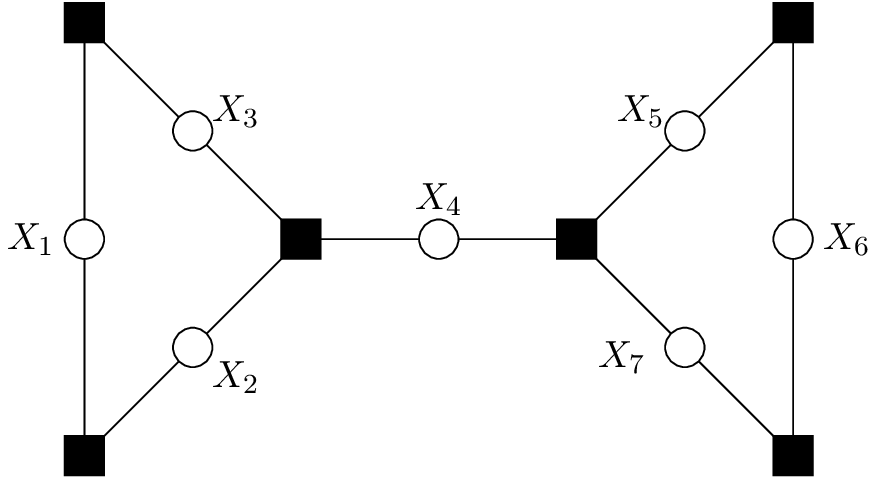, height=3.5cm}
    \quad\quad
    \epsfig{file=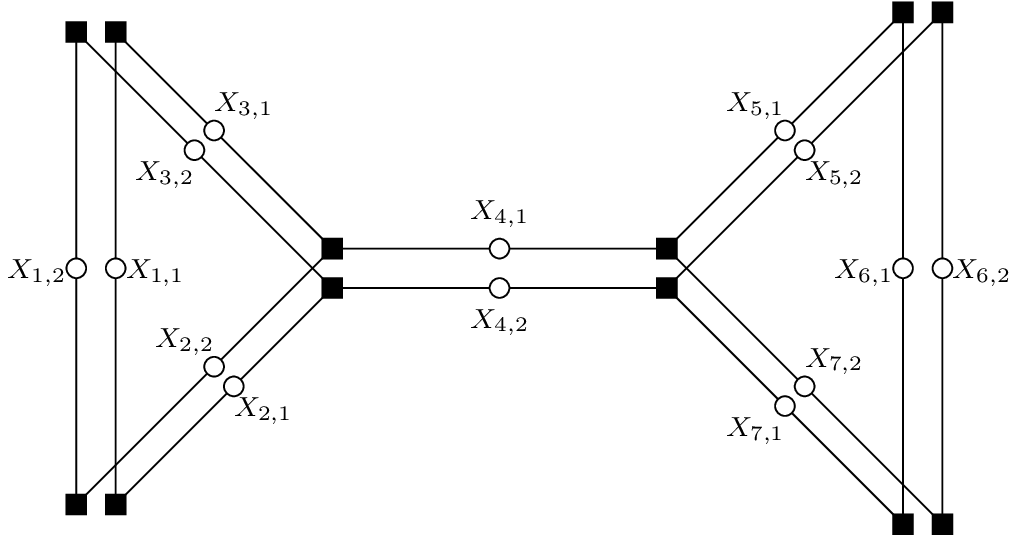, height=3.5cm}

    \caption{Left: Tanner graph $\tgraph{T}{H}$ for the parity-check matrix
    $\matr{H}$ in Ex.~\ref{ex:dumbbell:code:1}. Right: a (possible) double
    $\cgraph{T}$ cover of $\tgraph{T}{H}$.}
  \label{fig:dumbbell:code:1}
  \end{center}
\end{figure}

\begin{figure}[t]
  \begin{center}
    \epsfig{file=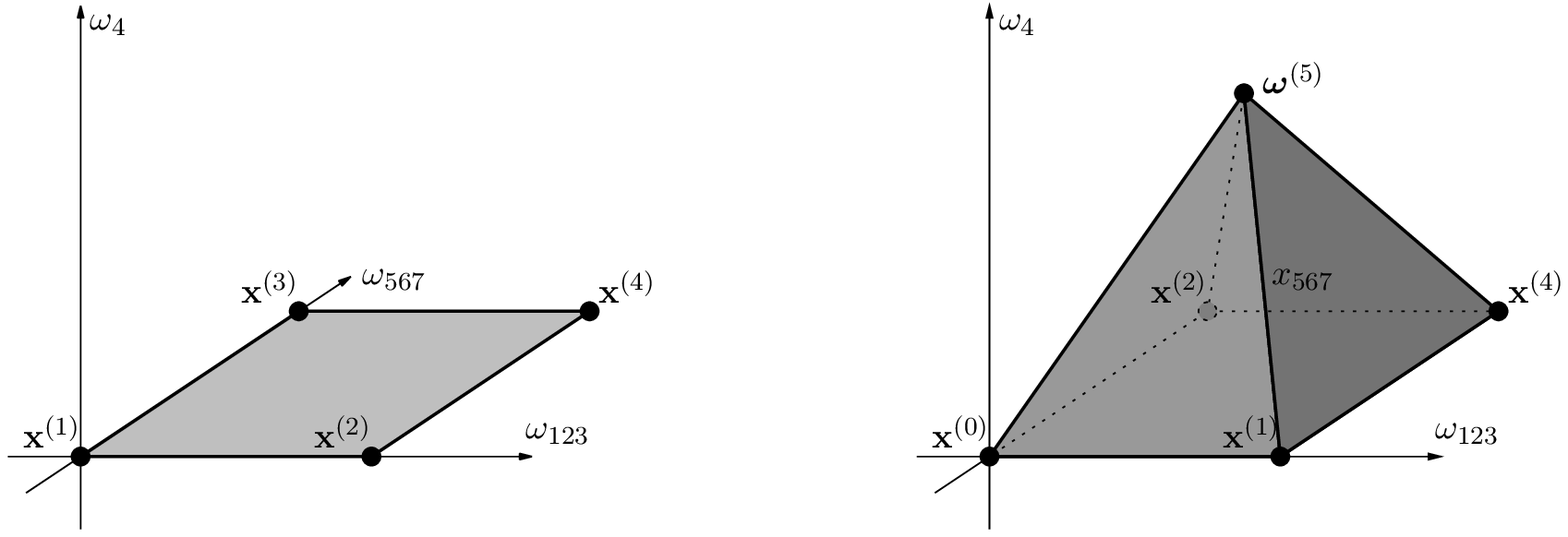, width=12cm}

    \caption{Left: codewords of $\code{C}$ and the polytope
      $\convhull(\code{C})$ for the code in
      Ex.~\ref{ex:dumbbell:code:1}. Right: fundamental polytope $\fph{P}{H}$.}
  \label{fig:dumbbell:code:2}
  \end{center}
\end{figure}

\begin{figure}
  \begin{center}
    \epsfig{file=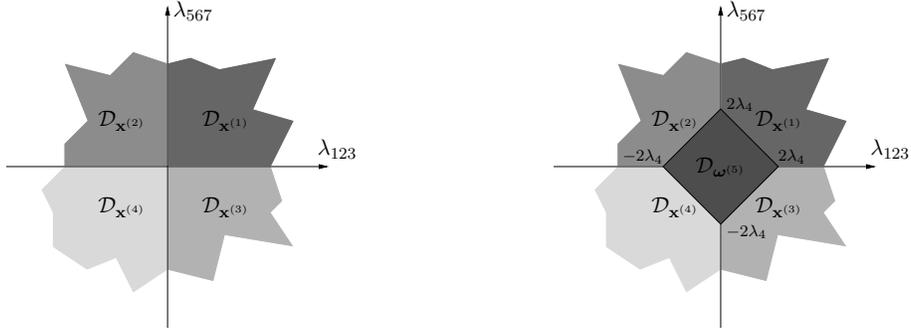, width=12cm}
  \end{center}

  \caption{Decision regions for the code $\code{C}$ described by the
      parity-check matrix ${\matr{H}}$ in Ex.~\ref{ex:dumbbell:code:1}. Left:
      Decision regions for MAPD/MLD (for any $\lambda_4$). These are also the
      decision regions for GCD and LPD if $\lambda_4 \geq 0$. Right: Decision
      regions for GCD and LPD if $\lambda_4 < 0$. (The decision region
      $\set{D}_{\vomega^{(5)}}$ is the square spanned by $(2\lambda_4, 0)$,
      $(0, 2\lambda_4)$, $(-2\lambda_4, 0)$, and $(0, -2\lambda_4, 0)$.)}
  \label{fig:dumbbell:code:3}
\end{figure}

\begin{figure}
  \begin{center}
    \epsfig{file=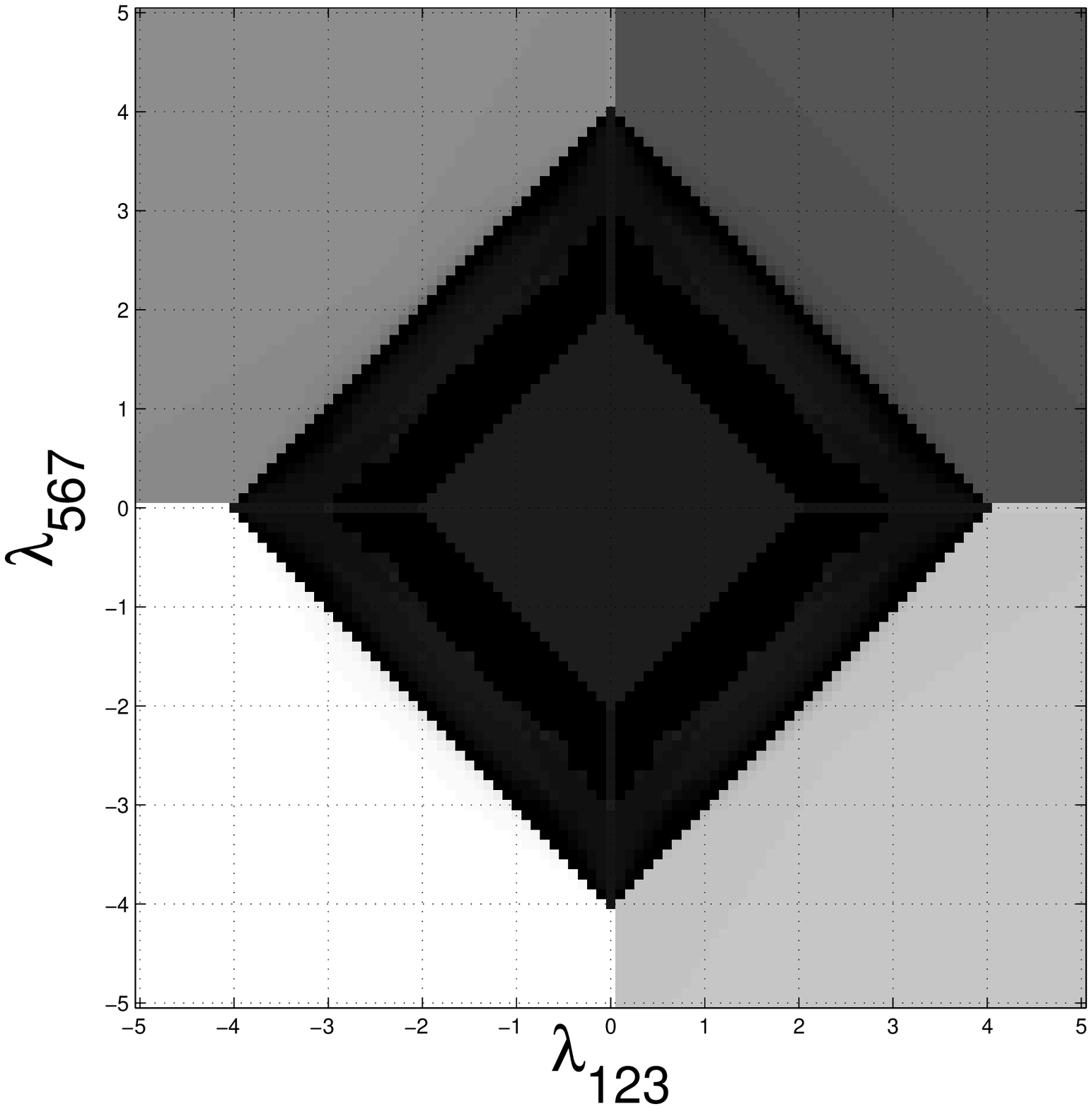, width=4cm}
    \quad\quad\quad
    \epsfig{file=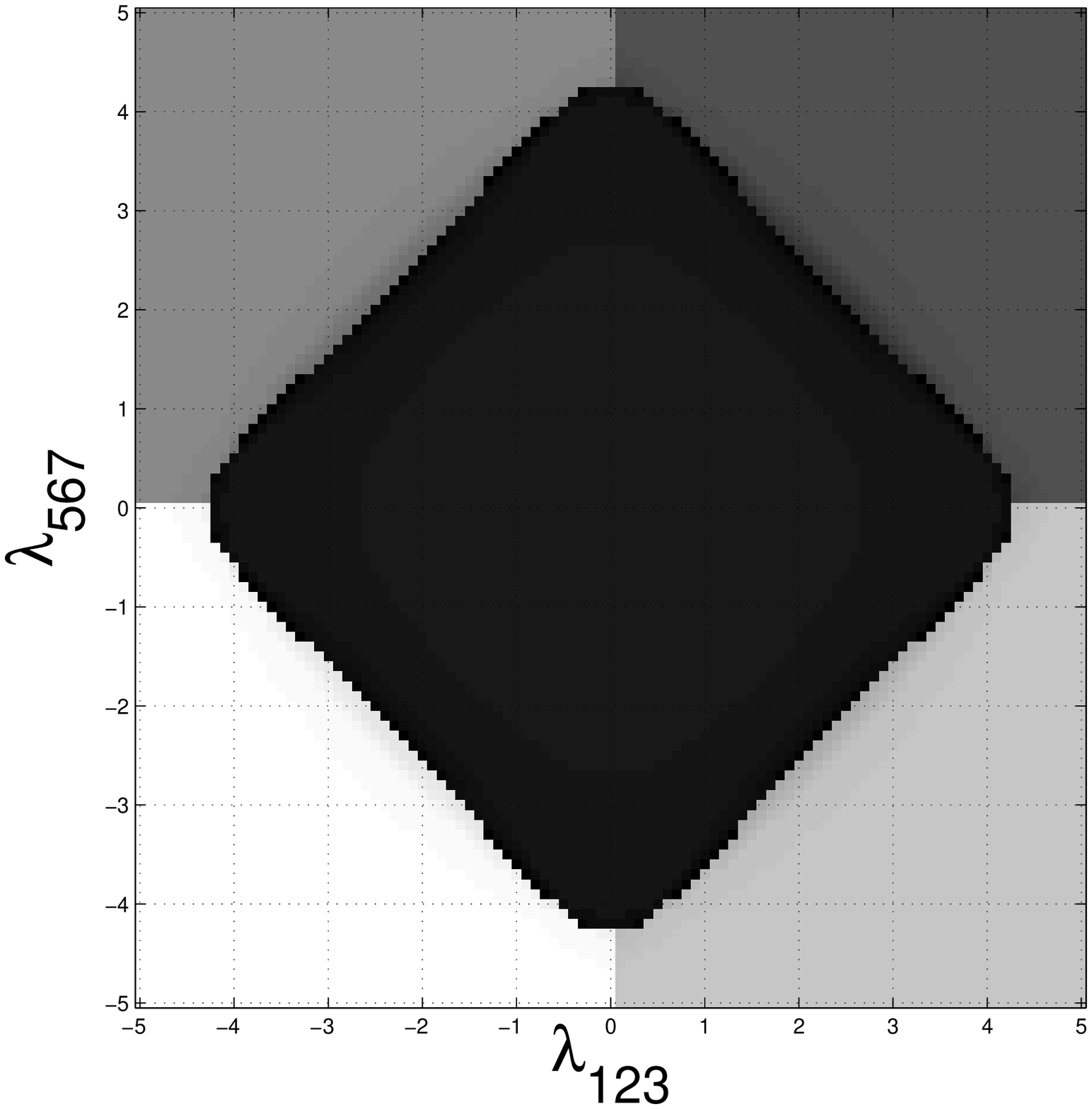, width=4cm}
  \end{center}

  \caption{Decision regions under iterative decoding for $\lambda_4 = -2$ for
      the code $\code{C}$ described by the parity-check matrix ${\matr{H}}$ in
      Ex.~\ref{ex:dumbbell:code:1}. For all simulated $\vlambda$-vectors $30$
      iterations were performed. The shade of the gray indicates the codeword
      decision; within the regions the light differences in the shade of gray
      indicate the convergence time. Note that in the middle square
      corresponding to $\set{D}_{\vomega^{(5)}}$ the decoders did not converge
      to a codeword. Left: Decision regions under MSA decoding. Right:
      Decision regions under SPA decoding.}
  \label{fig:dumbbell:code:4}
\end{figure}

\begin{Example}
  \label{ex:dumbbell:code:1}

  It is usually difficult to show a picture of the fundamental polytope
  because it is a polytope in $\R^n$ and even small codes have usually a block
  length $n$ that is larger than $3$. In this example we discuss a code of
  length $n = 7$ where all the essential features of the fundamental polytope
  can be shown in a three-dimensional space because the effective dimension of
  the fundamental polytope is three.

  The code $\code{C}$ under consideration is the $[7,2,3]$ binary linear code
  with parity-check matrix\footnote{Some of the features of this code were
  also discussed in~\cite{Koetter:Li:Vontobel:Walker:04:1,
  Koetter:Li:Vontobel:Walker:05:1:subm}.}
  \begin{align*}
    \matr{H}
      &= \begin{pmatrix}
           1 & 1 & 0 & 0 & 0 & 0 & 0\\
           1 & 0 & 1 & 0 & 0 & 0 & 0\\
           0 & 1 & 1 & 1 & 0 & 0 & 0\\
           0 & 0 & 0 & 1 & 1 & 0 & 1\\
           0 & 0 & 0 & 0 & 1 & 1 & 0\\
           0 & 0 & 0 & 0 & 0 & 1 & 1
         \end{pmatrix},
  \end{align*}
  whose Tanner graph $\tgraph{T}{H}$ is shown in
  Fig.~\ref{fig:dumbbell:code:1} (left). Because all bit nodes have degree two
  this is a so-called cycle code. It can easily be verified that the code
  $\code{C}$ consists of the four codewords
  \begin{align*}
    {\vect{x}}^{(1)}=(0 0 0 0 0 0 0),\ \ 
    {\vect{x}}^{(2)}=(1 1 1 0 0 0 0),\ \ 
    {\vect{x}}^{(3)}=(0 0 0 0 1 1 1),\ \ 
    {\vect{x}}^{(4)}=(1 1 1 0 1 1 1).
  \end{align*}
  Fig.~\ref{fig:dumbbell:code:1} (right) shows a possible double cover. One
  can check that $\tvx = (1{:}0, 1{:}0, 1{:}0, 1{:}1, 1{:}0, 1{:}0, 1{:}0)$ is
  an (unscaled) pseudo-codeword with $\vomega^{(5)} \defeq \vomega(\tvx) =
  (\frac{1}{2}, \frac{1}{2}, \frac{1}{2}, 1, \frac{1}{2}, \frac{1}{2},
  \frac{1}{2})$. Using Lemma~\ref{lemma:properties:fp:fc:2}, and applying some
  simplifications, the fundamental polytope can be expressed as
  \begin{align*}
    \fph{P}{H}
      = \left\{
          \vomega \in \R^n
            \left|
              \begin{array}{c}
                0 \leq \omega_i \leq 1 \ \forall i \in [7] \\
                \omega_1 = \omega_2 = \omega_3, \
                \omega_5 = \omega_6 = \omega_7 \\
                \omega_4 \leq 2\min(\omega_2, 1-\omega_2, 
                                    \omega_5, 1-\omega_5) 
              \end{array}
            \right.
        \right\}.
  \end{align*}
  It turns out that this fundamental polytope has five vertices: the four
  codewords listed above and the pseudo-codeword just mentioned. Because
  $\omega_1 = \omega_2 = \omega_3$ and $\omega_5 = \omega_6 = \omega_7$, the
  effective dimension of $\fph{P}{H}$ is three and it is sufficient to focus
  on the three-dimensional subspace spanned by $(\omega_{123}, \omega_4,
  \omega_{567})$ where $\omega_{123} \defeq \omega_1 = \omega_2 = \omega_3$
  and $\omega_{567} \defeq \omega_5 = \omega_6 =
  \omega_7$. Fig.~\ref{fig:dumbbell:code:2} (right) shows the fundamental
  polytope in this space. For comparison purposes,
  Fig.~\ref{fig:dumbbell:code:2} (left) shows the four codewords and the
  convex hull thereof (whose effective dimension is two).

  When drawing the decision regions for MAPD/MLD and LPD it turns out to be
  sufficient to consider the three-dimensional space spanned by
  $(\lambda_{123}, \lambda_4, \lambda_{567})$ where $\lambda_{123} \defeq
  \lambda_1 + \lambda_2 + \lambda_3$ and $\lambda_{567} \defeq \lambda_5 +
  \lambda_6 + \lambda_7$. This follows from the fact that $(\lambda_{123},
  \lambda_4, \lambda_{567})$ is a sufficient statistic for MAPD/MLD and LPD
  because $\sum_{i \in [7]} \omega_i \lambda_i = \omega_{123} \lambda_{123} +
  \omega_4 \lambda_4 + \omega_{567} \lambda_{567}$ for any $\vomega \in
  \fph{P}{H}$. For any $\lambda_4$ the MAPD/MLD the decision regions are shown
  in Fig.~\ref{fig:dumbbell:code:3} (left). It is not surprising that the
  value of $\lambda_4$ has no influence on the decision since $x_4$ is known
  to be equal to zero in all codewords. For LPD the decision regions are shown
  Fig.~\ref{fig:dumbbell:code:3} (left) when $\lambda_4 \geq 0$ and in
  Fig.~\ref{fig:dumbbell:code:3} (right) when $\lambda_4 < 0$. Finally, for
  MSA and SPA decoding the decision regions are shown in
  Fig.~\ref{fig:dumbbell:code:4} for $\lambda_4 = -2$. We note that in
  contrast to MAPD/MLD, MSA and SPA decoding cannot exploit that $x_4$ equals
  zero for all valid codewords since {\em no} locally-operating,
  message-passing algorithm can come to this conclusion. Because $\matr{H}$ is
  the parity-check matrix of a cycle code, MSA decoding should behave as
  predicted by GCD, which is indeed the case as shown in
  Fig.~\ref{fig:dumbbell:code:4} (left). Fig.~\ref{fig:dumbbell:code:4}
  (right) indicates that GCD gives also quite accurate predictions for SPA
  decoding for the present code.
\end{Example}

\begin{figure}
  \begin{center}

    \epsfig{file=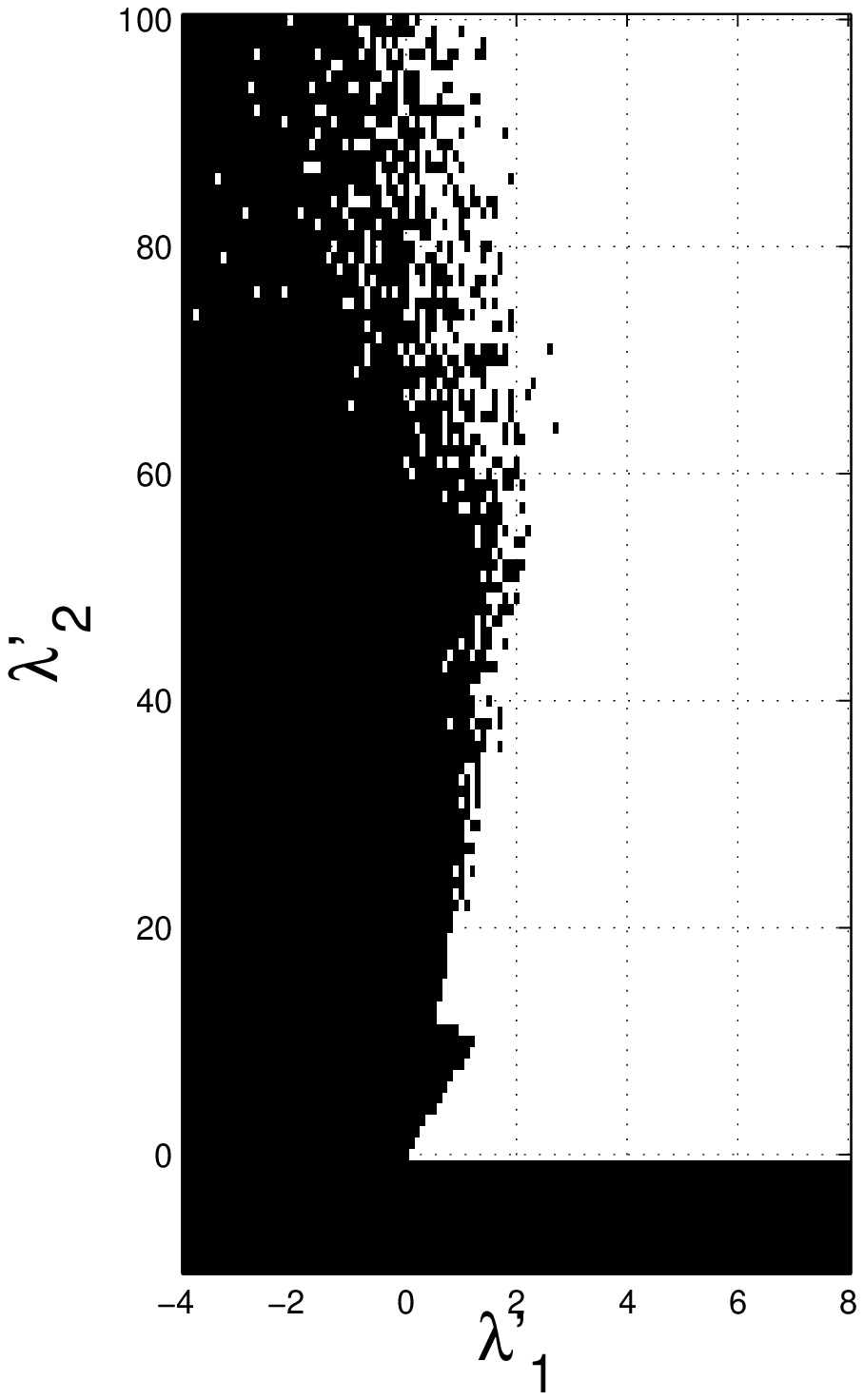,
            width=4cm}
    \quad\quad
    \epsfig{file=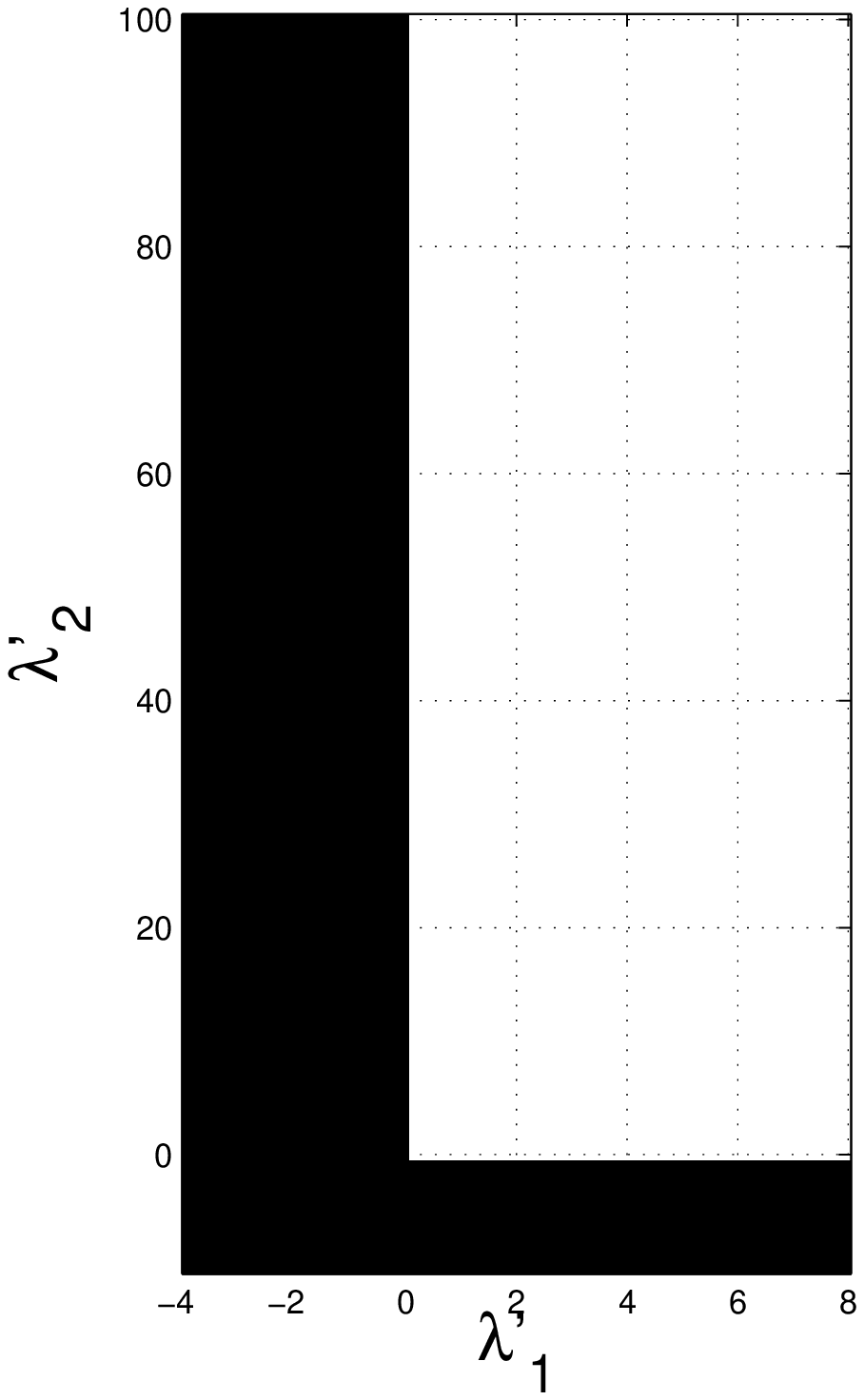, width=4cm}
  \end{center}
  \caption{Decision region plots for the $[155,62]$ binary linear LDPC code in
    Ex.~\ref{ex:code:155:62:1}. Shown is a slice in the plane spanned by
    $\vlambda'_1$ and $\vlambda'_2$ and with $\lambda'_3 = 0$. Observe that
    the $\lambda'_1$-axis is stretched compared to the $\lambda'_2$-axis. (See
    main text for more explanations.) Left: SPA decoding decision regions
    (max.~$100$ iterations). Right LPD decision regions (white: all-zeros
    codeword/black: non-zero (pseudo-)codeword).}
  \label{fig:code:155:62:1}
\end{figure}

\begin{figure}
  \begin{center}
    \epsfig{file=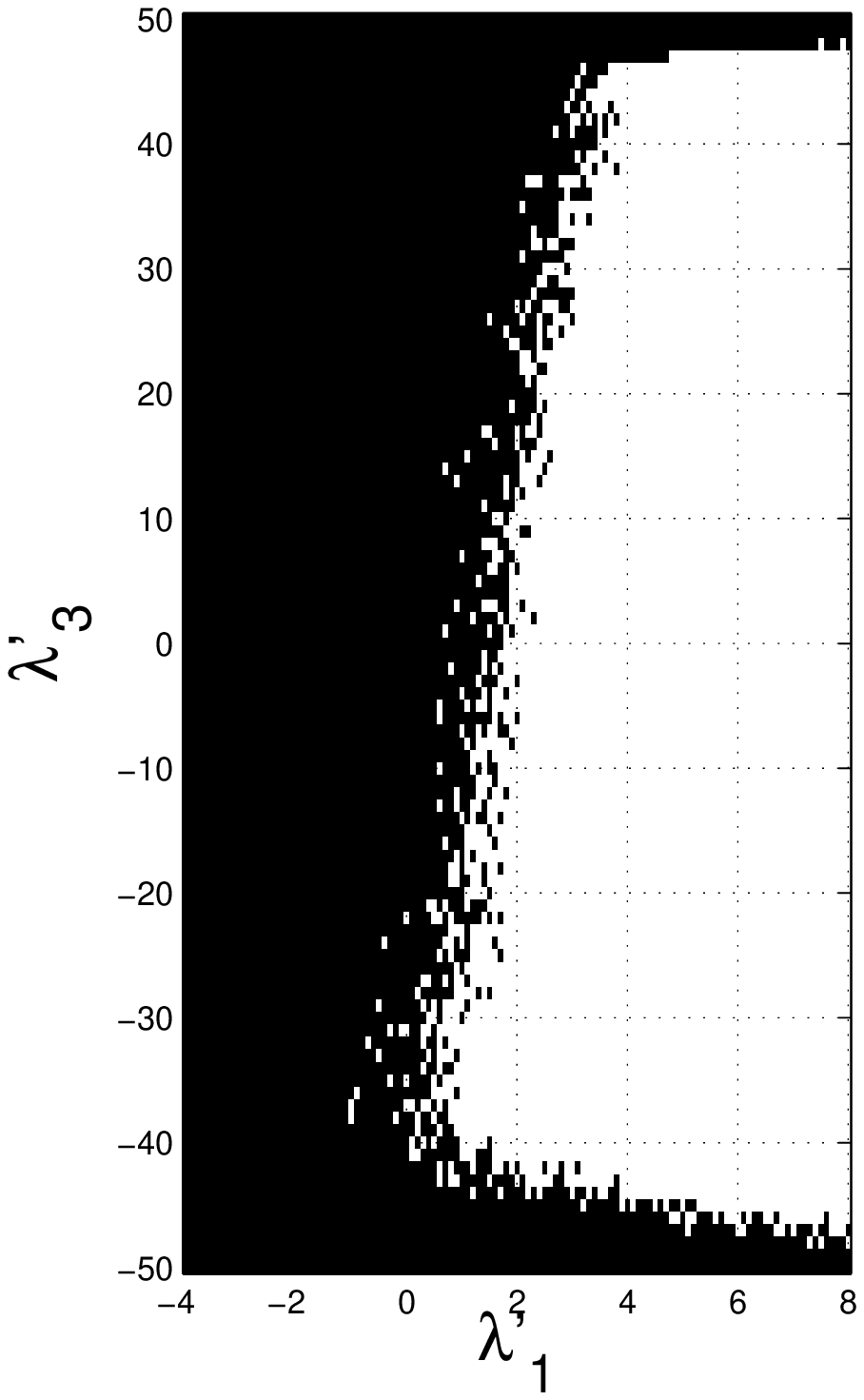, 
            width=4cm}
    \quad\quad
    \epsfig{file=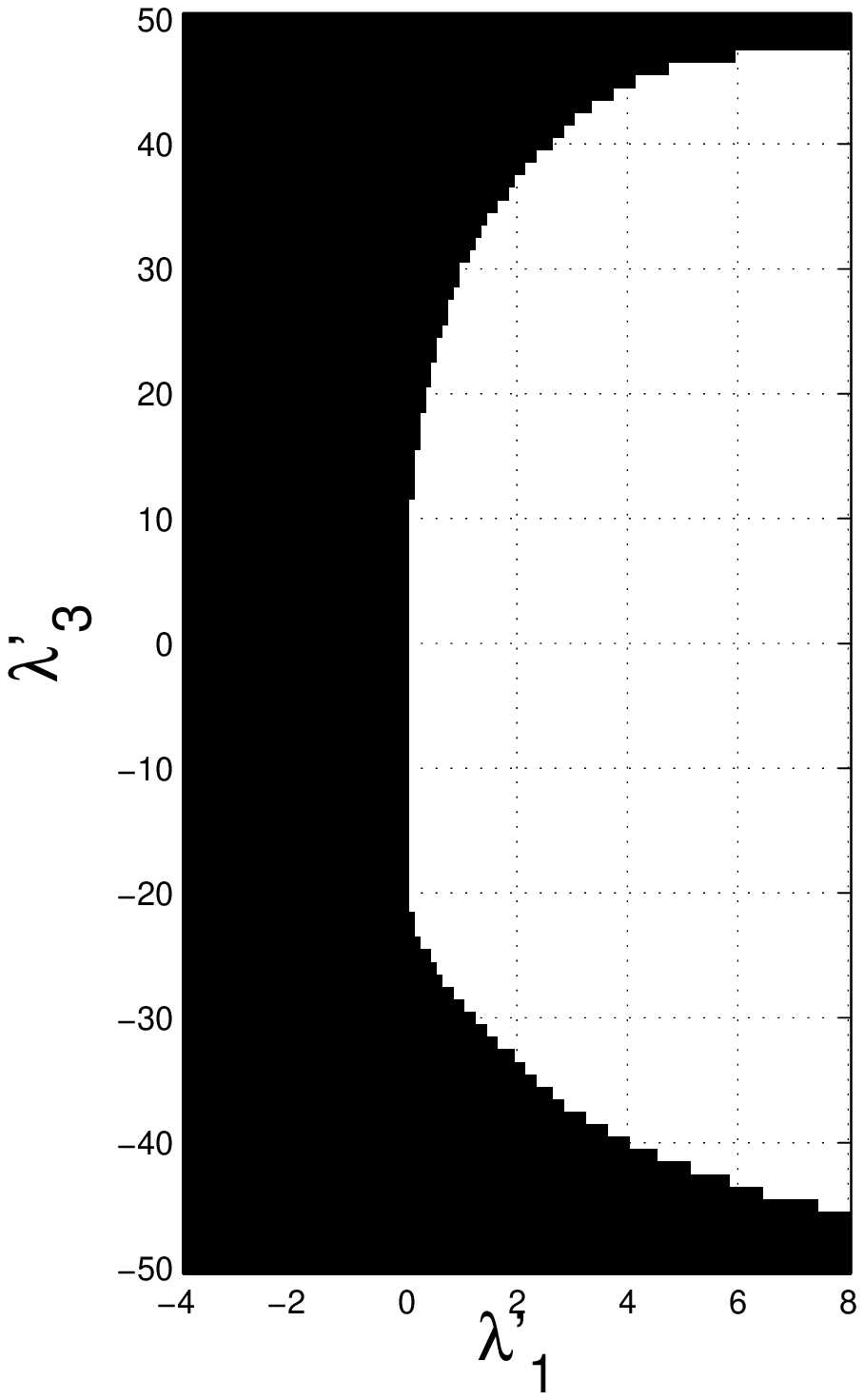, width=4cm}
  \end{center}
  \caption{Decision region plots for the $[155,62]$ binary linear LDPC code in
    Ex.~\ref{ex:code:155:62:1}. Shown is a slice with $\lambda'_2 = 50$ that
    is parallel to the plane spanned by $\vlambda'_1$ and
    $\vlambda'_3$. Observe that the $\lambda'_1$-axis is stretched compared to
    the $\lambda'_3$-axis. (See main text for explanations.) Left: SPA
    decoding decision regions (max.~$100$ iterations). Right LPD decision
    regions (white: all-zeros codeword/black: non-zero (pseudo-)codeword).}
  \label{fig:code:155:62:2}
\end{figure}

\begin{Example}
  \label{ex:code:155:62:1}

  We consider a $(3,5)$-regular $[155,62]$ binary LDPC code based on a
  parity-check matrix of size $93 \times 155$ for data transmission over an
  AWGNC. The parity-check matrix has been \emph{randomly} generated and
  four-cycles have been eliminated. Moreover, the matrix has full rank and so
  the code has rate is exactly $2/5$.

  The full space of LLR vectors is $155$-dimensional. However, for obvious
  practical problems we can only show a two-dimensional slice trough that
  space. Two interesting slices have been picked as follows. We first looked
  for a low-weight minimal pseudo-codeword in the fundamental cone: the one we
  selected has AWGNC pseudo-weight $13.65$. Next, we laid the unit vectors
  $\vlambda'_1$ and $\vlambda'_2$ such that the pairwise decision region
  boundary is the hyperplane defined by $\lambda'_1 = 0$ and such that
  $\Expec[\vLambda \ | \ \vX {=} \vect{0}]$ lies in the plane spanned by
  $\vlambda'_1$ and $\vlambda'_2$. Moreover, the unit vector $\vlambda'_3$ has
  been chosen randomly such that it is orthogonal to $\vlambda'_1$ and
  $\vlambda'_2$.  Given this setup, two slices are shown in
  Figs.~\ref{fig:code:155:62:1} and~\ref{fig:code:155:62:2}, respectively. In
  both cases we compare SPA decoding (with max.~$100$ iterations) and
  LPD. Both plots indicate that the decoding regions of LPD give a very good
  ``first-order'' approximation of SPA decoding.

  Some final comments:
  \begin{itemize}

    \item Using the results of Ex.~\ref{ex:awgnc:def:1} we see that for a
      signal-to-noise ratio of $\Eb/N_0 = 4.197 \mathrm{ dB}$ we have
      $\Expec[\lambda_1' \, | \, \vX {=} \vect{0}] = 15.54$,
      $\Expec[\lambda_2' \, | \, \vX {=} \vect{0}] = 50.00$, and
      $\Expec[\lambda_i' \, | \, \vX {=} \vect{0}] = 0$ for $i \in \set{I}
      \setminus \{ 1,2 \}$. Moreover, $\sqrt{\Var[\lambda_i' \, | \, \vX {=}
      \vect{0}]} = 2.90$ for $i \in \set{I}$.

    \item Let us briefly comment on the white triangle in
      Fig.~\ref{fig:code:155:62:1} in the rectangle $0 \leq \lambda'_1 \apprle
      1$ and $0 \leq \lambda'_2 \apprle 10$. It can easily be shown that for
      $\vlambda$ in the vicinity of the $\vect{0}$, the SPA decoder can only
      decode successfully if $\vlambda > 0$. The above-mentioned white
      triangle corresponds to the region where $\vlambda > 0$ and where
      $\twonorm{\vlambda}$ is small.

    \item Similar plots as in Figs.~\ref{fig:code:155:62:1}
      and~\ref{fig:code:155:62:2} can be obtained under MSA
      decoding. Similarly to SPA decoding, the closer $\vlambda$ lies to the
      decision boundary lies to the decision boundary, the more iterations are
      necessary. However, simulations show that the number of required
      iterations before convergence to the zero codeword increases much more
      in the case of MSA decoding.

  \end{itemize}
\end{Example}

Without going much into the details, let us mention some connections of the
fundamental polytope to concepts like the marginal polytope (and relaxations
thereof), Bethe free energy, and the cycle/metric polytope in matroid
theory. Marginal polytope: when translated to coding theory, the marginal
polytope~\cite{Wainwright:Jordan:03:1} is the polytope spanned by all
codewords, i.e.~$\convhull(\code{C})$; the fundamental polytope is then a
relaxation of this marginal polytope. Bethe free energy: consider the set of
all possible vectors $( \{ b_{X_i}(x_i) \}_{i \in \set{I}(\matr{H})},
\{b_{B_j}(b_j) \}_{j \in \set{J}(\matr{H})} )$ of beliefs on the variable and
check nodes of a Tanner graph. A vector in this set yields a
smaller-than-infinity Bethe free energy~\cite{Yedidia:Freeman:Weiss:05:1} if
and only if the sub-vector containing the beliefs $( \{ b_{X_i}(1) \}_{i \in
\set{I}(\matr{H})} )$ corresponds to a point in the fundamental
polytope. Cycle/metric polytope in matroid theory:\footnote{Here is a small
translation table from coding theory to matroid theory language: codes are
binary matroids, codewords are cycles, and cycle codes are graphic binary
matroids.} the cycle polytope of a binary matroid~\cite{Deza:Laurent:97:1} is
the polytope spanned by all codewords, i.e.~$\convhull(\code{C})$. The metric
polytope is then a certain relaxation of this cycle polytope. In fact, this
relaxation equals $\relaxationsetorder{r}{H}$ in
Def.~\ref{def:relaxation:of:order:r:1} for $r = |\set{J}(\matr{H})|$ and is
therefore the fundamental polytope of the parity-check matrix where all
codewords of the dual code are included. Equivalently, it can also be seen as
the intersection of all fundamental polytopes associated to all possible
parity-check matrices for the given code.


\section{Definition and Properties of Pseudo-Weights}
\label{sec:pseudo:weight:1}

After having seen different descriptions and properties of the fundamental
polytope and cone, we turn our attention now to the question of ``how bad'' a
certain pseudo-codeword is, i.e.~we want to quantify pairwise error
probabilities. Towards this end, let the pairwise error probability
$\PMLD_{\vx \to \vx'}$ between two codewords $\vx$ and $\vx'$ be the
probability that upon sending the codeword $\vx$, MLD decides in favor of
$\vx'$ (assuming that only $\vx$ and $\vx'$ are competing at the
decoder). Similarly, we let the pairwise error probability $\PGCDLPD_{\vx \to
\vomega}$ between a codeword $\vx$ and a pseudo-codeword $\vomega$ be the
probability that upon sending the codeword $\vx$, GCD/LPD decides in favor of
$\vomega$ (assuming that only $\vx$ and $\vomega$ are competing at the
decoder).

In the case of MLD of a binary code, the Hamming distance $\dH(\vx,\vx') =
\wH(\vx'-\vx)$ between two codewords $\vx$ and $\vx'$ is sufficient to deduce
the pairwise error probability $\PMLD_{\vx \to \vx'}$ when transmitting over
an AWGNC, a BSC, or a BEC. However, in the case of GCD/LPD we need different
measures for characterizing the pairwise error probability $\PGCDLPD_{\vx' \to
\vomega}$ of a codeword $\vx$ and a pseudo-codeword $\vomega$. Therefore, in
the following we will discuss the AWGNC, the BSC, and the BEC separately.

\subsection{AWGNC Pseudo-Weight}
\label{sec:pseudo:weight:awgnc:1}

\begin{figure}[t]
  \begin{center}
    \epsfig{file=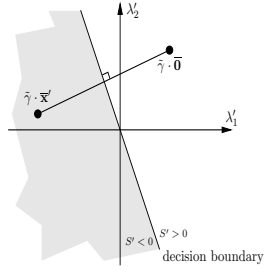, 
            height=3.5cm, width=3.4cm}
    \caption{Decision regions under MLD when only the zero
      codeword is competing against the codeword $\vx$. (See text for more
      details.)}
  \label{fig:MLD:pep:LLR:space:1}
  \end{center}
\end{figure}

\begin{figure}[t]
  \begin{center}
    \epsfig{file=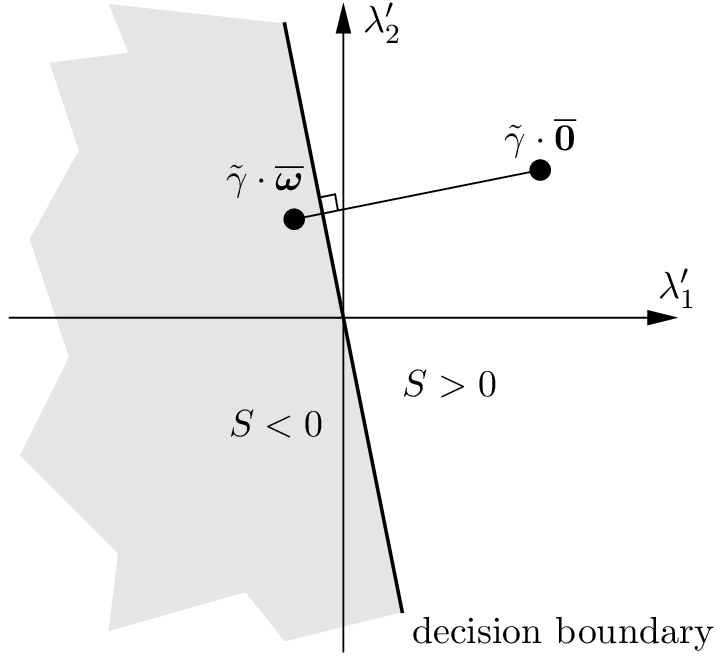, 
            height=3.5cm, width=3.2cm}
    \quad\quad\quad\quad\quad
    \epsfig{file=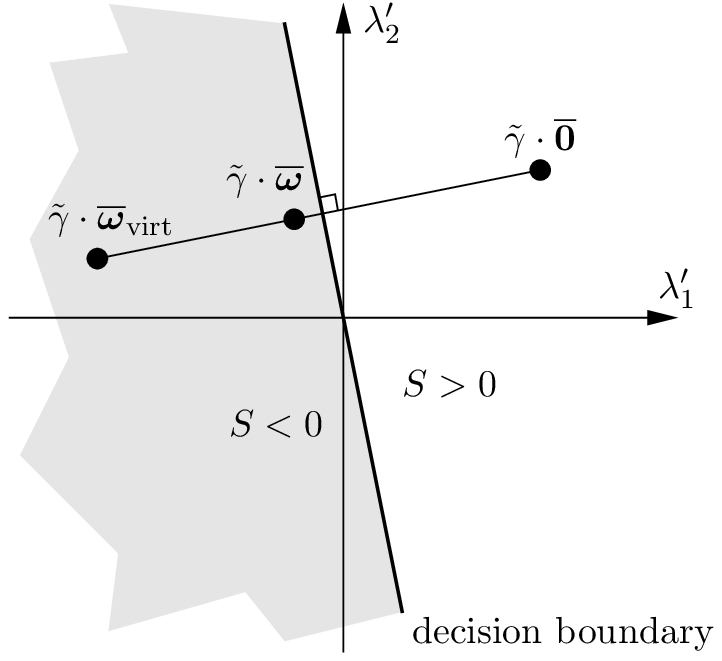, 
            height=3.5cm, width=3.2cm}
    \caption{Left: decision regions under GCD/LPD when only the zero codeword
      is competing against the pseudo-codeword $\vomega$. (See text for more
      details.) Right: same as left part, however, in order to obtain a setup
      similar to the MLD case in Fig.~\ref{fig:MLD:pep:LLR:space:1} we defined
      $\vomega_{\mathrm{virt}} \defeq
      \frac{\onenorm{\vomega}}{\twonorm{\vomega}^2} \cdot \vomega$ such that
      the decision hyperplane is at the same Euclidean distance from $\tilde
      \gamma \cdot \overline{\vect{0}}$ and from $\tilde \gamma \cdot
      \overline{\vomega}_{\mathrm{virt}}$.}
  \label{fig:GCD:MLD:pep:LLR:space:1}
  \end{center}
\end{figure}

We first consider the case of an AWGNC, where we will first study the MLD
pairwise error probability and then the GCD/LPD pairwise error
probability. So, let $\vx' \neq \vect{0}$ be a codeword and define the random
variable $S' \defeq \langle \vx', \vLambda \rangle - \langle \vect{0},
\vLambda \rangle = \sum_{i \in \set{I}:\, x'_i = 1} \Lambda_i$. Knowing that
the $\Lambda_i$'s are statistically independent given $\vX = \vect{0}$
(cf.~Footnote~\ref{footnote:lambda:vect:pdf:1}) and using the results of
Ex.~\ref{ex:awgnc:def:1}, we can easily find the distribution of $\vLambda$
given $\vX = \vect{0}$, i.e.
\begin{align*}
  S'|_{\vX = \vect{0}}
    &\sim
       \normaldistlr{4 R \frac{\Eb}{N_0} \wH(\vx')}
                    {\ 8 R \frac{\Eb}{N_0} \wH(\vx')}.
\end{align*}
Because MLD decides in favor of $\vx'$ and against $\vect{0}$ when $S' \leq 0$
(cf.~\eqref{eq:blockwise:MAPD:3}), the pairwise error probability turns out to
be\footnote{The case $S' = 0$ results in a tie. Depending on how ties are
resolved, MLD might actually decide in favor of $\vect{0}$. However, $P(S' {=}
0 \, | \, \vX {=} \vect{0}) = 0$. \label{footnote:resolving:ties:1}}
\begin{align}
  \PMLD_{\vect{0} \to \vx'}
    &= P(S' {\leq} 0 \, | \, \vX {=} \vect{0})
     = Q
       \left(
         \frac{4 R \frac{\Eb}{N_0} \wH(\vx')}
              {\sqrt{8 R \frac{\Eb}{N_0} \wH(\vx')}}
       \right)
     = Q
       \left(
         \sqrt
         {
           2 R \frac{\Eb}{N_0}
           \wH(\vx')
         }
       \right),
         \label{eq:MLD:pep:awgnc:1}
\end{align}
where $Q(\theta)$ is as usual the integral from $\theta$ to $\infty$ of the
normal distribution with mean $0$ and variance $1$. We see that it is
sufficient to know the Hamming weight of $\vx'$ in order to compute the MLD
pairwise error probability. (In the general case, we need only to know the
Hamming distance between $\vx$ and $\vx'$ in order to compute $\PMLD_{\vx \to
\vx'}$.)

Graphically, the pairwise error probability can be represented as
follows. First, let $\gamma \defeq \sqrt{\Ec} = \sqrt{R\Eb}$ and $\tilde
\gamma \defeq 4 \frac{\sqrt{\Ec}}{N_0} = 4 \frac{\sqrt{R \Eb}}{N_0}$ (note
that $\gamma \tilde \gamma = 4 \frac{\Ec}{N_0} = 4 R
\frac{\Eb}{N_0}$). Secondly, define $\overline{\vect{0}} \defeq \gamma \cdot
(\vect{1} - 2 \cdot \vect{0})$ and $\ovx' \defeq \gamma \cdot (\vect{1} - 2
\cdot \vx)$ (cf.~Ex.~\ref{ex:awgnc:def:1}). Fig.~\ref{fig:MLD:pep:LLR:space:1}
shows the plane of the LLR space that contains the origin, the point $\tilde
\gamma \cdot \overline{\vect{0}}$, and the point $\tilde \gamma \cdot
\ovx$. (The point $\tilde \gamma \cdot \overline{\vect{0}}$ corresponds to the
LLR vector that is obtained at the receiver if $\vect{X} = \vect{0}$ is
transmitted and no noise is added.) Rewriting $S'$ as
\begin{align}
  S'
    &\defeq
       \langle
         \vx', \vLambda
       \rangle
       -
       \langle
         \vect{0}, \vLambda
       \rangle
     = \langle
         \vx' - \vect{0}, \vLambda
       \rangle
     = \left\langle
         \overline{\vect{0}} - \ovx',
         \frac{\vLambda}{2\gamma}
       \right\rangle
     = \frac{1}{2 \gamma \tilde \gamma}
       \left\langle
         \tilde \gamma (\overline{\vect{0}} - \ovx'),
         \vLambda
       \right\rangle
         \label{eq:MLD:inner:product:1}
\end{align}
we see that $S'$ is proportional to the projection of $\vLambda$ onto the
vector connecting $\tilde \gamma \cdot \ovx$ to $\tilde \gamma \cdot
\overline{\vect{0}}$, that $S' = 0$ on the line labeled ``decision boundary'',
and that $S' < 0$ in the shaded area. It can easily be verified that the
squared Euclidean distance from $\tilde \gamma \cdot \overline{\vect{0}}$ to
the decision boundary is $\tilde \gamma^2 \cdot \wH(\vx)$. (The second-to-last
inner product in~\eqref{eq:MLD:inner:product:1} can be seen as doing the
projection in signal space, i.e.~$\vLambda/(2\gamma)$ is projected onto the
vector connecting the signal space point $\ovx$ to the signal space point
$\overline{\vect{0}}$.)

In general, MLD results in a decision hyperplane that consists of all points
that are equally far away from the two competing codewords and so the this
hyperplane does not need to go through the origin. However, when using binary
codes and BPSK signaling all signals have the same energy and so the decision
hyperplane goes through the origin as in Fig.~\ref{fig:MLD:pep:LLR:space:1}.

Now we want to compute the pairwise error probability in the case of
GCD/LPD. Let $\vomega \in \fph{P}{H}$ be a pseudo-codeword and define $S
\defeq \langle \vomega, \vLambda \rangle - \langle \vect{0}, \vLambda \rangle
= \sum_{i \in \set{I}} \omega_i \Lambda_i$. Again, because of the statistical
independence of the $\Lambda_i$'s given $\vX = \vect{0}$ we find that
\begin{align*}
  S|_{\vX = \vect{0}}
    &\sim
       \normaldistlr{4 R \frac{\Eb}{N_0} \sum_{i \in \set{I}} \omega_i}
                    {\ 8 R \frac{\Eb}{N_0} \sum_{i \in \set{I}} \omega_i^2}.
\end{align*}
Because GCD/LPD decides in favor of $\vomega$ and against $\vect{0}$ when $S
\leq 0$ (cf.~\eqref{eq:LPD:1}), the pairwise error probability turns out to
be\footnote{A comment similar to Footnote~\ref{footnote:resolving:ties:1}
applies here.}
\begin{align}
  \PGCDLPD_{\vect{0} \to \vomega}
    &= P(S {\leq} 0 \, | \, \vX {=} \vect{0})
     = Q
       \left(
         \frac{4 R \frac{\Eb}{N_0} \sum_{i \in \set{I}} \omega_i}
              {\sqrt{8 R \frac{\Eb}{N_0} \sum_{i \in \set{I}} \omega_i^2}}
       \right)
     = Q
       \left(
         \sqrt
         {
           2 R \frac{\Eb}{N_0}
           \frac{\left(
                   \sum_{i \in \set{I}} \omega_i
                 \right)^2}
                {\sum_{i \in \set{I}} \omega_i^2}
         }
       \right)
         \label{eq:GCD:LPD:pep:awgnc:1}
\end{align}
It was the idea of Wiberg~\cite{Wiberg:96} to define a generalization of the
Hamming weight such that~\eqref{eq:GCD:LPD:pep:awgnc:1} looks formally
like~\eqref{eq:MLD:pep:awgnc:1}.

\begin{Definition}[\cite{Wiberg:96,
  Forney:Koetter:Kschischang:Reznik:01:1}]
  \label{def:awgn:pseudo:weight:1} Let $\vomega \in \Rp^n$. The AWGNC
  pseudo-weight $\wpsAWGNC(\vomega)$ of $\vomega$ is given by
  \begin{align}
    \wpsAWGNC(\vomega)
      &\defeq
         \frac{\sonenorm{\vomega}^2}
              {\twonorm{\vomega}^2}
       = \frac{\left( \sum_{i \in [n]} \omega_i \right)^2}
              {\sum_{i \in [n]} \omega_i^2},
  \end{align}
  where we define $\wpsAWGNC(\vomega) \defeq 0$ if $\vomega = \vect{0}$.
  Wiberg~\cite[Ch.~6]{Wiberg:96} called this quantity the ``generalized
  weight'', whereas Forney et
  al.~\cite{Forney:Koetter:Kschischang:Reznik:01:1} called it the ``effective
  weight''. (Note that in contrast to the Hamming weight, the AWGNC
  pseudo-weight is not a norm.)
\end{Definition}

With this, Eq.~\eqref{eq:GCD:LPD:pep:awgnc:1} can be written as
\begin{align*}
  \PGCDLPD_{\vect{0} \to \vomega}
    &= Q
       \left(
         \sqrt
         {
           2 R \frac{\Eb}{N_0}
           \wpsAWGNC(\vomega)
         }
       \right)
\end{align*}
which indeed looks formally like~\eqref{eq:MLD:pep:awgnc:1}. With suitable
definitions, the general case $\PGCDLPD_{\vx \to \vomega}$ can also be
formulated by using a generalization of Hamming distance. However, in
contrast to the Hamming distance, the resulting generalization of the Hamming
distance will not be a distance in the mathematical sense.

Similar to the MLD case we can also give a graphical interpretation of the
decision regions in the GCD/LPD case. Fig.~\ref{fig:GCD:MLD:pep:LLR:space:1}
shows the plane through the origin, the point $\tilde \gamma \cdot
\overline{\vect{0}}$, and the point $\tilde \gamma \cdot
\overline{\vect{x}}$. Rewriting $S$ as
\begin{align}
  S
    &\defeq
       \langle
         \vomega, \vLambda
       \rangle
       -
       \langle
         \vect{0}, \vLambda
       \rangle
     = \langle
         \vomega - \vect{0}, \vLambda
       \rangle
     = \left\langle
         \overline{\vect{0}} - \ovomega,
         \frac{\vLambda}{2\gamma}
       \right\rangle
     = \frac{1}{2 \gamma \tilde \gamma}
       \left\langle
         \tilde \gamma (\overline{\vect{0}} - \ovomega),
         \vLambda
       \right\rangle
         \label{eq:GCD:LPD:inner:product:1}
\end{align}
we see that $S$ is proportional to the projection of $\vLambda$ onto the
vector connecting $\tilde \gamma \cdot \ovomega$ to $\tilde \gamma \cdot
\overline{\vect{0}}$, that $S = 0$ on the line labeled ``decision boundary'',
and that $S < 0$ in the shaded area. (The second-to-last inner product
in~\eqref{eq:GCD:LPD:inner:product:1} can be seen as doing the projection in
signal space, i.e.~$\vLambda/(2\gamma)$ is projected onto the vector
connecting the signal space point $\ovx$ to the signal space point
$\overline{\vect{0}}$.) In contrast to MLD, the two points $\tilde \gamma
\cdot \overline{\vect{0}}$ and $\tilde \gamma \cdot \ovomega$ do not have the
same distance from the decision boundary in general; in fact, it can even
happen that the two points lie on the same side of the decision
boundary. Finally, note that the squared Euclidean distance of $\tilde \gamma
\cdot \overline{\vect{0}}$ to the decision boundary is now given by $\tilde
\gamma^2 \cdot \wpsAWGNC(\vomega)$, which looks formally like the formula that
we obtained in the case of MLD.

It is clear that these geometrical observations can be connected to the
discussion on linear programming at the end of
Sec.~\ref{sec:channels:and:MAPD:LPD:1}; the details of this connection are
left to the reader as an exercise.

\subsection{BSC Pseudo-Weight}
\label{sec:pseudo:weight:bsc:1}

We first discuss MLD. Defining $S'$ as in Sec.~\ref{sec:pseudo:weight:awgnc:1}
for a codeword $\vx' \neq \vect{0}$, we see that a necessary condition for
$S'|_{\vX = \vect{0}}$ to be non-positive is that the number of bit flips on
the channel is at least $\frac{1}{2} \wH(\vx')$. The BSC pseudo-weight is
defined such that we can formally make the same statement for GCD/LPD.

\begin{Definition}[\cite{Forney:Koetter:Kschischang:Reznik:01:1}]
  \label{def:bsc:pseudo:weight:1}

  Let $\vomega \in \Rp^n$. Let $\vomega'$ be a vector of length $n$ with the
  same components as $\vomega$ but in non-increasing order. Introducing
  \begin{align*}
    f(\xi)
      &\defeq \omega'_i \quad (i-1 < \xi \leq i,\ 0 < \xi \leq n), \\
    F(\xi)
      &\defeq \int_{0}^{\xi} f(\xi') \dint{\xi'}, \\
    e
      &\defeq F^{-1} \left( \frac{F(n)}{2} \right),
  \end{align*}
  the BSC pseudo-weight $\wpsBSC(\vomega)$ is defined to be $\wpsBSC(\vomega)
  \defeq 2e$.\footnote{Note that the quantity $e$ is obviously related to the
  median of the ``pdf'' given by $f(\xi)/F(n)$. However, let us remark that
  this is a different ``distribution'' than used later on in
  Lemma~\ref{lemma:pseudo:weight:variance:interpretation:1} when
  characterizing the AWGNC pseudo-weight.}
\end{Definition}

With this definition and $S$ defined as in
Sec.~\ref{sec:pseudo:weight:awgnc:1} we see that a necessary condition for
$S|_{\vX = \vect{0}}$ to be non-positive is that the number of bit flips on
the channels is at least $\wpsBSC(\vomega)/2$. Note however that the BSC
pairwise error probability formulas for GCD/LPD are not simply obtained from
the BSC pairwise error probability formulas for MLD by replacing the Hamming
weight by the BSC pseudo-weight. Namely, whereas in the case of MLD it only
matters how many channel bit flips correspond to positions in $\supp(\vx')$,
in the case of GCD/LPD it not only matters how many channel bit flips
correspond to positions in $\supp(\vomega)$ but also at which position these
bit flips are.

Another way to generalize the Hamming weight in the case of the BSC is given
by the fractional and max-fractional weight.

\begin{Definition}[\cite{Feldman:03:1}]
  \label{def:fractional:and:max:fractional:weight:1}

  The fractional and max-fractional weight of a vector $\vomega \in \Rp^n$ are
  defined to be, respectively,
  \begin{align}
    \wfr(\vomega)
      &= \sonenorm{\vomega}. \\
    \wmaxfr(\vomega)
      &\defeq
         \frac{\wfr(\vomega)}
              {\infnorm{\vomega}}
       = \frac{\sonenorm{\vomega}}
              {\infnorm{\vomega}}.
  \end{align}
  For $\vomega = \vect{0}$ we define $\wmaxfr(\vomega) \defeq 0$. We actually
  use a slightly different notation than~\cite{Feldman:03:1}. Here, $\wfr$ and
  $\wmaxfr$ are defined for any vector in $\Rp^n$, whereas
  in~\cite{Feldman:03:1}, $\wfr$ and $\wmaxfr$ already denote the minimum of
  these values over all nonzero vertices of the fundamental polytope.
\end{Definition}

Fix some non-zero vector $\vomega \in [0,1]^n$. Using the above definition, it
can be seen that a necessary condition for $S|_{\vX = \vect{0}}$ to be
non-positive is that the number of bit flips on the channel is at least
$\frac{1}{2} \wfr(\vomega)$. Similarly, fix some non-zero vector $\vomega \in
\Rp^n$. Then, a necessary condition for $S|_{\vX = \vect{0}}$ to be
non-positive is that the number of bit flips on the channel is at least
$\frac{1}{2} \wmaxfr(\vomega)$. (The details of these two statements can be
found in
Sec.~\ref{sec:proof:after:def:fractional:and:max:fractional:weight:1}.)

\subsection{BEC Pseudo-Weight}
\label{sec:pseudo:weight:bec:1}

We first discuss the MLD. Defining $S'$ as in
Sec.~\ref{sec:pseudo:weight:awgnc:1} for a codeword $\vx' \neq \vect{0}$, we
see that a necessary condition for $S'|_{\vX = \vect{0}}$ to be
non-positive\footnote{Because of special properties of the BEC, $S$ can never
be negative.} is that the number of erasures on the channel is at least
$\wH(\vx')$. The BEC pseudo-weight is defined such that we can formally make
the same statement for GCD/LPD.

\begin{Definition}[\cite{Forney:Koetter:Kschischang:Reznik:01:1}]
  \label{def:bec:pseudo:weight:1}

  Let $\vomega \in \Rp^n$. The BEC pseudo-weight $\wpsBEC(\vomega)$ is defined
  to be
  \begin{align*}
    \wpsBEC(\vomega)
      &= \card{\supp(\vomega)}.
  \end{align*}
\end{Definition}

With this definition and $S$ defined as in
Sec.~\ref{sec:pseudo:weight:awgnc:1} we see that a necessary condition for
$S|_{\vX = \vect{0}}$ to be non-positive is that the number of bit flips on
the channels is at least $\wpsBEC(\vomega)$. In contrast to the BSC, the BEC
pairwise error probability formulas for GCD/LPD are simply obtained from the
BEC pairwise error probability formulas for MLD by replacing the Hamming
weight by the BEC pseudo-weight. (Note that the exact formulas depend on how
ties are resolved.)

\subsection{Pseudo-Weight Properties}

This section collects different lemmas that characterize the different
pseudo-weights and the fractional and max-fractional weights.

\begin{Lemma}
  \label{lemma:weight:scaling:invariance:1}

  The AWGNC, BSC, and BEC pseudo-weights and the max-fractional weight are
  invariant under scaling by a positive scalar, i.e.
  \begin{align*}
    \wpsAWGNC(\alpha \cdot \vomega)
      &= \wpsAWGNC(\vomega), \\
    \wpsBSC(\alpha \cdot \vomega)
      &= \wpsBSC(\vomega), \\
    \wpsBEC(\alpha \cdot \vomega)
      &= \wpsBEC(\vomega), \\
    \wmaxfr(\alpha \cdot \vomega)
      &= \wmaxfr(\vomega),
  \end{align*}
  for any $\alpha \in \Rpp$ and any $\vomega \in \Rp^n$. Note that the
  fractional weight is not scaling-invariant.
\end{Lemma}

\begin{Proof}
  Follows easily from the definitions.
\end{Proof}

\begin{Lemma}
  If $\vomega \in \{ 0, 1 \}^n$ then the AWGNC, the BSC, and the BEC
  pseudo-weights and the fractional and max-fractional weight reduce to the
  Hamming weight, i.e.~$\wpsAWGNC(\vomega) = \wH(\vomega)$, etc.
\end{Lemma}

\begin{Proof}
  This is straightforward. E.g.~in the case of an AWGNC the result follows
  from observing that $\sonenorm{\vomega} = \wH(\vomega)$ and that
  $\twonorm{\vomega} = \sqrt{\wH(\vomega)}$ which implies that
  $\wpsAWGNC(\vomega) = \sonenorm{\vomega}^2/\twonorm{\vomega}^2 =
  \wH(\vomega)^2/\wH(\vomega) = \wH(\vomega)$.
\end{Proof}

\mbox{}

The following definitions generalize the notion of the minimum Hamming weight
of a binary linear code.

\begin{Definition}
  \label{def:minimum:pseudo:weight:1}

  The minimum AWGNC, BSC, and BEC pseudo-weight and the minimum fractional and
  max-fractional weights are defined to be, respectively,
  \begin{align*}
    \wpsAWGNCminh{H}
      &\defeq
         \min_{\vomega \in \setV(\fph{P}{H}) \setminus \{ 0 \}}
           \wpsAWGNC(\vomega), \\
    \wpsBSCminh{H}
      &\defeq
         \min_{\vomega \in \setV(\fph{P}{H}) \setminus \{ 0 \}}
           \wpsBSC(\vomega), \\
    \wpsBECminh{H}
      &\defeq
         \min_{\vomega \in \setV(\fph{P}{H}) \setminus \{ 0 \}}
           \wpsBEC(\vomega), \\
    \wfrminh{H}
      &\defeq
         \min_{\vomega \in \setV(\fph{P}{H}) \setminus \{ 0 \}}
           \wfr(\vomega), \\
    \wmaxfrminh{H}
      &\defeq
         \min_{\vomega \in \setV(\fph{P}{H}) \setminus \{ 0 \}}
           \wmaxfr(\vomega),
  \end{align*}
  where $\setV(\fph{P}{H}) \setminus \{ 0 \}$ is the set of all non-zero
  vertices of the fundamental polytope $\fph{P}{H}$.
\end{Definition}

It is important to note that the above minimal weights depend on the choice of
parity-check matrix $\matr{H}$, i.e.~different parity-check matrices for the
same code can lead to different minimal weights. This is in contrast to the
minimal Hamming weight of a code which is independent of the specific choice
of parity-check matrix by which a binary linear code is represented.

\begin{Lemma}
  \label{lemma:minimum:pseudo:weight:1}

  \begin{alignat*}{2}
    \wpsAWGNCminh{H}
       &= \min_{\vomega \in \fph{P}{H} \setminus \{ 0 \}}
           \wpsAWGNC(\vomega)
     &&= \min_{\vomega \in \fch{K}{H} \setminus \{ 0 \}}
           \wpsAWGNC(\vomega), \\
    \wpsBSCminh{H}
      &= \min_{\vomega \in \fph{P}{H} \setminus \{ 0 \}}
           \wpsBSC(\vomega)
     &&= \min_{\vomega \in \fch{K}{H} \setminus \{ 0 \}}
           \wpsBSC(\vomega), \\
    \wpsBECminh{H}
      &= \min_{\vomega \in \fph{P}{H} \setminus \{ 0 \}}
           \wpsBEC(\vomega)
     &&= \min_{\vomega \in \fch{K}{H} \setminus \{ 0 \}}
           \wpsBEC(\vomega), \\
    \wmaxfrminh{H}
      &= \min_{\vomega \in \fph{P}{H} \setminus \{ 0 \}}
           \wmaxfr(\vomega)
     &&= \min_{\vomega \in \fch{K}{H} \setminus \{ 0 \}}
           \wmaxfr(\vomega).
  \end{alignat*}
  Note that there is no such statement for the fractional weight.
\end{Lemma}

\begin{Proof}
  These are simple consequences of the fact that the AWGNC, BSC, and BEC
  pseudo-weights and the max-fractional weight are scaling-invariant, that
  Lemma~\ref{lemma:convex:combination:pseudo:distance:lower:bound:1} holds,
  and that $\fch{K}{H} \setminus \{ 0 \} = \conichull(\fph{P}{H}) \setminus \{
  0 \}$.
\end{Proof}

\mbox{}

In the following, our standard channel will be the AWGNC. Therefore, when
nothing else is specified, pseudo-weight will mean AWGNC pseudo-weight and we
will write $\wps(\vomega)$ and $\wpsminh{H}$ instead of $\wpsAWGNC(\vomega)$
and $\wpsAWGNCminh{H}$, respectively.

\begin{Lemma}
  \label{lemma:pseudo:weight:variance:interpretation:1}

  Let $\vomega \in \Rp^n$ and let $\set{S} \defeq \supp(\vomega)$ be its
  support. Consider the non-zero entries of $\vomega$ to be $|\set{S}|$
  samples of a positive random variable $\Omega$. Introducing the empirical
  first moment (mean) $\hat \Expec[\Omega] = (1/|\set{S}|) \sum_{i \in
  \set{S}} \omega_i = (1/|\set{S}|) \sonenorm{\vomega}$, the empirical second
  moment $\hat \Expec[\Omega^2] = (1/|\set{S}|) \sum_{i \in \set{S}}
  \omega_i^2 = (1/|\set{S}|) \twonorm{\vomega}^2$, and the empirical variance
  $\widehat \Var[\Omega] = \hat \Expec[\Omega^2] - (\hat \Expec[\Omega])^2$,
  we can rewrite the AWGNC pseudo-weight as
  \begin{align}
    \wps(\vomega)
      &= |\set{S}|
         \cdot
         \frac{(\hat \Expec[\Omega])^2}
              {\hat \Expec[\Omega^2]}.
  \end{align}
  In the case that $\vomega$ is scaled such that $\hat \Expec[\Omega] = 1$
  (i.e.~$\sonenorm{\vomega} = |\set{S}|$), we can write
  \begin{align}
    \wps(\vomega)
      &= |\set{S}|
         \cdot
         \frac{1}
              {\widehat \Var[\Omega] + 1}.
  \end{align}
  Therefore, the more the non-zero components of $\vomega$ are apart, the
  smaller is the AWGNC pseudo-weight.
\end{Lemma}

\begin{Proof}
  See Sec.~\ref{sec:proof:lemma:pseudo:weight:variance:interpretation:1}.
\end{Proof}

\begin{Lemma}
  \label{lemma:pseudo:weight:angle:interpretation:1}

  Let $\vomega \in \Rp^n$ and let $\angle(\vomega,\vect{1})$ be the angle
  between the vectors $\vomega$ and $\vect{1}$. Interestingly,
  $\wps(\vomega)$ is only a function of $n$ and the angle
  $\angle(\vomega,\vect{1})$:
  \begin{align}
    \wps(\vomega)
      &= n
         \cdot
         \cos
           \big(
             \angle(\vomega, \vect{1})
           \big)^2
  \end{align}
  We see that the larger the angle $\angle(\vomega,\vect{1})$ becomes, the
  smaller is $\wps(\vomega)$. Alternatively, if we let $\vect{1}_{\vomega}$ be
  the indicator vector of $\vomega$, i.e.~the $i$-the position is $1$ if
  $\omega_i$ is non-zero and it is $0$ otherwise, then
  \begin{align}
    \wps(\vomega)
      &= |\supp(\vomega)|
         \cdot
         \cos
           \big(
             \angle(\vomega, \vect{1}_{\vomega})
           \big)^2
  \end{align}
\end{Lemma}

\begin{Proof}
  See Sec.~\ref{sec:proof:lemma:pseudo:weight:angle:interpretation:1}.
\end{Proof}

\begin{Lemma}
  \label{lemma:convex:combination:pseudo:distance:lower:bound:1}

  For any positive integer $L$, let $\{ \vomega^{(\ell)} \}_{\ell \in [L]}$ be
  a set of vectors where $\vomega^{(\ell)} \in \Rp^n$, $\ell \in [L]$. Then,
  \begin{align*}
    \wpsAWGNC
      \left(
        \sum_{\ell \in [L]}
          \alpha_{\ell} \vomega^{(\ell)}
      \right)
      &\geq
         \min_{\ell \in [L]}
           \wpsAWGNC(\vomega^{(\ell)}) \\
    \wpsBSC
      \left(
        \sum_{\ell \in [L]}
          \alpha_{\ell} \vomega^{(\ell)}
      \right)
      &\geq
         \min_{\ell \in [L]}
           \wpsBSC(\vomega^{(\ell)}), \\
    \wpsBEC
      \left(
        \sum_{\ell \in [L]}
          \alpha_{\ell} \vomega^{(\ell)}
      \right)
      &\geq
         \min_{\ell \in [L]}
           \wpsBEC(\vomega^{(\ell)}), \\
    \wmaxfr
      \left(
        \sum_{\ell \in [L]}
          \alpha_{\ell} \vomega^{(\ell)}
      \right)
      &\geq
         \min_{\ell \in [L]}
           \wmaxfr(\vomega^{(\ell)}),
  \end{align*}
  for any $\alpha_{\ell} \geq 0$, $\ell \in [L]$ where not all $\alpha_i$ are
  zero. This means that the AWGNC pseudo-weight of any conic combination of an
  arbitrary set of vectors in $\Rp^n$ is at least as large as the smallest
  AWGNC pseudo-weight of any of these vectors. This property is intuitively
  clear from the geometrical meaning of the AWGNC pseudo-weight. (Similar
  statements can be made for the BSC and BEC pseudo-weight and for the
  max-fractional weight.)
\end{Lemma}

\begin{Proof}
  See
  Sec.~\ref{sec:proof:lemma:convex:combination:pseudo:distance:lower:bound:1}.
\end{Proof}

\begin{Lemma}
  \label{lemma:convex:combination:pseudo:distance:lower:bound:2}

  For any positive integer $L$, let $\{ \vomega^{(\ell)} \}_{\ell \in [L]}$ be
  a set of vectors where $\vomega^{(\ell)} \in \Rp^n$, $\ell \in [L]$. If
  $\sonenorm{\vomega^{(\ell)}} = 1$ for all $\ell \in [L]$ then
  \begin{align}
    \frac{1}
         {\sqrt{
            \wps
              \left(
                \sum_{\ell \in [L]}
                  \alpha_{\ell} \vomega^{(\ell)}
              \right)
               }
         }
      &\leq
         \sum_{\ell \in [L]}
           \frac{\alpha_{\ell}}
                {\sqrt{\wps(\vomega^{(\ell)})}}
  \end{align}
  for any $\alpha_{\ell} \geq 0$, $\ell \in [L]$, such that
  $\sum_{\ell \in [L]} \alpha_{\ell} = 1$.
\end{Lemma}

\begin{Proof}
  See
  Sec.~\ref{sec:proof:lemma:convex:combination:pseudo:distance:lower:bound:2}.
\end{Proof}

\begin{Lemma}
  \label{lemma:pseudo:weight:derivatives:1}

  Let $\vomega \in \Rp^n$. Then
  \begin{align*}
    \frac{\partial}{\partial \omega_i}
      \wps(\vomega)
      &\begin{cases}
         > 0 & \text{if $\omega_i < \sonenorm{\vomega} / \wps(\vomega)$} \\
         = 0 & \text{if $\omega_i = \sonenorm{\vomega} / \wps(\vomega)$} \\
         < 0 & \text{if $\omega_i > \sonenorm{\vomega} / \wps(\vomega)$}
       \end{cases}.
  \end{align*}
\end{Lemma}

\begin{Proof}
  See Sec.~\ref{sec:proof:lemma:pseudo:weight:derivatives:1}.
\end{Proof}

\mbox{}

Roughly speaking, the above lemma means that if we are given a vector $\vomega
\in \Rp^n$ and want to decrease its AWGNC pseudo-weight then we must either
decrease the small components or increase the large components. In both cases
the empirical variance increases which is in agreement with the observations
in Lemma~\ref{lemma:pseudo:weight:variance:interpretation:1}.

\begin{Lemma}
  \label{lemma:pseudoo:weight:inequalities:1}

  Let $\vomega \in \Rp^n$ with $\vect{0} \leq \vomega \leq \vect{1}$. Remember
  that $\wps(\vomega) = \wpsAWGNC(\vomega)$ by definition. Then
  \begin{alignat}{2}
    \wfr(\vomega)
      &\leq
         \wmaxfr(\vomega)
      &\leq
         \wps(\vomega)
      &\leq
         \wpsBEC(\vomega),
           \label{eq:pseudo:weight:inequalities:1} \\
    \wfr(\vomega)
      &\leq
         \wmaxfr(\vomega)
      &\leq
         \wpsBSC(\vomega)
      &\leq
         \wpsBEC(\vomega),
           \label{eq:pseudo:weight:inequalities:2}
  \end{alignat}
  and
  \begin{alignat}{2}
    \wfrminh{H}
      &\leq
         \wmaxfrminh{H}
      &\leq
         \wpsminh{H}
      &\leq
         \wpsBECminh{H},
           \label{eq:pseudo:weight:inequalities:3} \\
    \wfrminh{H}
      &\leq
         \wmaxfrminh{H}
      &\leq
         \wpsBSCminh{H}
      &\leq
         \wpsBECminh{H},
           \label{eq:pseudo:weight:inequalities:4}
  \end{alignat}
\end{Lemma}

\begin{Proof}
  See Sec.~\ref{sec:proof:lemma:pseudoo:weight:inequalities:1}.
\end{Proof}

\mbox{}

Note that there is no hierarchy between $\wps(\vomega)$ and
$\wpsBSC(\vomega)$, i.e.~one can find $\vomega$'s such that either one is
larger. Consider for example $\vomega = (1, 1, \frac{1}{2}, \frac{1}{2},
\frac{1}{2}, \frac{1}{2})$ for which the AWGNC pseudo-weight is larger:
$\wps(\vomega) = \frac{4^2}{3} = \frac{16}{3} = 5.333 > \wpsBSC(\vomega) = 2
\cdot 2 = 4$. However, the vector $\vomega = (1, \frac{1}{4}, \ldots,
\frac{1}{4})$ of length $65$ is an example where the BSC pseudo-weight is
larger: $\wps(\vomega) = \frac{17^2}{5} = 57.8 < \wpsBSC(\vomega) = 2 \cdot 31
= 62$.

Asymptotically, i.e.~for $n \to \infty$, the AWGNC and BSC pseudo-weight can
vary drastically in the following sense. In
Prop.~\ref{prop:min:ps:weight:upper:bound:1} we will show that
$\wps(\,\cdot\,)$ always grows sub-linearly for an ensemble of
$(\wcol,\wrow)$-regular LDPC codes where $3 \leq \wcol < \wrow$. However, for
properly chosen families of $(\wcol,\wrow)$-regular LDPC codes one can
guarantee a linear behavior of $\wpsBSC(\,\cdot\,)$ as $n \to
\infty$~\cite{Feldman:Malkin:Stein:Servedio:Wainwright:04:1}. Some of the
reasons and implications of this fact are also discussed
in~\cite{Feldman:Koetter:Vontobel:05:1}.

The above considerations have also implications for the fractional and
max-fractional weight (see
Def.~\ref{def:fractional:and:max:fractional:weight:1}) that was introduced
in~\cite{Feldman:03:1} to analyze the decoding behavior when transmitting over
a BSC. Using Lemma~\ref{lemma:pseudoo:weight:inequalities:1} we see that when
considering the limit $n \to \infty$ the fractional and the max-fractional
weight can grow at best like the AWGNC pseudo-weight. However, the comments in
the previous paragraph show that the AWGNC and BSC pseudo-weight can behave
quite differently for $n \to \infty$, therefore the fractional/max-fractional
weight and the BSC pseudo-weight can also behave quite differently for $n \to
\infty$. Note though that from an analysis point of view, the fractional
weight might sometimes be a more manageable quantity since it is a linear
function of the argument whereas the BSC pseudo-weight is more complicated
function. Indeed,~\cite[Sec.~4.4.3]{Feldman:03:1} shows an efficient procedure
for computing the minimal fractional weight of a code with given parity-check
matrix.


\section{A Simple Upper Bound on the Minimum AWGNC Pseudo-Weight}
\label{sec:pw:upper:bound:1}

In this section we investigate the asymptotic behavior of the minimum
pseudo-weight of families of $(\wcol,\wrow)$-regular LDPC codes, i.e.~codes
whose parity-check matrices have a fixed column and row
weight.\footnote{Although similar methods can be devised for irregular LDPC
codes, we focus on the regular case only.} Our main result will be that the
relative\footnote{In the same way as the relative Hamming weight of a vector
is the Hamming weight of the vector divided by $n$, we can define relative
pseudo-weights for all the pseudo-weights that were introduced in
Sec.~\ref{sec:pseudo:weight:1}.} minimum AWGNC pseudo-weight of any
$(\wcol,\wrow)$-regular code, $3 \leq \wcol < \wrow$, approaches zero as $n
\to \infty$, a behavior which is in sharp contrast to the observation made by
Gallager~\cite{Gallager:63} that the relative minimum Hamming weight of a
randomly generated $(\wcol,\wrow)$-regular LDPC code, $3 \leq \wcol < \wrow$,
is lower bounded by a nonzero number with probability one for $n \to \infty$~.

In the following, we associate the Tanner graph $\graph{T} \defeq
\tgraph{T}{H}$ to the parity-check matrix $\matr{H}$ and denote its girth and
diameter by $\girthg{T}$ and $\diamg{T}$, respectively.

\begin{Definition}
  \label{def:tanner:graph:ordering:1}

  Let $\graph{T}$ be a Tanner graph of an arbitrary code (not necessarily
  $(\wcol,\wrow)$-regular). We let an arbitrary variable node $V$ of
  $\graph{T}$ to be the root. We classify the remaining variable and check
  nodes according to their (graph) distance from the root, i.e.~all nodes at
  distance $1$ from the root will be called nodes of tier $1$, all nodes at
  distance $2$ from the root node will be called nodes of tier $2$, etc. We
  call this ordering ``breadth-first spanning-tree ordering with root $V$.''
  Because of the bipartite-ness of $\graph{T}$, it follows easily that the
  nodes of the even tiers are variable nodes whereas the nodes of the odd
  tiers are check nodes. Furthermore, a check node at tier $2t+1$ can only be
  connected to variable nodes in tier $2t$ and possibly to variable nodes in
  tier $2t+2$. Note that the last tier is tier $\diamg{T}$ and that the symbol
  nodes are at tiers $0, 2, \ldots, 2 \lfloor \diamg{T}/2 \rfloor$.
\end{Definition}

Let us upper bound the number of nodes for each tier when we perform
breadth-first spanning-tree ordering according to
Def.~\ref{def:tanner:graph:ordering:1} with respect to an arbitrary node $V$
of the Tanner graph $\graph{T}$ of an arbitrary $(\wcol,\wrow)$-regular LDPC
code. Let $N_{V,t}(\graph{T})$ be the number of nodes at tier $t$ and let
$\Nmax_{V,t} \defeq \Nmax_{V,t,\wcol,\wrow}$ be the maximal number of nodes
possible at tier $t$ for any $(\wcol,\wrow)$-regular LDPC code. It is not
difficult to see that $\Nmax_{V,0} = 1$, $\Nmax_{V,1} = \wcol$, $\Nmax_{V,2} =
\wcol(\wrow-1)$, $\Nmax_{V,3} = \wcol(\wrow-1)(\wcol-1)$, $\Nmax_{V,4} =
\wcol(\wrow-1)(\wcol-1)(\wrow-1)$. In general, $\Nmax_{V,2t} = \wcol
(\wcol-1)^{t-1} (\wrow-1)^t$ for $t > 0$ and $\Nmax_{V,2t+1} = \wcol
(\wcol-1)^t (\wrow-1)^t$ for $t \geq 0$.

\begin{Definition}
  \label{def:canonical:completion:1}

  Let $\graph{T}$ be the Tanner graph of a code whose parity-check matrix
  $\matr{H}$ has uniform row weight $\wrow$. After performing the
  breadth-first spanning-tree ordering with an arbitrary variable node $V$ as
  root we construct a pseudo-codeword $\vomega$ in the following way. If bit
  $i$ corresponds to a variable node in tier $2t$, then
  \begin{align}
    \omega_i
      &\defeq \frac{1}{(\wrow-1)^t}.
  \end{align}
  We call this the canonical completion with root $V$. It will be shown in
  Lemma~\ref{lemma:canonical:completion:properties:1} that $\vomega \in
  \fch{K}{H}$, i.e.~$\vomega$ is a pseudo-codeword.
\end{Definition}

\begin{figure}
  \begin{center}
    \epsfig{file=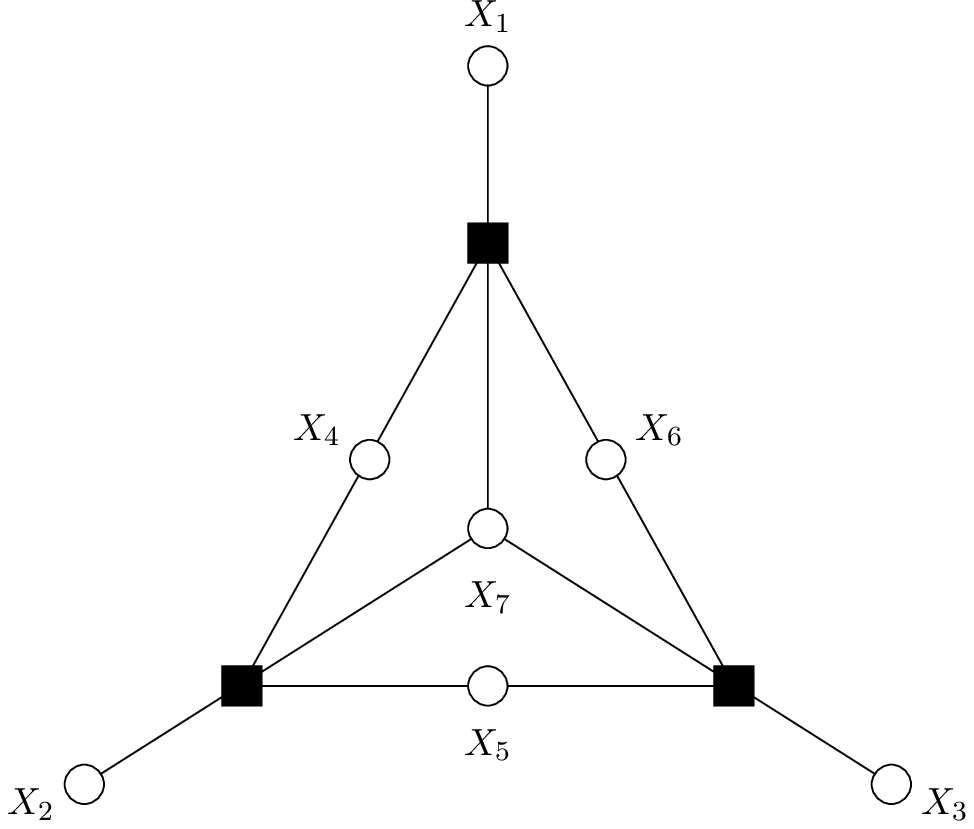, width=4cm}
    \quad\quad
    \epsfig{file=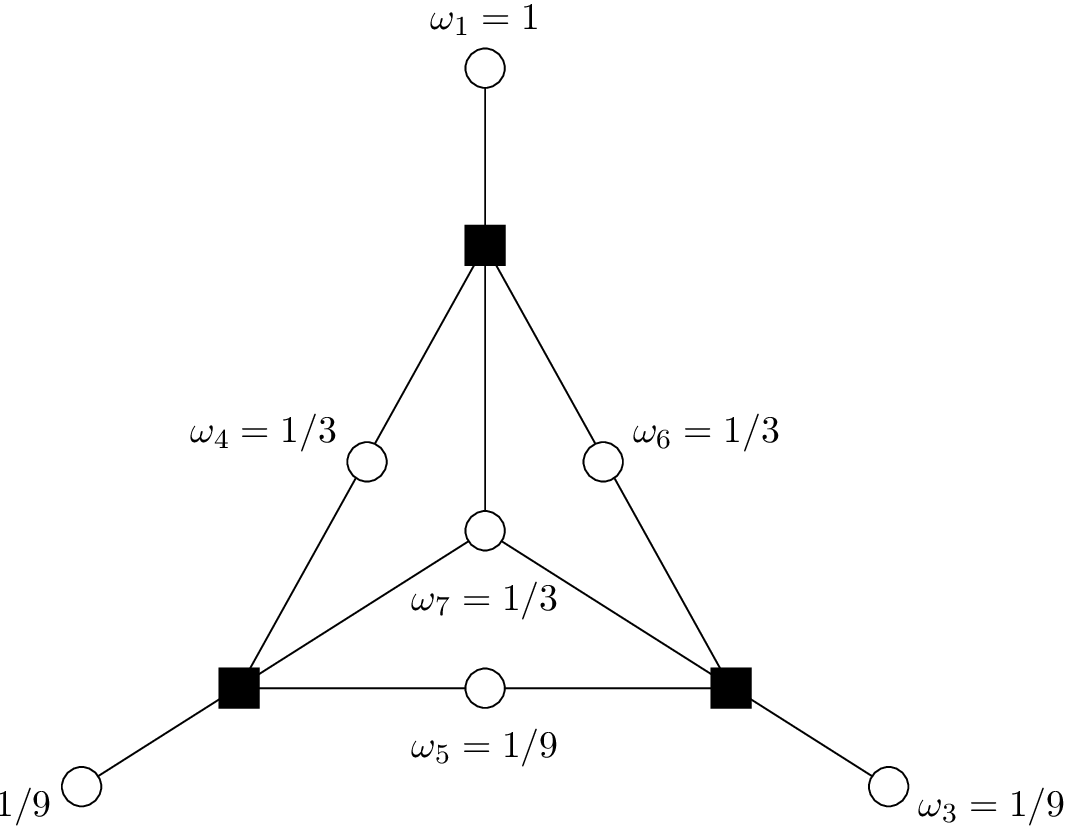, width=4cm}
    \quad\quad
    \epsfig{file=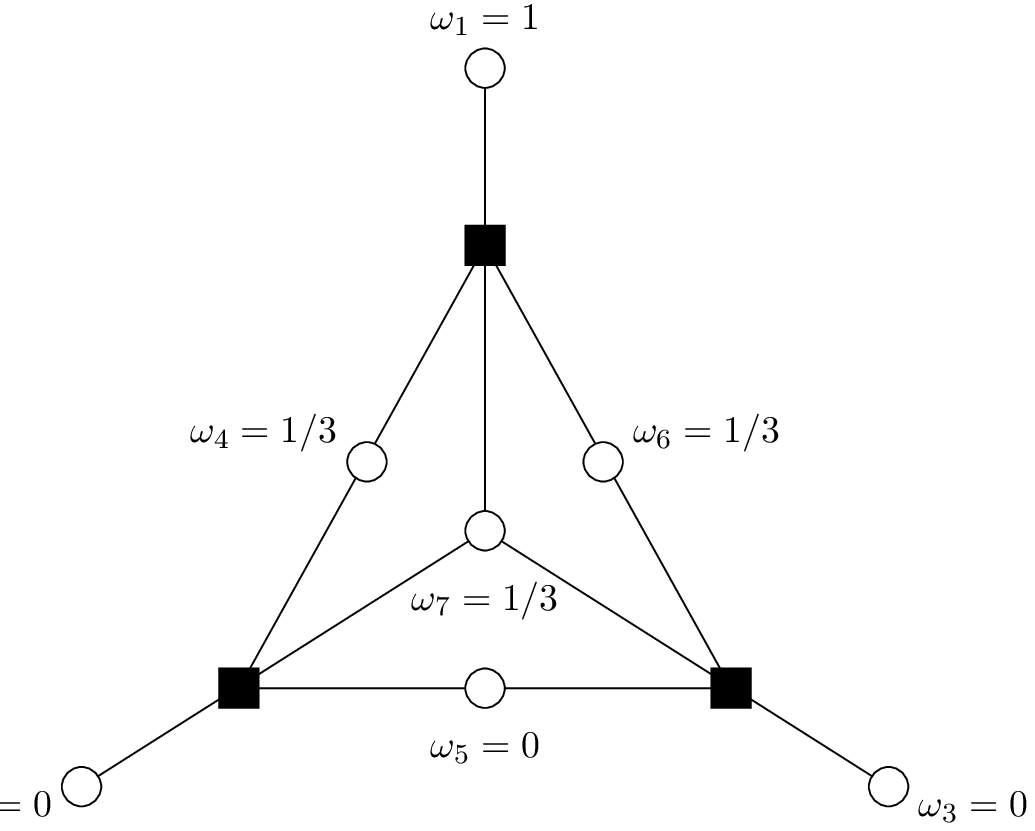, width=4cm}
  \end{center}
  \caption{Left: Tanner graph for the [7,4,3] code in
    Ex.~\ref{ex:hamming:code:canonical:completion:1}. Middle: Canonical
    completion with respect to node $X_1$. Right: Another pseudo-codeword.}
  \label{fig:hamming:code:canonical:completion:1}  
\end{figure}

\begin{Example}
  \label{ex:hamming:code:canonical:completion:1}

  Fig.~\ref{fig:hamming:code:canonical:completion:1} (left) shows the Tanner
  graph of a $[7,4,3]$ binary linear code. (It is the length-$7$ Hamming
  code.) Note that in this Tanner graph, all check nodes have degree four,
  i.e.~$\wrow = 4$. Performing breadth-first spanning-tree ordering with root
  $X_1$ we see that tier $0$ consists of $\{ X_1 \}$, tier $2$ consists of $\{
  X_4, X_6, X_7 \}$, and tier $4$ consists of $\{ X_2, X_3, X_5
  \}$. Correspondingly, the canonical completion with root $X_1$ yields the
  vector $\vomega = \big(1, \frac{1}{9}, \frac{1}{9}, \frac{1}{3},
  \frac{1}{9}, \frac{1}{3}, \frac{1}{3} \big)$ shown in
  Fig.~\ref{fig:hamming:code:canonical:completion:1} (middle). It is easy to
  check that $\vomega$ is inside the fundamental cone for this graph and is
  therefore a pseudo-codeword. The AWGNC pseudo-weight for $\vomega$ equals
  \begin{align*}
    \wps(\vomega)
      &= \frac{\left(
                 1 + \frac{1}{9} + \frac{1}{9} + \frac{1}{3} + \frac{1}{9}
                   + \frac{1}{3} + \frac{1}{3}
               \right)^2}
              {1 + \frac{1}{81} + \frac{1}{81} + \frac{1}{9} + \frac{1}{81}
                 + \frac{1}{9} + \frac{1}{9}}
       = 3.973.
  \end{align*}
  (As an aside, we note that the Tanner graph in
  Fig.~\ref{fig:hamming:code:canonical:completion:1} (left) also supports a
  pseudo-codeword $\vomega'$ of type
  $\vomega'=(1,0,0,\frac{1}{3},0,\frac{1}{3},\frac{1}{3})$ whose AWGNC
  pseudo-weight equals only three and is thus at ``minimum distance'' for this
  code, see Fig.~\ref{fig:hamming:code:canonical:completion:1} (right).)
\end{Example}

Without going into the details, let us mention that
Def.~\ref{def:canonical:completion:1} can be generalized in the following way:
instead of doing a canonical completion with respect to a single variable
node, one might do a canonical completion with respect to a set of variable
nodes. The entries of the pseudo-vector will then be defined according to the
graph distance to this set of nodes. This generalized notion of canonical
completion was e.g.~used in~\cite{Haley:Grant:05:1,
Vontobel:Smarandache:05:1}.

\begin{Lemma}
  \label{lemma:canonical:completion:properties:1}

  Let $\graph{T}$ be the Tanner graph of a code whose parity-check matrix
  $\matr{H}$ has uniform row weight $\wrow$. The canonical completion with an
  arbitrary codeword symbol node $V$ as root yields a vector $\vomega$ such
  that $\vomega$ is in the fundamental cone $\fch{K}{H}$. The vector $\vomega$
  has AWGNC pseudo-weight $\wps(\vomega{}) = \sonenorm{\vomega}^2 /
  \twonorm{\vomega}^2$, where
  \begin{align}
    \sonenorm{\vomega}
      &= \sum_{t=0}^{\lfloor \diamg{T}/2 \rfloor}
           N_{V,2t}(\graph{T})
           \frac{1}{(\wrow-1)^t},
             \label{eq:canonical:completion:sonenorm:1} \\
    \twonorm{\vomega}^2
      &= \sum_{t=0}^{\lfloor \diamg{T}/2 \rfloor}
           N_{V,2t}(\graph{T})
           \left(
             \frac{1}{(\wrow-1)^t}
           \right)^2.
             \label{eq:canonical:completion:twonorm:1}
  \end{align}
\end{Lemma}

\begin{Proof}
  See Sec.~\ref{sec:proof:lemma:canonical:completion:properties:1}.
\end{Proof}

For a given $\graph{T}$, one can numerically calculate the pseudo-weight of
the pseudo-codeword given by the canonical completion for any given root; this
will always yield an upper bound on $\wpsminc{C}$. In the next proposition we
will see that the canonical-completion approach is powerful enough to show
that $\wpsminc{C}$ can at best only grow sub-linearly for
$(\wcol,\wrow)$-regular LDPC codes with $3 \leq \wcol < \wrow$.

\begin{Proposition}
  \label{prop:min:ps:weight:upper:bound:1}

  Let $\matr{H}$ be the $(\wcol,\wrow)$-regular parity-check matrix of a
  length-$n$ LDPC code $\code{C}$ with $3 \leq \wcol < \wrow$. Then the
  minimum pseudo-weight is upper bounded by
  \begin{align}
    \wpsminh{H}
      &\leq \beta'
            \cdot
            n^{\beta},
  \end{align}  
  where {\small
  \begin{align}
    \beta'
      &\defeq
         \beta'(\wcol, \wrow)
       \defeq
         \left( \frac{\wcol (\wcol-1)} {\wcol-2} \right)^2,
    \quad
    \beta
       \defeq
         \beta(\wcol, \wrow)
       \defeq
         \frac{\log\left( (\wcol-1)^2 \right)}
              {\log\big( (\wcol-1)(\wrow-1) \big)} < 1.
  \end{align}
}
\end{Proposition}

\begin{Proof}
  See Sec.~\ref{sec:proof:prop:min:ps:weight:upper:bound:1}.
\end{Proof}

\mbox{}

Note that this proposition excludes two type of $(\wcol,\wrow)$-regular
codes. The first type is the family of codes where $\wcol = 2$, also known as
cycle codes. In that case a much better upper bound can be given: the minimum
distance, and therefore also the minimal AWGNC pseudo-weight, grow at best
only logarithmically in the block length $n$.

The second type of codes that where excluded were families of codes where
$\wcol \geq \wrow$. Note however that randomly generated
$(\wcol,\wrow)$-regular LDPC codes are not too interesting since the dimension
of the code will be zero or near-zero with high probability. Nevertheless, let
us mention that there are interesting and practically useful families of
algebraically constructed $(\wcol,\wrow)$-regular codes where the rate does
not vanish, e.g.~\cite{Kou:Lin:Fossorier:01:1}.

\begin{Corollary}
  \label{cor_rel_min_ps_weight_1}

  Consider a sequence of $(\wcol,\wrow)$-regular LDPC codes, $3 \leq \wcol <
  \wrow$, whose length goes to infinity. The relative minimum AWGNC
  pseudo-weight (i.e.~the fraction of minimum pseudo-weight to code length)
  must go to zero. This is in sharp contrast to the fact that the relative
  minimum Hamming weight of a randomly generated $(\wcol,\wrow)$-regular LDPC
  code, $3 \leq \wcol < \wrow$, is lower bounded by a nonzero number with
  probability one for $n \to \infty$~\cite{Gallager:63}.
\end{Corollary}

Let us finish this section with two observation. The first observation is
about the ``strange'' shape of the fundamental cone. Using
Lemma~\ref{lemma:pseudo:weight:angle:interpretation:1} we see that
Prop.~\ref{prop:min:ps:weight:upper:bound:1} says that for families of
$(\wcol,\wrow)$-regular LDPC codes there are pseudo-codewords (i.e.~vectors in
the fundamental cone) whose angle with the all-ones vector goes to
$90^{\circ}$ for $n \to \infty$. However, none of the polytopes associated to
this family of codes contains the vector $(\wrow-1+\varepsilon, 1, \ldots,
1)$, where $\varepsilon > 0$, yet the angle of this vector with the all-ones
vector goes to $0^{\circ}$ for $n \to \infty$.

The second observation is that the BEC pseudo-weight of the canonical
completion with respect to any variable node equals the block length. This
means that although the fundamental cone characterizes the pseudo-codewords
for the AWGNC and the BEC, the worst-case pseudo-codewords within the
fundamental cone might be quite different depending on the channel.


\section{The Relationship of the Fundamental Polytope to other Concepts that
            Explain the Behavior of Iterative Decoding}

\label{sec:relationship:fundamental:polytope:other:concepts:1}

As we mentioned in the introduction to the paper, a variety of concepts have
been introduced in the past that try to explain the behavior of MPID. In this
section we would like to show how some of these are related to the fundamental
polytope and the various pseudo-weights.

\subsection{Stopping Sets}

Let us recall the definition of a stopping
set~\cite{Di:Proietti:Telatar:Richardson:Urbanke:02:1} for a Tanner graph
$\graph{T}$. A subset $\set{S}$ of the variable nodes of $\graph{T}$ is called
a stopping set if and only if every check node in $\neighborhood(\set{S})$ is
connected to at least two variable nodes in $\set{S}$. Stopping sets are a
means to understand the suboptimal behavior of iterative decoding techniques
for the BEC, in fact they completely characterize iterative decoding in that
case. It has been observed later that stopping sets seem to also reflect, to
some degree, the performance of iteratively decoded codes for other channels.

\begin{Proposition}
  \label{prop:stopping:set:vs:fp:1}

  On the one hand, if $\vomega \in \fph{P}{H}$ then $\supp(\vomega)$ is a
  stopping set of $\tgraph{T}{H}$. On the other hand, if $\set{S}$ is a
  stopping set of $\tgraph{T}{H}$ then there exists a vector $\vomega \in
  \fph{P}{H}$ such that $\supp(\vomega) = \set{S}$.
\end{Proposition}

\begin{Proof}
  See Sec.~\ref{sec:proof:prop:stopping:set:vs:fp:1}.
\end{Proof}

\mbox{}

In the light of Prop.~\ref{prop:stopping:set:vs:fp:1} it seems quite intuitive
that the BEC pseudo-weight of a vector $\vomega \in \fph{P}{H}$ is defined to
be $\wpsBEC(\vomega) = \card{\supp(\vomega)}$, see
Def.~\ref{def:bec:pseudo:weight:1}, but we will not go into the details here.

\begin{figure}
  \begin{center}
    \epsfig{file=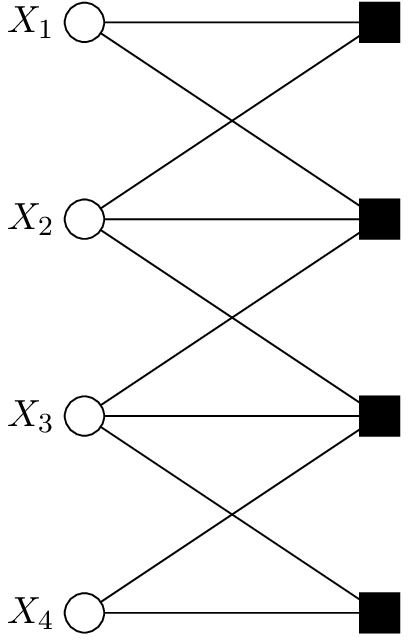, width=3cm}
  \end{center}
  \caption{Tanner graph $\graph{T}$.}
  \label{fig:tanner:graph:two:codes:1}
\end{figure}

While the notion of stopping set is well suited to the BEC it is not refined
enough to capture the situation for the AWGN channel. Consider the
parity-check matrix $\matr{H}$ whose Tanner graph $\graph{T} \defeq
\tgraph{T}{H}$ is shown in Fig.~\ref{fig:tanner:graph:two:codes:1} and whose
fundamental cone is
\begin{align*}
  \fch{K}{H}
   &= \{ \alpha_1 \cdot (2,2,1,1) + 
         \alpha_2 \cdot (1,1,2,2)
         \ | \
         \alpha_1, \alpha_2 \in \Rp \}.
\end{align*}
While all the non-zero vectors in $\fch{K}{H}$ have BEC pseudo-weight $4$
(i.e.~their supports yield stopping sets of size $4$), the AWGNC
pseudo-weight is usually smaller than $4$, e.g.~the two minimal
pseudo-codewords $(2,2,1,1)$ and $(1,1,2,2)$ have AWGNC pseudo-weight $3.6$.

\subsection{Near Codewords and Trapping Sets}

Near-codewords were introduced by MacKay and Postol~\cite{MacKay:Postol:03:1}:
a vector $\vx \in \GF{2}^n$ is called a $(w,w')$ near-codeword in a Tanner
graph $\graph{T} \defeq \tgraph{T}{H}$ with $n$ variable nodes if $\wH(\vx) =
w$ and $\wH(\vs) = w'$ where $\vs = \vx \cdot \matr{H}^\tr \text{ (in
$\GF{2}$)}$ is the syndrome of $\vx$ with respect to $\matr{H}$. In other
words, the graph induced by the $w$ non-zero components of $\vx$ contains $w'$
check nodes of odd degree. Richardson's definition of trapping sets is
essentially identical~\cite{Richardson:03:1}: $\vx \in \GF{2}^n$ is a $(w,w')$
near-codeword if and only if $\supp(\vx)$ is a $(w,w')$ trapping set. 

As was remarked in~\cite{MacKay:Postol:03:1}: ``near codewords with small $w'$
tend to be error states from which the sum-product decoding algorithm cannot
escape.'' Therefore it is important to understand the $(w,w')$ near-codewords
that have low $w$ and low $w'$. To exemplify this with a simple, albeit
extreme, example, consider an LDPC code $\code{C}$ represented by a
parity-check matrix $\matr{H}$. Fix some $i' \in \set{I}$ and let $\vx \in
\GF{2}^n$ be a vector where $x_{i'} = 1$ and $x_{i} = 0$ for $i \in \set{I}
\setminus \{ i' \}$. It is easy to check that $\vx$ is a $(1,w')$
near-codeword where $w'$ equals the Hamming weight of the $i'$ column of
$\matr{H}$. In fact, it can cause problems when transmitting over an
AWGNC. Assume that the all-zeros codeword is transmitted ($+ \sqrt{\Ec}
\vect{1}$ after modulation) and that the noise vector is the all-zeros vector
except for the $i'$-th position that is negative. If it is negative enough
then MPID will decide wrongly.

A connection between near-codewords and trapping sets on the one hand and
pseudo-codewords on the other hand can be made in the following way. One way
is to find the pseudo-codeword in the fundamental cone that is the closest to
a $(w,w')$ near-codeword $\vx$. If $w'$ is small, only small changes have to
be applied to the components of the vector $\vx$ to get a pseudo-codeword.
Alternatively, when trying to assign a pseudo-codeword to a near-codeword one
might want to apply the canonical completion that is rooted at the
near-codeword.

\subsection{Why Four-Cycles are Potentially Bad}

\label{sec:why:four:cycles:potentiall:bad:1}

Already people like Wiberg realized that for MPID to work well one should have
Tanner graphs that look locally tree-like which means that the girth of a
graph should be reasonably large. A first step in that direction is to avoid
four-cycles.\footnote{Note though that some researchers have studied
algebraically-constructed Tanner graphs with girth four, see
e.g.~\cite{Lucas:Bossert:Breitbach:98:1, Xu:Tang:Kou:Lin:AbdelGhaffar:02:1,
Fossorier:Palanki:Yedidia:03:1}, and exhibited some codes which work very well
under iterative decoding.} In this subsection we would like to explore what
the fundamental-polytope view can contribute to this topic.

A simple observation towards this goal is the following: considering the proof
of Prop.~\ref{prop:min:ps:weight:upper:bound:1} we see that the smaller the
girth of the graph is the smaller can be made the AWGNC pseudo-weight of the
canonical completion.

A different avenue is pursued by the following lemma and its corollaries which
explore the effect of girth on the fundamental polytope upon adding redundant
rows to a parity-check matrix.

\begin{Lemma}
  \label{lemma:adding:one:row:1}

  Let $\code{C}$ be a code with parity-check matrix $\matr{H}$. Basic coding
  theory tells us that the modified parity-check matrix
  \begin{align*}
    \matr{H}' \defeq
        \begin{pmatrix}
          \matr{H} \\
          \vect{a} \cdot \matr{H}
        \end{pmatrix}
           \text{ (in $\GF{2}$)},
  \end{align*}
  where $\vect{a} \in \GF{2}^{\card{\set{J}(\matr{H})}}$ is an arbitrary
  vector, defines the same code $\code{C}$. If the Tanner graph
  $\tgraph{T}{H}$ of $\matr{H}$ is a forest, i.e.~cycle-free, then $\fph{P}{H}
  = \fp{P}(\matr{H}')$. 
\end{Lemma}

\begin{Proof}
  See Sec.~\ref{sec:proof:lemma:adding:one:row:1}.
\end{Proof}

\mbox{}

Note that in the absence of cycle-freeness of $\tgraph{T}{H}$ one can easily
exhibit a vector $\vect{a}$ where $\fp{P}(\matr{H}') \subsetneq \fph{P}{H}$.

\begin{Corollary}
  \label{cor:adding:one:row:2}

  Similar to Lemma~\ref{lemma:adding:one:row:1}, consider a code $\code{C}$
  with parity-check matrix $\matr{H}$ and a modified parity-check matrix
  $\matr{H}'$, where $\vect{a} \in \GF{2}^n$ is an arbitrary vector. However,
  now we do not require that $\tgraph{T}{H}$ is a forest. Let $\matr{H}_1$ be
  the $|\supp(\vect{a})| \times n$ submatrix of $\matr{H}$ where we include
  the $j$-th row if and only if $a_j \neq 0$. If the Tanner graph
  $\graph{T}(\matr{H}_1)$ of $\matr{H}_1$ is a forest, i.e.~cycle-free, then
  $\fph{P}{H} = \fp{P}(\matr{H}')$.
\end{Corollary}

\begin{Proof}
  See Sec.~\ref{sec:proof:cor:adding:one:row:2}.
\end{Proof}

\begin{Corollary}
  \label{cor:adding:rows:3}

  Let $\code{C}$ be a code with parity-check matrix $\matr{H}$. Basic coding
  theory tells us that the modified parity-check matrix
  \begin{align*}
    \matr{H}' \defeq
        \begin{pmatrix}
          \matr{H} \\
          \matr{A} \cdot \matr{H}
        \end{pmatrix}
           \text{ (in $\GF{2}$)},
  \end{align*}
  where $\matr{A}$ is an arbitrary matrix over $\GF{2}$ with
  $\card{\set{J}(\matr{H})}$ columns, defines the same code $\code{C}$. For
  each row $r$ of $\matr{A}$, let $\matr{H}_r$ be the submatrix of $\matr{H}$
  where we include the $j$-th row of $\matr{H}$ if $[\matr{A}]_{r,j} = 1$. If
  $\graph{T}(\matr{H}_r)$ is a cycle-free Tanner graph for all rows $r$ of
  $\matr{A}$, then $\fph{P}{H} = \fp{P}(\matr{H}')$.
\end{Corollary}

\begin{Proof}
  See Sec.~\ref{sec:proof:cor:adding:rows:3}.
\end{Proof}

\mbox{}

Lemma~\ref{lemma:adding:one:row:1} and its corollaries have some important
consequences.\footnote{Similar observations were also made by
Wainwright~\cite{Wainwright:05:1}.}
\begin{itemize}

  \item Let $\matr{H}$ be a parity-check matrix of a code $\code{C}$ where the
    Tanner graph $\tgraph{T}{H}$ has girth six. We can create a new
    parity-check matrix $\matr{H}'$ that describes the same code in the
    following way: let $\matr{H}'$ consist of all rows of $\matr{H}$ and the
    modulo-$2$ sums of all pairs of rows of $\matr{H}$. Then $\fph{P}{H} =
    \fp{P}(\matr{H}')$. (This observation follows from the fact that girth six
    for $\tgraph{T}{H}$ implies that $\graph{T}( \bigl( \begin{smallmatrix}
    \vect{h}_{j'} \\ \vect{h}_{j''} \end{smallmatrix} \bigr) )$ is cycle-free
    for all pairs of rows $j', j''$ of $\matr{H}$.) Note that applying the
    same procedure to Tanner graphs $\tgraph{T}{H}$ with girth four will
    usually lead to $\fp{P}(\matr{H}') \subsetneq \fph{P}{H}$.

  \item More generally, let $\matr{H}$ be a parity-check matrix of a code
    $\code{C}$ where the Tanner graph $\tgraph{T}{H}$ has girth $g$. We can
    create a new parity-check matrix $\matr{H}'$ that describes the same code
    in the following way: let $\matr{H}'$ consist of all rows of $\matr{H}$,
    the modulo-$2$ sums of all pairs of rows of $\matr{H}$, \ldots, the
    modulo-$2$ sums of all $(g-2)/2$-tuples of rows of $\matr{H}$. Then
    $\fph{P}{H} = \fp{P}(\matr{H}')$.

  \item The above observations have some interesting consequences for
    $\relaxationsetorder{r}{H}$ as defined in
    Def.~\ref{def:relaxation:of:order:r:1}: if $\tgraph{T}{H}$ has girth $g$
    then $\relaxationsetorder{r}{H} = \fph{P}{H}$ for $r \leq (g-2)/2$. This
    means that the larger the girth of the Tanner graph $\tgraph{T}{H}$ is,
    the more codewords from the dual code have to be added to the parity-check
    matrix so that the fundamental polytope changes. Parity-check matrices
    whose Tanner graphs have large girth therefore possess a good
    complexity-approximation tradeoff: it takes much more effort to get a
    better approximation of $\convhull(C)$.

\end{itemize}

The above considerations show that large girth seems to be a desirable design
criterion when construction LDPC codes. This supports for example the type of
random LDPC code constructions as presented by Hu et
al.~in~\cite{Hu:Eleftheriou:Arnold:05:1}. It is certainly also a desirable
criterion when designing algebraically constructed LDPC codes, nevertheless
one has to be careful beyond having simply a large girth: a Tanner graph with
a cycle structure that is "too nice" can lead to either low-weight codewords
(which is very bad) or low-weight pseudo-codewords (which might potentially be
detected and avoided in a decoder). E.g.~in the case of the Margulis
construction with Ramanujan graphs one has large girth but also a minimum
distance of $24$ for $n = 4896$~\cite{Rosenthal:Vontobel:00,
MacKay:Postol:03:1}. Obviously, adding any possible better constraints does
not help as this minimum codeword will always be included. Although the
original Margulis codes~\cite{Margulis:82} do not seem to have low-weight
codewords they exhibit some near-codewords~\cite{MacKay:Postol:03:1}. These
near-codewords might be avoided using better relaxations.

Another word of caution: when adding redundant rows to a parity-check matrix
it is clear that the decoding performance of GCD and LPD can only become
better. A question remains as how far GCD is still a good model of MPID when
the parity-check matrix contains many more rows than columns. (Some initial
explorations in this direction were presented
in~\cite{Kelley:Sridhara:05:1:subm}.)


\section{Conclusions}
\label{sec:conclusions:1}

We have introduced graph-cover decoding, a theoretical tool that helps to
establish a bridge between linear-programming decoding and message-passing
iterative decoding and explains why they perform similarly. The central object
behind these decoding algorithms is the fundamental polytope which is a
function of the graphical representation of the code (and not of the
channel). Therefore, different representations of the same code yield
(potentially) different fundamental polytopes. Vectors inside the fundamental
polytope are called pseudo-codewords and their influence is measured by the
pseudo-weight, a function that depends on the pseudo-codeword and the channel
law. For all the cases where the behavior of message-passing decoding is known
analytically, the graph-cover decoder gives the correct predictions and for
the other cases the graph-cover decoder seems to be a good model of the
behavior of message-passing decoding. Moreover, there are connections to Bethe
free energy, the marginal polytope, and the metric polytope.

Some of the questions for future research that should be addressed are as
follows. First, given a code and its representation, what analytical and
computational tools can be used to characterize the fundamental polytope?
(Some initial work in this direction was presented
in~\cite{Vontobel:Koetter:04:1, Chaichanavong:Siegel:05:1} where a lower bound
on the AWGNC pseudo-weight was given.) Secondly, how can one construct codes
on graphs whose fundamental polytopes have good properties? Thirdly, one can
always change the Tanner graph of a code, e.g.~by repeating a check many
times, so that the fundamental polytope and therefore also the linear
programming decoding performance remains the same whereas the iterative
decoding performance will change. So, up to what degree is the graph-cover
decoding a good model for message-passing decoding? (Some initial work in this
direction was presented in~\cite{Vontobel:Koetter:04:2}
and~\cite{Kelley:Sridhara:05:1:subm}.)

\appendix


\section{Proofs}
\label{sec:app:proofs:1}

This appendix contains a variety of proofs that were used in the main text.

\subsection{Proof of Proposition~\ref{prop:image:in:fp:2}}
\label{sec:proof:prop:image:in:fp:2}

We prove Prop.~\ref{prop:image:in:fp:2} in three major steps. First,
Lemma~\ref{lemma:image:in:fp:help:1} will show that $\GCDset{H}$ is a subset of
$\fph{P}{H}$. Secondly, Lemma~\ref{lemma:image:in:fp:help:2} will prove that
if a point in $\fph{P}{H}$ has only rational entries then it must also be in
$\GCDset{H}$. Thirdly, Lemma~\ref{lemma:image:in:fp:help:3} will prove that all
vertices of $\fph{P}{H}$ are vectors with rational
entries. Eq.~\eqref{eq:sets:s:and:fp:1} is then a simple consequence of these
first two lemmas, \eqref{eq:sets:s:and:fp:2} is a simple consequence
of~\eqref{eq:sets:s:and:fp:1}, and the statement that all vertices of
$\fph{P}{H}$ are in $\GCDset{H}$ is a consequence of the third lemma.

\begin{Lemma}
  \label{lemma:image:in:fp:help:1}

  It holds that
  \begin{align*}
    \GCDset{H}
      &\subseteq
         \fph{P}{H}.
  \end{align*}
\end{Lemma}

\begin{Proof}
  Let $\cgraph{T}$ be any $M$-fold cover of $\tgraph{T}{H}$ and let $\ccode{C}
  \defeq \ctgcode{C}{T}$. Because of~\eqref{eq:fp:def:1}, if we can show that
  $\vomega(\cvx) \in \convhull(\code{C}_j)$ for all $\cvx \in \ctgcode{C}{T}$
  and for all $j \in \set{J}$ we are done. Fix some $\cvx \in \ctgcode{C}{T}$
  and some $j \in \set{J}$. As we saw in the remarks after
  Ex.~\ref{ex:simple:code:1}, the Tanner graph $\cgraph{T}$ defines some
  permutations $\pi_{j,i}$ for all $i \in \set{I}_j$ and so $\cvx$ fulfills
  \begin{align}
    \sum_{i \in \set{I}_j}
      \cover{x}_{i,\pi_{j,i}(m)}
      &= 0 
           \quad \text{(in $\GF{2}$)}
             \label{eq:cover:code:codeword:sum:property:1}
  \end{align}
  for all $m \in [M]$. In order to simplify the following expressions, let us
  introduce some dummy permutations $\pi_{j,i}$ for all $i \in \set{I}
  \setminus \set{I}_j$. Then, for $m \in [M]$, let us define the vectors
  ${\vx'}^{(m)} \in \R^n$ with
  \begin{align*}
    {x'}^{(m)}_{i}
      &\defeq
         \cover{x}_{i,\pi_{j,i}(m)}
  \end{align*}
  for all $i \in \set{I}$. Rewriting
  \eqref{eq:cover:code:codeword:sum:property:1} as
  \begin{align}
    \sum_{i \in \set{I}_j}
      {x'_i}^{(m)}
        &= 0 \quad \text{(in $\GF{2}$)}
  \end{align}
  we see that ${\vx'}^{(m)} \in \code{C}_j(\matr{H})$ for all $m \in [M]$.  A
  convex sum of these $M$ vectors must obviously lie in
  $\convhull(\code{C}_j(\matr{H}))$:
  \begin{align}
    \sum_{m \in [M]}
      \frac{1}{M}
        {\vx'}^{(m)}
      &\in \convhull(\code{C}_j(\matr{H})).
             \label{eq:cover:code:codeword:sum:property:2}
  \end{align}
  Observing that the $i$-th position of the left-hand side in
  \eqref{eq:cover:code:codeword:sum:property:2} takes on the value
  \begin{align}
    \sum_{m \in [M]}
      \frac{1}{M}
        {x'_i}^{(m)}
      &= \frac{1}{M}
           \sum_{m \in [M]}
             {x'_i}^{(m)}
       = \frac{1}{M}
           \sum_{m \in [M]}
             \cover{x}_{i,\pi_{j,i}(m)}
       = \frac{1}{M}
           \sum_{m' \in [M]}
             \cover{x}_{i,m'}
       = \omega_i(\cvx),
           \label{eq:cover:code:codeword:sum:property:3}
  \end{align}
  we conclude that $\vomega(\cvx) \in
  \convhull(\code{C}_j(\matr{H}))$. Because $\cgraph{T}$, $\cvx$, and $j$ were
  arbitrary, this finishes the proof.

  Note that when $\fph{P}{H}$ contains more than one point in $\R^n$ then the
  subset relationship between $\GCDset{H}$ and $\fph{P}{H}$ is strict:
  $\GCDset{H} \subsetneq \fph{P}{H}$. To prove this, simply choose a point
  $\vomega$ in $\fph{P}{H}$ where at least one component is irrational:
  because all points in $\GCDset{H}$ have rational components it follows that
  $\vomega \notin \GCDset{H}$. (Note that the case where $\fph{P}{H}$ contains
  only one point in $\R^n$ can only happen for block length $n = 1$ and
  parity-check matrices like $\matr{H} = (1)$.)
\end{Proof}

\begin{Lemma}
  \label{lemma:image:in:fp:help:2}

  If a point in $\fph{P}{H}$ has only rational entries then it must also be in
  $\GCDset{H}$.
\end{Lemma}

The main part of the following proof will consist of an algorithm;
Ex.~\ref{ex:simple:code:1:getting:the:cover:graph:1} (which can be found in
the text after this proof) illustrates the involved concepts with the help of
a code that we have already used earlier on.

\begin{Proof}
  We will prove this lemma as follows: for an arbitrary point $\vnu \in
  \fph{P}{H} \cap \Q^n$ we will show that there is an $M$-cover $\cgraph{T}$
  of $\tgraph{T}{H}$ such that we can exhibit a codeword $\cvx \in
  \ctgcode{C}{T}$ such that $\vomega(\cvx) = \vnu$.
  
  So, let $\vnu \in \fph{P}{H} \cap \Q^n$. Because $\vnu \in \fph{P}{H}$ we
  have $\vnu \in \convhull(\code{C}_j(\matr{H}))$ for $j \in \set{J}$. Using
  Carath\'eodory's Theorem (see e.g.~\cite[p.~10]{Barvinok:02:1}), we can
  conclude that for all $j \in \set{J}$ we can write
  \begin{align*}
    \vnu
      &= \valpha^{(j)} \matr{P}^{(j)},
  \end{align*}
  where $\matr{P}^{(j)}$ is an $(n+1) \times n$ matrix where the rows
  represent some vertices of $\convhull(\code{C}_j(\matr{H}))$, i.e.~codewords
  of $\code{C}_j(\matr{H})$, and where $\valpha^{(\ell)}$ is a vector of
  length $n+1$ where all entries are nonzero and sum to one. For each $j \in
  \set{J}$ these statements can be reformulated to
  \begin{align*}
    \begin{pmatrix}
      \vnu & 1
    \end{pmatrix}
      &= \valpha^{(j)}
         \begin{pmatrix} 
           \matr{P}^{(j)} & \vect{1}^\tr
         \end{pmatrix}.
  \end{align*}
  This is a system of $n+1$ equations with $n+1$ unknowns. Because $\vnu \in
  \Q^n$ and because all entries of $\matr{P}^{(j)}$ are either $0$ or $1$, we
  can conclude with the help of Cram\'er's rule for solving systems of linear
  equations (see e.g.~\cite{Horn:Johnson:90}) that all entries of
  $\valpha^{(j)}$ must be rational.

  Now we proceed to construct a finite cover $\cgraph{T}$ of $\tgraph{T}{H}$
  and a codeword $\cvx \in \ctgcode{C}{T}$. Let $M$ be a common denominator of
  all the entries of all the vectors $\valpha^{(j)}$, $j \in \set{J}$: from
  this we have that not only $M \valpha^{(j)} \in \Intp^n$, $j \in \set{J}$,
  but also that $M \vnu \in \Intp^n$. The graph $\cgraph{T}$ shall be an
  $M$-cover of $\tgraph{T}{H}$ with symbol nodes $X_{i,m}$, $(i,m) \in \set{I}
  \times [M]$ and check nodes $B_{j,m}$, $(j,m) \in \set{J} \times [M]$. The
  entries of the codeword $\cvx$ shall be
  \begin{align*}
    \cover{x}_{i,m}
      &= \begin{cases}
           1 & \text{($i \in \set{I}$, $m \in [M \nu_i]$)} \\
           0 & \text{(otherwise)}
         \end{cases}.
  \end{align*}
  It now remains to specify the connection pattern of $\cgraph{T}$, i.e.~what
  symbol node is connected to what check node. Once this pattern is specified,
  it will be easy to see that $\cgraph{T}$ is indeed an $M$-cover of
  $\tgraph{T}{H}$ and that $\cvx$ is a codeword in $\ctgcode{C}{T}$. We use
  the following algorithm:
  \begin{itemize}
    
    \item For all $j \in \set{J}$ do: 

    \begin{itemize}

      \item Let $m_j \defeq 1$. For all $i \in \set{I}_j$, let $m'_i \defeq 1$
        and $m''_i \defeq M \nu_i + 1$.

      \item For $\ell$ from $1$ to $n+1$ do: for $s$ from $1$ to $M
        \alpha^{(j)}_{\ell}$ do:

        \begin{itemize}

          \item For all $i \in \set{I}_j$ do:

            \begin{itemize}

              \item If $[\matr{P}^{(j)}]_{\ell,i} = 1$ then connect
                $X_{i,m'_i}$ to $B_{j,m_j}$ and let $m'_i \defeq m'_i + 1$.

              \item If $[\matr{P}^{(j)}]_{\ell,i} = 0$ then connect
                $X_{i,m''_i}$ to $B_{j,m_j}$ and let $m''_i \defeq m''_i + 1$.

            \end{itemize}

          \item Let $m_j \defeq m_j + 1$.

        \end{itemize}

      \end{itemize}

  \end{itemize}
  We leave it to the reader to check that this construction indeed yields the
  desired graph cover and codeword.  

\end{Proof}

\begin{figure}
  \begin{center}
    \epsfig{figure=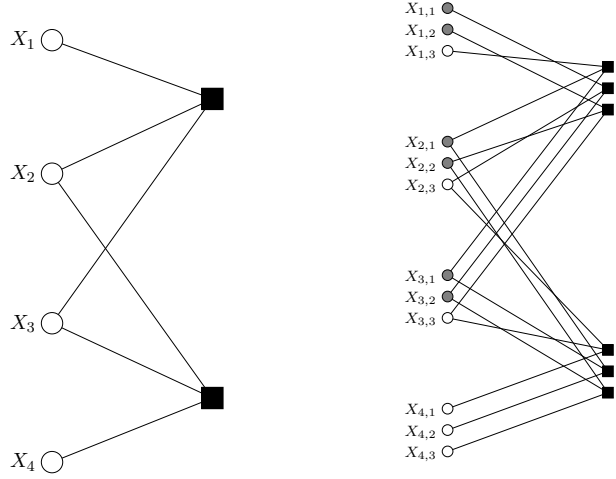,width=8cm}
  \end{center}
  \caption{Left: Tanner graph $\tgraph{T}{H}$ of the simple binary linear code
           in Ex.~\ref{ex:simple:code:1}. Right: $3$-cover of $\tgraph{T}{H}$
           as found in
           Ex.~\ref{ex:simple:code:1:getting:the:cover:graph:1}. The shading
           of the symbol nodes indicates the codeword found in this example.}
  \label{fig:simple:code:1:special:3:cover:1}
\end{figure}

\begin{Example}
  \label{ex:simple:code:1:getting:the:cover:graph:1}

  We continue Ex.~\ref{ex:simple:code:1}. In Exs.~\ref{ex:simple:code:2}
  and~\ref{ex:simple:code:1:pseudo:codeword:1} we saw that the vector $\vnu =
  (\frac{2}{3}, \frac{2}{3}, \frac{2}{3}, 0)$ is a pseudo-codeword. Let us
  show how the algorithm in the proof of Lemma~\ref{lemma:image:in:fp:help:2}
  handles this vector. First of all, we must check that $\vnu \in \fph{P}{H}
  \cap \Q^n$. This is indeed true. Next, we have to find the matrices
  $\matr{P}^{(1)}$ and $\matr{P}^{(2)}$. Note that the codes $\code{C}_1$ and
  $\code{C}_2$ are the sets
  \begin{align*}
    \code{C}_1
      &= \begin{Bmatrix}
           (0, 0, 0) \\
           (0, 1, 1) \\
           (1, 0, 1) \\
           (1, 1, 0)
         \end{Bmatrix}
         \times
         \begin{Bmatrix}
           (0) \\
           (1)
         \end{Bmatrix}
       = \begin{Bmatrix}
           (0, 0, 0, 0) \\
           (0, 0, 0, 1) \\
           (0, 1, 1, 0) \\
           (0, 1, 1, 1) \\
           (1, 0, 1, 0) \\
           (1, 0, 1, 1) \\
           (1, 1, 0, 0) \\
           (1, 1, 0, 1)
         \end{Bmatrix},
    \quad
    \code{C}_2
       = \begin{Bmatrix}
           (0) \\
           (1)
         \end{Bmatrix}
         \times
         \begin{Bmatrix}
           (0, 0, 0) \\
           (0, 1, 1) \\
           (1, 0, 1) \\
           (1, 1, 0)
         \end{Bmatrix}
       = \begin{Bmatrix}
           (0, 0, 0, 0) \\
           (0, 0, 1, 1) \\
           (0, 1, 0, 1) \\
           (0, 1, 1, 0) \\
           (1, 0, 0, 0) \\
           (1, 0, 1, 1) \\
           (1, 1, 0, 1) \\
           (1, 1, 1, 0)
         \end{Bmatrix}.
  \end{align*}
  For the given vector $\vnu$ it turns out that $\vnu = \valpha^{(1)}
  \matr{P}^{(1)}$ and $\vnu = \valpha^{(2)} \matr{P}^{(2)}$
  with\footnote{Other choices for $\valpha^{(1)}$, $\valpha^{(2)}$,
  $\matr{P}^{(1)}$, and $\matr{P}^{(2)}$ can also yield $\vnu$.}
  \begin{alignat*}{2}
    \valpha^{(1)}
     &&= \begin{pmatrix}
           0 & 0 & \frac{1}{3} & \frac{1}{3} & \frac{1}{3}
         \end{pmatrix},
    \quad\quad\quad
    \valpha^{(2)}
      &= \begin{pmatrix}
           \frac{1}{3} & 0 & 0 & 0 & \frac{2}{3}
         \end{pmatrix}, \\
    \matr{P}^{(1)}
     &&= \begin{pmatrix}
           0 & 0 & 0 & 0 \\
           0 & 0 & 0 & 1 \\
           0 & 1 & 1 & 0 \\
           1 & 0 & 1 & 0 \\
           1 & 1 & 0 & 0
         \end{pmatrix},
    \quad\quad\quad
    \matr{P}^{(2)}
      &= \begin{pmatrix}
           0 & 0 & 0 & 0 \\
           0 & 0 & 1 & 1 \\
           0 & 1 & 0 & 1 \\
           0 & 1 & 1 & 0 \\
           1 & 1 & 1 & 0
         \end{pmatrix}
  \end{alignat*}
  Note that the first two lines of $\matr{P}^{(1)}$ and the three middle lines
  of $\matr{P}^{(2)}$ are dummy lines so that $\valpha^{(1)}$ and
  $\valpha^{(2)}$ have $n+1 = 5$ entries.

  We see that $M = 3$ is the smallest common denominator of all the entries in
  $\valpha^{(1)}$ and $\valpha^{(2)}$, therefore let us find a $3$-cover of
  $\tgraph{T}{H}$ that has a codeword $\cvx \in \ctgcode{C}{T}$ such that
  $\vomega(\cvx) = \vnu$. Applying the rest of the algorithm in
  Lemma~\ref{lemma:image:in:fp:help:2} we find the $3$-cover graph
  $\cgraph{T}$ in Fig.~\ref{fig:simple:code:1:special:3:cover:1} (right) and
  the codeword $\cvx = (1{:}1{:}0, \ 1{:}1{:}0, \ 1{:}1{:}0, \ 0{:}0{:}0) \in
  \ctgcode{C}{T}$.
\end{Example}

\begin{Lemma}
  \label{lemma:image:in:fp:help:3}

  All vertices of $\fph{P}{H}$ are vectors with rational entries.
\end{Lemma}

\begin{Proof}
  Remember that $\fph{P}{H} = \cap_{j \in \set{J}(\matr{H})}
  \convhull(\code{C}_j(\matr{H}))$ is defined as the intersection of
  $\card{\set{J}(\matr{H})}$ polytopes. However, all polytopes
  $\convhull(\code{C}_j(\matr{H}))$, $j \in \set{J}(\matr{H})$ can be defined
  with linear inequalities that involve only integer coefficients,
  cf.~Lemmas~\ref{lemma:properties:fp:fc:1}
  and~\ref{lemma:properties:fp:fc:2}. Therefore, also $\fph{P}{H}$ can be
  defined with linear inequalities that involve only integer coefficients. Now,
  any vertex of $\fph{P}{H}$ is a point in $\fph{P}{H}$ where $n$ inequalities
  hold with equality and where these $n$ equalities form a system of linear
  equations with full rank. Using Cram\'er's rule for solving systems of
  linear equations (see e.g.~\cite{Horn:Johnson:90}) we see that indeed all
  vertices of $\fph{P}{H}$ are vectors with rational entries.
\end{Proof}

\subsection{Proof of Proposition~\ref{prop:relationship:gc:dec:lp:dec:1}}
\label{sec:proof:prop:relationship:gc:dec:lp:dec:1}

Let $\lambda_i \defeq \lambda_i(y_i)$ be defined as
in~\eqref{eq:log:likelihood:definition:1}. Let us first prove the following
lemma.

\begin{Lemma}
  \label{lemma:gc:dec:property:1}

  \begin{align}
    \hvomega^{\GCD{H}}(\vy)
      &= \arg \min_{\vomega \in \GCDset{H}}
           \sum_{i \in \set{I}}
             \omega_i \lambda_i.
               \label{eq:gc:dec:property:1}
  \end{align}
\end{Lemma}

\begin{Proof}
  Let us first rewrite the right-hand side
  of~\eqref{eq:graph:cover:decoder:1}. Because
  $P_{\cvY|\cvX}(\lift{\vy}{M}|\lift{\vect{0}}{M})$ is a constant for a given
  $\vy$, instead of maximizing $(1/M) \log P_{\cvY|\cvX}(\lift{\vy}{M}|\cvx)$
  in~\eqref{eq:graph:cover:decoder:1} we can also maximize
  \begin{align*}
    \frac{1}{M}
      \log \frac{P_{\cvY|\cvX}(\lift{\vy}{M}|\cvx)}
                {P_{\cvY|\cvX}(\lift{\vy}{M}|\lift{\vect{0}}{M})}
      &= \frac{1}{M}
           \sum_{i \in \set{I}}
             \sum_{m \in [M]}
               \log
                 \frac{P_{Y_{i,m}|X_{i,m}}(y_i|\cover{x}_{i,m})}
                      {P_{Y_{i,m}|X_{i,m}}(y_i|0)} \\
      &= \frac{1}{M}
           \sum_{i \in \set{I}}
             \sum_{m \in [M]}
               \log
                 \frac{P_{Y_{i}|X_{i}}(y_i|\cover{x}_{i,m})}
                      {P_{Y_{i}|X_{i}}(y_i|0)} \\
      &= -
         \frac{1}{M}
           \sum_{i \in \set{I}}
             \sum_{m \in [M]}
               \cover{x}_{i,m}
               \lambda_i \\
      &= - 
         \sum_{i \in \set{I}}
           \left(
             \frac{1}{M}
               \sum_{m \in [M]}
                 \cover{x}_{i,m}
           \right)
           \lambda_i \\
      &\overset{(*)}{=}
         -
         \sum_{i \in \set{I}}
           \omega_i(\cvx)
           \cdot
           \lambda_i,
  \end{align*}
  where at step $(*)$ we used~\eqref{eq:def:pseudo:codeword:1:1}. With this we
  can extend~\eqref{eq:graph:cover:decoder:1} to read
  \begin{align*}
    (\hM, \hcT, \hcvx)^{\GCD{H}}(\vy) 
      &= \arg \max_{(M, \cgraph{T}, \cvx) \in \cGCDset{H}}
           \frac{1}{M}
             \log
               P_{\cvY|\cvX}(\lift{\vy}{M}|\cvx) \\
      &= \arg \min_{(M, \cgraph{T}, \cvx) \in \cGCDset{H}}
           \sum_{i \in \set{I}}
             \omega_i(\cvx)
             \cdot
             \lambda_i.
  \end{align*}
  Remembering the relationship between $\cGCDset{H}$ and $\GCDset{H}$ as
  defined in~\eqref{eq:cgcs:def:1} and~\eqref{eq:gcs:def:1}, respectively, we
  can write
  \begin{align*}
    \hvomega^{\GCD{H}}(\vy)
      &\defeq
         \vomega
           \left(
             \hcvx^{\GCD{H}}(\vy)
           \right)
       = \arg \min_{\vomega \in \GCDset{H}}
           \sum_{i \in \set{I}}
             \omega_i
             \cdot
             \lambda_i,
  \end{align*}
  which proves the lemma.
\end{Proof}

Lemma~\ref{lemma:gc:dec:property:1} allows us now to prove
Prop.~\ref{prop:relationship:gc:dec:lp:dec:1}. Using the convexity of
$\fph{P}{H}$ and a result that we found in Prop.~\ref{prop:image:in:fp:2},
namely that all vertices of $\fph{P}{H}$ are in $\GCDset{H}$, we can
extend~\eqref{eq:gc:dec:property:1} to read
\begin{align*}
  \hvomega^{\GCD{H}}(\vy)
    &= \arg \min_{\vomega \in \GCDset{H}}
         \sum_{i \in \set{I}}
           \omega_i \lambda_i
     = \arg \min_{\vomega \in \fph{P}{H}}
         \sum_{i \in \set{I}}
           \omega_i \lambda_i,
\end{align*}
which proves the proposition.

\subsection{Proof of Lemma~\ref{lemma:integrality:1}}
\label{sec:proof:lemma:integrality:1}

Let $\cgraph{T}$ be an $M$-cover of $\tgraph{T}{H}$ and let $\cvx \in
\ctgcode{C}{T}$. We know that $M \vomega(\cvx) \in \Intp^n$ and from
Prop.~\ref{prop:image:in:fp:2} we know that $\vomega(\cvx) \in \fph{P}{H} \cap
\Q^n$. Because $\fch{K}{H} = \conichull(\fph{P}{H})$ we conclude that $M
\vomega(\cvx) \in \fch{K}{H}$. Therefore, $M \vomega(\cvx) \in \fch{K}{H} \cap
\Int^n$, which proves the first statement.

Similar to the proof of Lemma~\ref{lemma:image:in:fp:help:1}, let us fix some
$j \in \set{J}$ and let us associate the vectors ${\vx'}^{(m)}$, $m \in [M]$
to $\cvx$. There it was shown that ${\vx'}^{(m)} \in \code{C}_j(\matr{H})$ for
all $m \in [M]$. Rewriting~\eqref{eq:cover:code:codeword:sum:property:3} to
read $M \vomega(\cvx) = \sum_{m \in [M]} {\vx'}^{(m)}$, we see that $M
\vomega(\cvx) \in \code{C}_j \text{ (in $\GF{2}$)}$. Because $j$ was arbitrary
and because $\code{C} = \cap_{j \in \set{J}} \code{C}_j(\matr{H})$, we have
$\set{Z}(\matr{H}) \subseteq \code{C} \text{ (in $\GF{2}$)}$. Moreover, it is
clear that $\set{Z}(\matr{H}) \supseteq \code{C} \text{ (in
$\GF{2}$)}$. Combining these two results proves the second statement.

Let $\vomega$ be a minimal pseudo-codeword and consider the half-ray given by
$\{ \alpha \vomega \, | \, \alpha \in \Rp \}$. Because the fundamental cone
$\fch{K}{H}$ is the conic hull of the fundamental polytope $\fph{P}{H}$ we
know that there is a non-zero vertex of the fundamental polytope lying on this
half-ray. However, in Prop.~\ref{prop:image:in:fp:2} we have seen that all
vertices of $\fph{P}{H}$ have rational entries and are therefore also in
$\GCDset{H}$. Looking at one of the pre-images $(M,\cgraph{T},\cvx) \in
\cGCDset{H}$ of this vertex we finally see that there must be an $\alpha \in
\Rpp$ such that $\alpha \cdot \vomega = \cvx \in \set{Z}(\matr{H})$. This
proves the third statement.

\subsection{Proof of
  Lemma~\ref{lemma:fp:tree:1}}
\label{sec:proof:lemma:fp:tree:1}

Let us study the set $\GCDset{H}$ as defined in~\eqref{eq:gcs:def:1}
and~\eqref{eq:gcs:def:2}; Prop.~\ref{prop:image:in:fp:2}, which shows a
connection between $\GCDset{H}$ and $\fph{P}{H}$, will then give the desired
result. (Note that we only discuss the case where $\tgraph{T}{H}$ is a
tree. The case where $\tgraph{T}{H}$ is a forest, i.e.~a collection of trees,
is a straightforward extension.)

So, let $\cgraph{T}$ be an $M$-cover $\cgraph{T}$ of $\tgraph{T}{H}$. Because
$\tgraph{T}{H}$ is a tree it is easy to see that $\cgraph{T}$ is a collection
of $M$ disjoint trees that are copies of $\tgraph{T}{H}$. With suitable
labeling of the vertices of $\cgraph{T}$ we have $\ctgcode{C}{T} = \big\{ \cvx
\in \GF{2}^{nM} \, | \, (\tilde x_{1,m}, \ldots, \tilde x_{n,m}) \in \code{C}
\text{ for all $m \in [M]$} \big\}$ and it follows that
\begin{align*}
  \GCDset{H}
    &= \bigcup_{\cgraph{T}: \ 
                \cgraph{T} \text{ is a finite-cover graph of }
                \tgraph{T}{H}}
         \vomega \big( \ctgcode{C}{T} \big)
\end{align*}
equals $\convhull(\code{C}) \cap \Q^n$. Using Prop.~\ref{prop:image:in:fp:2}
we see that $\fph{P}{H} = \overline{\GCDset{H}} = \overline{\convhull(\code{C})
\cap \Q^n} = \convhull(\code{C})$ as promised.

\subsection{Proof of Statements after
  Definition~\ref{def:fractional:and:max:fractional:weight:1}}
\label{sec:proof:after:def:fractional:and:max:fractional:weight:1}

The first statement is proven as follows. Note that $\vomega \in [0,1]^n$. Let
$\set{E} \subseteq \set{I}$ be set of positions were the channel bit flips
happened. $S|_{\vX = \vect{0}}$ is non-negative if and only if $\sum_{i \in
\set{I}} \omega_i \lambda_i \leq 0$ if and only if $-\sum_{i \in \set{E}}
\omega_i + \sum_{i \in \set{I} \setminus \set{E}} \omega_i \leq 0$ if and only
if $-2 \sum_{i \in \set{E}} \omega_i + \sum_{i \in \set{I}} \omega_i \leq
0$. Therefore, a necessary condition for $S|_{\vX = \vect{0}}$ to be
non-positive is that $|\set{E}| \geq \frac{1}{2} \wfr(\vomega)$. This follows
by observing that $|\set{E}| \geq \sum_{i \in \set{E}} \omega_i \geq
\frac{1}{2} \sum_{i \in \set{I}} \omega_i = \frac{1}{2}
\wfr(\vomega)$.

The second statement follows by replacing $\vomega$ by $\vomega /
\infnorm{\vomega}$ in the above argument and by observing that $\vomega /
\infnorm{\vomega} \in [0,1]^n$.

\subsection{Proof of
  Lemma~\ref{lemma:pseudo:weight:variance:interpretation:1}}
\label{sec:proof:lemma:pseudo:weight:variance:interpretation:1}

The expressions in the lemma are obtained doing the following manipulations:
\begin{align}
  \wps(\vomega)
    &\defeq
       \frac{\sonenorm{\vomega}^2}
            {\twonorm{\vomega}^2}
     = \frac{|\set{S}|^2 \cdot (\hat \Expec[\Omega])^2}
            {|\set{S}| \cdot \hat \Expec[\Omega^2]}
     = |\set{S}|
       \cdot
       \frac{(\hat \Expec[\Omega])^2}
            {\hat \Expec[\Omega^2]}
     = |\set{S}|
       \cdot
       \frac{1}
            {\frac{\hat \Expec[\Omega^2] - (\hat \Expec[\Omega])^2}
                  {(\hat \Expec[\Omega])^2}
             + 1}
     = |\set{S}|
       \cdot
       \frac{1}
            {\frac{\widehat \Var[\Omega]}
                  {(\hat \Expec[\Omega])^2}
             + 1}.
\end{align}

\subsection{Proof of
  Lemma~\ref{lemma:pseudo:weight:angle:interpretation:1}}
\label{sec:proof:lemma:pseudo:weight:angle:interpretation:1}

From vector analysis it is well known that $\langle \vomega, \vect{1} \rangle
= \twonorm{\vect{1}} \twonorm{\vomega} \cos\big( \angle(\vomega, \vect{1})
\big)$. With this, we can write
\begin{align}
  \wps(\vomega)
    &= \frac{\langle \vomega, \vect{1} \rangle^2}
            {\twonorm{\vomega}^2}
     = \frac{\twonorm{\vect{1}}^2 \twonorm{\vomega}^2
             \cos\big( \angle(\vomega,\vect{1}) \big)^2}
            {\twonorm{\vomega}^2}
     = n
       \cdot
       \cos
         \big( \angle(\vomega,\vect{1}) \big)^2.
\end{align}
The proof of the second part of the lemma statement is analogous.

\subsection{Proof of 
            Lemma~\ref{lemma:convex:combination:pseudo:distance:lower:bound:1}}
\label{sec:proof:lemma:convex:combination:pseudo:distance:lower:bound:1}

We only consider the AWGNC pseudo-weight case, the other cases are left to the
reader as an exercise. The proof for the AWGNC pseudo-weight case is done in
two steps: first we prove a simplified statement
(Lemma~\ref{lemma:aux:lemma:convex:combination:pseudo:distance:lower:bound:1}),
then we prove the general case.

\begin{Lemma}
  \label{lemma:aux:lemma:convex:combination:pseudo:distance:lower:bound:1}

  Consider the same setup as in
  Lemma~\ref{lemma:convex:combination:pseudo:distance:lower:bound:1}.
  Assuming additionally that $\sonenorm{\vomega^{(\ell)}} = 1$ for all $\ell
  \in [L]$ and that $\sum_{\ell \in [L]} \alpha_{\ell} = 1$ we have
  \begin{align}
    \wps
      \left(
        \sum_{\ell \in [L]}
          \alpha_{\ell} \vomega^{(\ell)}
      \right)
      &\geq
         \min_{\ell \in [L]}
           \wps(\vomega^{(\ell)})
  \end{align}
\end{Lemma}

\begin{Proof}
  Let $\vnu \defeq \sum_{\ell \in [L]} \alpha_{\ell} \vomega^{(\ell)}$. Using
  the assumptions, it is easy to see that $\sonenorm{\vnu} = 1$. Moreover,
  \begin{align}
    \twonorm{\vnu}^2
      &= \twonorm{
           \sum_{\ell \in [L]}
             \alpha_{\ell} \vomega^{(\ell)}
         }^2
       = \sum_{i \in [n]}
           \left(
             \sum_{\ell_1 \in [L]}
               \alpha_{\ell_1} \omega^{(\ell_1)}_i
           \right)
           \left(
             \sum_{i_2 \in [L]}
               \alpha_{\ell_2} \omega^{(\ell_2)}_i
           \right) \nonumber \\
      &= \sum_{\ell_1 \in [L]}
           \sum_{\ell_2 \in [L]}
             \alpha_{\ell_1} \alpha_{\ell_2}
             \sum_{i \in [n]}
               \omega^{(\ell_1)}_i \omega^{(\ell_2)}_i \nonumber \\
      &\overset{(*)}{\leq}
         \sum_{\ell_1 \in [L]}
           \sum_{\ell_2 \in [L]}
             \alpha_{\ell_1} \alpha_{\ell_2}
             \sqrt
             {
               \left(
                 \sum_{i \in [n]}
                   \big( \omega^{(\ell_1)}_i \big)^2
               \right)
             }
             \sqrt
             {
               \left(
                 \sum_{i \in [n]}
                   \big( \omega^{(\ell_2)}_i \big)^2
               \right)
             } \nonumber \\
      &= \sum_{\ell_1 \in [L]}
           \sum_{i_2 \in [L]}
             \alpha_{\ell_1} \alpha_{\ell_2}
             \twonorm{\vomega^{(\ell_1)}} \twonorm{\vomega^{(\ell_2)}}
       = \left(
           \sum_{\ell \in [L]}
             \alpha_{\ell}
               \twonorm{\vomega^{(\ell)}}
         \right)^2
           \label{eq:wps:lower:bound:two:norm:y:2} \\
      &\leq
         \left(
           \max_{\ell' \in [L]}
             \twonorm{\vomega^{(\ell')}}
           \cdot
           \sum_{\ell \in [L]}
             \alpha_{\ell}
         \right)^2
       = \max_{\ell \in [L]}
           \twonorm{\vomega^{(\ell)}}^2,
             \label{eq:wps:lower:bound:two:norm:y:1}
  \end{align}
  where step $(*)$ follows from the Cauchy-Schwarz inequality. Concluding,
  \begin{align}
    \wps(\vnu)
      &= \frac{\sonenorm{\vnu}^2}
              {\twonorm{\vnu}^2}
       = \frac{1}
              {\twonorm{\vnu}^2}
       \overset{(*)}{\geq}
         \frac{1}
              {\max_{\ell \in [L]}
               \twonorm{\vomega^{(\ell)}}^2
              }
       = \min_{\ell \in [L]}
           \frac{1}{\twonorm{\vomega^{(\ell)}}^2}
       = \min_{\ell \in [L]}
           \frac{\sonenorm{\vomega^{(\ell)}}^2}
                {\twonorm{\vomega^{(\ell)}}^2}
       = \min_{\ell \in [L]}
           \wps(\vomega^{(\ell)}),
  \end{align}
  where step $(*)$ follows from (\ref{eq:wps:lower:bound:two:norm:y:1}).

\end{Proof}

Now we prove
Lemma~\ref{lemma:convex:combination:pseudo:distance:lower:bound:1}. For $\ell
\in [L]$, let $\alpha'_{\ell} \defeq \alpha_{\ell} \sonenorm{\vomega^{(\ell)}}
/ \left( \sum_{\ell' \in [L]} \alpha_{\ell'} \sonenorm{\vomega^{(\ell')}}
\right)$ and let $\vomega'_{\ell} \defeq \vomega^{(\ell)} /
\sonenorm{\vomega^{(\ell)}}$.  Note that $\sum_{\ell \in [L]} \alpha'_{\ell} =
1$ and that $\sonenorm{\vomega'_{\ell}} = 1$, $\ell \in [L]$. Then
\begin{align}
  \wps(\vnu)
    &\overset{(*)}{=}
       \wps
         \left(
           \frac{\vnu}
             {\sum_{\ell' \in [L]}
                \alpha_j \sonenorm{\vomega^{(\ell')}}
             }
         \right)
     = \wps
         \left(
           \sum_{\ell \in [L]}
             \frac{\alpha_{\ell} \sonenorm{\vomega^{(\ell)}}}
                  {\sum_{\ell' \in [L]} \alpha_{\ell'}
                     \sonenorm{\vomega^{(\ell')}}}
             \cdot
             \frac{\vomega^{(\ell)}}{\sonenorm{\vomega^{(\ell)}}}
         \right) \\
    &= \wps
         \left(
           \sum_{\ell \in [L]}
             \alpha'_{\ell} \vomega'_{\ell}
         \right)
     \overset{(**)}{\geq}
         \min
           \left(
             \wps(\vomega^{(1)}), \ldots, \wps(\vomega^{(L)})
           \right).
\end{align}
where at step $(*)$ we used the scaling-invariance of $\wps(\,\cdot\,)$ and at
step $(**)$ we used the above lemma and the fact that $\wps(\vomega^{(\ell)})
= \wps(\vomega'_{\ell})$ for $\ell \in [L]$.

\subsection{Proof of 
            Lemma~\ref{lemma:convex:combination:pseudo:distance:lower:bound:2}}
\label{sec:proof:lemma:convex:combination:pseudo:distance:lower:bound:2}

\begin{Proof}
  Let $\vnu \defeq \sum_{\ell \in [L]} \alpha_{\ell} \vomega^{(\ell)}$. Note
  that the assumptions in the lemma statement imply that $\sonenorm{\vnu} =
  1$. The inequality follows then by using partial results of the proof of
  Lemma~\ref{lemma:aux:lemma:convex:combination:pseudo:distance:lower:bound:1}.
  Specifically, we use (\ref{eq:wps:lower:bound:two:norm:y:2}) which says that
  \begin{align}
    \twonorm{\vnu}^2
      &\leq
         \left(
           \sum_{\ell \in [L]}
             \alpha_{\ell}
               \twonorm{\vomega^{(\ell)}}
         \right)^2
    \quad
    \text{or, equivalently,}
    \quad
    \twonorm{\vnu}
       \leq
         \sum_{\ell \in [L]}
           \alpha_{\ell}
             \twonorm{\vomega^{(\ell)}}.
  \end{align}
  For $\sonenorm{\vnu} = 1$ and $\sonenorm{\vomega}^{(\ell)} = 1$ we have
  $\twonorm{\vnu} = 1/\sqrt{\wps(\vnu)}$ and $\twonorm{\vomega^{(\ell)}} =
  1/\sqrt{\wps(\vomega^{(\ell)})}$, respectively, and the result follows then
  immediately from the assumptions in the lemma statement and the above
  considerations.
\end{Proof}

\subsection{Proof of 
            Lemma~\ref{lemma:pseudo:weight:derivatives:1}}
\label{sec:proof:lemma:pseudo:weight:derivatives:1}

\begin{Proof}
  We have
  \begin{align*}
    \frac{\partial}{\partial \omega_i}
      \wps(\vomega)
      &= \frac{\partial}{\partial \omega_i}
           \frac{\left(
                   \sum_{{i'} \in [n]} \omega_{i'}
                 \right)^2}
                {\sum_{{i'} \in [n]} \omega_{i'}^2}
       = \frac{2\left(
                  \sum_{{i'} \in [n]} \omega_{i'}
                \right)}
              {\sum_{{i'} \in [n]} \omega_{i'}^2}
         -
         \frac{\left(
                 \sum_{{i'} \in [n]} \omega_{i'}
               \right)^2
               2 \omega_i}
              {\left(
                 \sum_{{i'} \in [n]} \omega_{i'}^2
               \right)^2} \\
       &= 2
          \frac{\left(
                  \sum_{{i'} \in [n]} \omega_{i'}
                \right)^2}
              {\left(
                 \sum_{{i'} \in [n]} \omega_{i'}^2
               \right)^2}
              \left[
                \left(
                  \sum_{{i'} \in [n]} \omega_{i'}
                \right)
		\frac{\sum_{{i'} \in [n]} \omega_{i'}^2}
                     {\left(
                        \sum_{{i'} \in [n]} \omega_{i'}
                      \right)^2}
                -
                \omega_i
              \right].
  \end{align*}
  The lemma follows then by analyzing the expression in the square brackets.
\end{Proof}

\subsection{Proof of 
            Lemma~\ref{lemma:pseudoo:weight:inequalities:1}}
\label{sec:proof:lemma:pseudoo:weight:inequalities:1}

\begin{Proof}
  For $\vomega = \vect{0}$ the statement is trivial. So, assume that $\vomega
  \neq \vect{0}$. Because we assume in the lemma that $\vomega \leq \vect{1}$
  we must have $\max_{i \in [n]} \omega_i \leq 1$, which proves the first
  inequality in~\eqref{eq:pseudo:weight:inequalities:1}. The second inequality
  in~\eqref{eq:pseudo:weight:inequalities:1} follows upon observing that
  \begin{align*}
    \wmaxfr(\vomega)
      &= \frac{\sum_{i \in [n]} \omega_i}
              {\max_{i \in [n]} \omega_i}
       = \frac{\left(
                 \sum_{i \in [n]} \omega_i
               \right)^2}
              {\left(
                 \max_{i \in [n]} \omega_i
               \right)
               \left(
                 \sum_{i \in [n]} \omega_i
               \right)}
       = \frac{\left(
                 \sum_{i \in [n]} \omega_i
               \right)^2}
              {\sum_{i \in [n]}
                 (\omega_i \max_{i' \in [n]} \omega_{i'})} \\
      &\leq
          \frac{\left(
                 \sum_{i \in [n]} \omega_i
               \right)^2}
              {\sum_{i \in [n]} \omega_i^2}
       = \wps(\vomega).
  \end{align*}
  The third inequality in~\eqref{eq:pseudo:weight:inequalities:1} can be
  proven as follows. Let $\vect{1}_{\vomega}$ be the indicator vector of
  $\vomega$, i.e.~the $i$-the position is $1$ if $\omega_i$ is non-zero and it
  is $0$ otherwise. Then, using the Cauchy-Schwarz inequality we see that
  $\sonenorm{\vomega}^2 = \langle \vomega, \vect{1} \rangle^2 = \langle
  \vomega, \vect{1}_{\vomega} \rangle^2 \leq \twonorm{\vomega}^2 \cdot
  \twonorm{\vect{1}_{\vomega}}^2 = \twonorm{\vomega}^2 \cdot
  \card{\supp(\vomega)} = \twonorm{\vomega}^2 \cdot \wpsBEC(\vomega)$ and
  dividing by $\twonorm{\vomega}^2$ yields the desired expression.

  The inequalities in~\eqref{eq:pseudo:weight:inequalities:3} follow from the
  inequalities in~\eqref{eq:pseudo:weight:inequalities:1} by observing that
  $\wps(\,\cdot\,)$ and $\wpsBEC(\,\cdot\,)$ are scaling-invariant and
  therefore, when finding $\wpsminh{H}(\,\cdot\,)$ and
  $\wpsBECminh{H}(\,\cdot\,)$, it is sufficient to minimize over the non-zero
  vertices of the fundamental polytope.

  The first inequality in~\eqref{eq:pseudo:weight:inequalities:2} is the same
  as the first inequality in~\eqref{eq:pseudo:weight:inequalities:1}. In order
  to prove the second inequality in~\eqref{eq:pseudo:weight:inequalities:2}
  consider the functions $f(\,\cdot\,)$ and $F(\,\cdot\,)$ and the value $e$
  in Def.~\ref{def:bsc:pseudo:weight:1}. On the one hand, the area under
  $f(\,\cdot\,)$ from $0$ to $e$ equals $(1/2) \cdot F(n) = (1/2) \cdot
  \sonenorm{\vomega}$ by definition. On the other hand, because $f(\,\cdot\,)$
  is non-increasing, the same area is upper bounded by $e \cdot
  \infnorm{\vomega}$. Solving for $2e$ we obtain $2e \geq \sonenorm{\vomega} /
  \infnorm{\vomega}$. The third inequality
  in~\eqref{eq:pseudo:weight:inequalities:2} is obtained as follows. First,
  note that $F(\card{\supp(\vomega)}) = F(n)$. Secondly, consider the chord
  from $(0,F(0)=0)$ to $(\card{\supp(\vomega)}, F(\card{\supp(\vomega)}) =
  \sonenorm{\vomega})$. Because $F(\,\cdot\,)$ is concave, the cord is always
  below $F(\,\cdot\,)$ in the domain of interest. Therefore, $F(e)$, which by
  definition must be equal to $(1/2) \cdot \sonenorm{\vomega}$, is not smaller
  than $(\sonenorm{\vomega} / \card{\supp(\vomega)}) \cdot e$. Combining these
  observations we obtain $2e \leq \card{\supp(\vomega)}$.

  The inequalities in~\eqref{eq:pseudo:weight:inequalities:4} follow from the
  inequalities in~\eqref{eq:pseudo:weight:inequalities:2} by observing that
  $\wps(\,\cdot\,)$ and $\wpsBSC(\,\cdot\,)$ are scaling-invariant and
  therefore, when finding $\wpsminh{H}(\,\cdot\,)$ and
  $\wpsBSCminh{H}(\,\cdot\,)$, it is sufficient to minimize over the non-zero
  vertices of the fundamental polytope.
\end{Proof}

\subsection{Proof of 
            Lemma~\ref{lemma:canonical:completion:properties:1}}
\label{sec:proof:lemma:canonical:completion:properties:1}

The expressions in~\eqref{eq:canonical:completion:sonenorm:1}
and~\eqref{eq:canonical:completion:twonorm:1} for the AWGNC pseudo-weight are
an immediate consequence of Defs.~\ref{def:tanner:graph:ordering:1}
and~\ref{def:canonical:completion:1}. Our main task is therefore to show that
$\vomega \in \fch{K}{H}$. To that end, let us use the fundamental cone
description of Lemma~\ref{lemma:properties:fp:fc:2}. It is obvious that
$\omega_i \geq 0$ for all $i \in \set{I}(\matr{H})$. Now, consider a check
node $B_j$ at tier $2t+1$ for some $t \geq 0$ that is connected to variable
nodes at tier $2t$ and possibly some variable nodes at tier $2t+2$. We
distinguish two cases:
\begin{itemize}

  \item The check node $B_j$ is connected to only one variable node, say
    $X_{i_1}$ at tier $2t$, and $\wrow-1$ variable nodes, say $X_{i_2},
    \ldots, X_{i_{\wrow}}$, at tier $2t+2$. From
    Def.~\ref{def:canonical:completion:1} it follows that $\omega_{i_1} =
    1/(\wrow-1)^t$ and that $\omega_{i_2} = \cdots = \omega_{i_{\wrow}} =
    1/(\wrow-1)^{t+1}$. It is easy to check that $\omega_{i'} \leq \sum_{i \in
    \set{I}_j \setminus \{ i' \}} \omega_i$ is satisfied for all $i' \in
    \set{I}_j = \{ i_1, \ldots, i_{\wrow} \}$. Indeed, the most crucial of
    them being for $i' = i_1$ where we have the inequality $1/(\wrow-1)^t \leq
    (\wrow-1) \cdot 1/(\wrow-1)^{t+1}$ that is satisfied with equality.

  \item The check node $B_j$ is connected to at least two variable nodes, say
    $X_{i_1}, \ldots, X_{i_h}$ at tier $2t$, and $\wrow-h$ variable nodes, say
    $X_{i_{h+1}}, \ldots, X_{i_{\wrow}}$, at tier $2t+2$ where $2 \leq h \leq
    \wrow$. From Def.~\ref{def:canonical:completion:1} it follows that
    $\omega_{i_1} = \cdots \omega_{i_h} = 1/(\wrow-1)^t$ and that
    $\omega_{i_{h+1}} = \cdots = \omega_{i_{\wrow}} = 1/(\wrow-1)^{t+1}$. It
    is easy to check that $\omega_{i'} \leq \sum_{i \in \set{I}_j \setminus \{
    i' \}} \omega_i$ is satisfied for all $i' \in \set{I}_j = \{ i_1, \ldots,
    i_{\wrow} \}$. Actually, unless $\wrow = 2$, none of them is satisfied
    with equality.

\end{itemize}
Because the check node $B_j$ was arbitrary, this concludes the proof that
$\vomega \in \fch{K}{H}$.

\subsection{Proof of Proposition~\ref{prop:min:ps:weight:upper:bound:1}}
\label{sec:proof:prop:min:ps:weight:upper:bound:1}

Let $\graph{T} \defeq \tgraph{T}{H}$ be the Tanner graph corresponding to
$\matr{H}$. To prove the upper bound on $\wpsminh{H}$ we proceed as
follows. By definition, the AWGNC pseudo-weight of any non-zero
pseudo-codeword is larger than or equal to $\wpsminh{H}$. Therefore, any upper
bound on the pseudo-weight of any non-zero pseudo-codeword will yield an upper
bound on $\wpsminh{H}$.

Our choice for a non-zero pseudo-codeword is a pseudo-codeword that was
obtained by the canonical completion rooted at an arbitrary variable node $V$,
see Def.~\ref{def:canonical:completion:1}. Its AWGNC pseudo-weight was
established in Lemma~\ref{lemma:canonical:completion:properties:1}. To get an
upper bound on $\wps(\vomega)$, we need a lower bound on $\twonorm{\vomega}^2$
and an upper bound on $\sonenorm{\vomega}$. We start with the lower bound on
$\twonorm{\vomega}^2$. We have\footnote{In order to shorten the the notation
used in this proof we will use $j \defeq \wcol$ and $k \defeq \wrow$.}
\begin{align}
  \twonorm{\vomega}^2
    &= \sum_{t=0}^{\lfloor \diamg{T}/2 \rfloor}
         N_{V,2t}(\graph{T})
           \left(
             \frac{1}{(k-1)^t}
           \right)^2
     \geq
       \sum_{t=0}^{0}
         N_{V,2t}(\graph{T}) \left( \frac{1}{(k-1)^t} \right)^2
     = 1,
         \label{eq:vomega:twonorm:upper:bound:1}
\end{align}
where we used $N_{V,0}(\graph{T}) = 1$. A side note: if we can assume that the
girth $\girthg{T}$ of $\graph{T}$ is at least six, we have $N_{V,0}(\graph{T})
= \Nmax_{V,0}(\graph{T})$ and $N_{V,2}(\graph{T}) = \Nmax_{V,2}(\graph{T})$,
and therefore we get the better lower bound
\begin{align*}
  \twonorm{\vomega}^2
    &= \sum_{t=0}^{\lfloor \diamg{T}/2 \rfloor}
         N_{V,2t}(\graph{T}) \left( \frac{1}{(k-1)^t} \right)^2
     \geq
       \sum_{t=0}^{1}
         N_{V,2t}(\graph{T}) \left( \frac{1}{(k-1)^t} \right)^2 \\
    &= \sum_{t=0}^{1}
         \Nmax_{V,2t}(\code{C}) \left( \frac{1}{(k-1)^t} \right)^2
     = 1 + j (k-1) \frac{1}{(k-1)^2}
     = 1 + \frac{j}{k-1}.
\end{align*}
For even larger girth, we could give even better lower bounds, but we will not
pursue this any further.

Now we turn to the problem of obtaining an upper bound on $\sonenorm{\vomega}
= \sum_{t=0}^{\lfloor \diamg{T}/2 \rfloor} N_{V,2t}(\graph{T})
\frac{1}{(k-1)^t}$. Because $N_{V,2t}(\graph{T}) \leq \Nmax_{V,2t}$ for all $t
\geq 0$, this sum is clearly upper bounded by the same sum for a Tanner graph
which has the same number of variable nodes but which has maximal expansion,
i.e.,
\begin{align*}
  \sonenorm{\vomega}
    &= \sum_{t=0}^{\lfloor \diamg{T}/2 \rfloor}
         N_{V,2t}(\graph{T}) \frac{1}{(k-1)^t}
     \leq
       \sum_{t=0}^{t'}
         N'_{V,2t} \frac{1}{(k-1)^t},
\end{align*}
where we introduced $N'_{V,2t} \defeq \Nmax_{V,2t}$ for $0 \leq t \leq t'$
where $t'$ is some constant such that $\sum_{t=0}^{t'-1} N'_{V,2t} < n =
\sum_{t=0}^{\lfloor \diamg{T}/2 \rfloor} N_{V,2t}(\graph{T}) \leq
\sum_{t=0}^{t'} N'_{V,2t}$. By construction, $t'$ will fulfill $t' \leq
\lfloor \diamg{T}/2 \rfloor$. Continuing,
\begin{align}
  \sonenorm{\vomega}
    &\leq
       \sum_{t=0}^{t'}
         N'_{V,2t} \frac{1}{(k-1)^t}
     = 1
       +
       \sum_{t=1}^{t'}
         j (j-1)^{t-1} (k-1)^t \frac{1}{(k-1)^t} \nonumber \\
    &= 1
       +
       \sum_{t=1}^{t'}
         j (j-1)^{t-1}
     = 1
       +
       \frac{j}{j-2}
       \cdot
       \left(
         (j-1)^{t'} - 1
       \right) \nonumber \\
    &\leq 
       \frac{j}{j-2}
       \cdot
       (j-1)^{t'}.
         \label{eq:vomega:onenorm:upper:bound:1}
\end{align}
Combining~\eqref{eq:vomega:twonorm:upper:bound:1}
and~\eqref{eq:vomega:onenorm:upper:bound:1} we obtain
\begin{align}
  \wpsminh{H}
    &\leq
       \wps(\vomega)
     =
       \frac{\sonenorm{\vomega}^2}{\twonorm{\vomega}^2}
     \leq
       \frac{\left(
               \frac{j}{j-2}
               \cdot
               (j-1)^{t'}
             \right)^2
            }
            {1}
     \leq
       \left(
         \frac{j}{j-2}
       \right)^2
       \cdot
         (j-1)^{2t'}
\end{align}
In order to complete the proof, we need an upper bound (in function of the
code size $n$) on $t'$. Remembering the definition of $t'$, such a bound can
be obtained as follows:
\begin{align}
  n
    &\geq
       \sum_{t=0}^{t'-1}
         \Nmax_{V,2t}
     = 1
       +
       \sum_{t=1}^{t' - 1}
         j (j-1)^{t-1} (k-1)^t
     = 1
       +
       j(k-1) \frac{\gamma_{j,k}^{t' - 1} - 1}
                   {\gamma_{j,k} - 1} \\
    &= 1
       +
       \underbrace
       {\left(
          \frac{jk-j}
               {jk-j-k+1-1}
        \right)
       }_{\geq 1}
       \left(
         \gamma_{j,k}^{t' - 1} - 1
       \right)
     \geq 
       \gamma_{j,k}^{t' - 1},
\end{align}
where $\gamma_{j,k} = (j-1)(k-1)$. Therefore,
\begin{align}
  t'
    &\leq
       1
       +
       \frac{\log(n)}{\log(\gamma_{j,k})}.
\end{align}
Finally,
\begin{align}
  \wpsminh{H}
    &\leq
       \left(
         \frac{j}{j-2}
       \right)^2
       \cdot
         (j-1)^{2 t'}
     \leq
       \left(
         \frac{j}{j-2}
       \right)^2
       \cdot
         (j-1)^{2 + 2\frac{\log(n)}{\log(\gamma_{j,k})}} \\
    &= \left(
         \frac{j (j-1)}
              {j-2}
       \right)^2
       \cdot
       (j-1)^{2\frac{\log(n)}{\log(\gamma_{j,k})}}
     = \beta'(j,k)
       \cdot
       n^{\beta(j,k)},
\end{align}
where
\begin{align}
  \beta'(j,k)
    &\defeq
       \left(
         \frac{j (j-1)}
              {j-2}
       \right)^2,
  \quad\quad
  \beta(j,k)
     \defeq
       2
       \frac{\log(j-1)}
            {\log(\gamma_{j,k})}
     = \frac{\log\left( (j-1)^2 \right)}
            {\log\big( (j-1)(k-1) \big)}.
\end{align}
For $k > j$ we have $\beta(j,k) < 1$.

\subsection{Proof of Proposition~\ref{prop:stopping:set:vs:fp:1}}
\label{sec:proof:prop:stopping:set:vs:fp:1}

Let us prove the first statement. Because $\vomega$ is in the fundamental
polytope $\fph{P}{H}$, it is also in the fundamental cone $\fch{K}{H}$ and so,
for each $j \in \set{J}$ and for each $i' \in \set{I}_j$ it fulfills (see
Def.~\ref{lemma:reformulations:fc:1}): $\sum_{i \in \set{I}_j \setminus \{ i'
\}} \omega_i \geq \omega_{i'}$. This means that for all $j \in \set{J}$, if
there is an $i'_j \in \set{I}_j$ such that $\omega_{i'_j} > 0$ then there are
at least two distinct $i'_j, i''_j \in \set{I}_j$ such that $\omega_{i'_j} >
0$ and $\omega_{i''_j} > 0$. But this is equivalent to the condition that each
check node in $\neighborhood(\supp(\vomega))$ is connected to at least two
variable nodes in $\supp(\vomega)$.

Let us now prove the second statement. Let $\set{S}$ be a stopping set and let
$\vnu \in \Rp^n$ be a vector where $\nu_i \defeq 1$ if $i \in \set{S}$ and
$\nu_i \defeq 0$ otherwise. It can easily be seen that this vector fulfills
all the conditions for being in the fundamental cone $\fch{K}{H}$, using
e.g.~the inequalities in Lemma~\ref{lemma:properties:fp:fc:2}. Following the
comment after Def.~\ref{def:fundamental:cone:1}, there is an $\alpha \in \Rpp$
(in fact, a whole interval of $\alpha$'s) such that $\vomega \defeq \alpha
\vnu$ is in the fundamental polytope $\fph{P}{H}$.

\subsection{Proof of Lemma~\ref{lemma:adding:one:row:1}}
\label{sec:proof:lemma:adding:one:row:1}

It follows from the definition of the fundamental polytope
(Def.~\ref{def:fundamental:polytope:1}) and the discussion before and
after~\eqref{eq:canonical:relaxation:def:1} that $\convhull(\code{C})
\subseteq \fp{P}(\matr{H}') \subseteq \fph{P}{H}$. However, using
Lemma~\ref{lemma:fp:tree:1} we can conclude that $\fph{P}{H} =
\convhull(\code{C})$ which proves that $\fph{P}{H} = \fp{P}(\matr{H}')$ as
desired.

An alternative proof would be to show that (under the conditions in the lemma
statement) $\vomega \in \fph{P}{H}$ implies $\vomega \in \fp{P}(\matr{H}')$
where for $\fph{P}{H}$ and $\fp{P}(\matr{H}')$ we use the description given in
Lemma~\ref{lemma:properties:fp:fc:2}. Some manipulations of the involved
inequalities lead to the desired result. We leave the details to the reader.

\subsection{Proof of Corollary~\ref{cor:adding:one:row:2}}
\label{sec:proof:cor:adding:one:row:2}

Let $\overline{\matr{H}}_1$ be the matrix that contains the rows of $\matr{H}$
that are not included in $\matr{H}_1$. We have
\begin{align*}
  \fph{P}{H}
     &= \fp{P}\left( \matr{H}_1 \right)
        \cap
        \fp{P}\left( \overline{\matr{H}}_1 \right), \\
  \fp{P}(\matr{H}')
     &= \fp{P}
          \left(
            \begin{pmatrix}
              \matr{H}_1 \\
              \vect{a} \cdot \matr{H}
             \end{pmatrix}
          \right)
        \cap
        \fp{P}\left( \overline{\matr{H}}_1 \right).
\end{align*}
Using Lemma~\ref{lemma:adding:one:row:1} we conclude that $\fp{P} \left(
\begin{pmatrix} \matr{H}_1 \\ \vect{a} \cdot \matr{H} \end{pmatrix}
\right)$ equals $\fp{P}(\matr{H}_1)$ and that therefore $\fp{P}(\matr{H}')$
equals $\fph{P}{H}$.

\subsection{Proof of Corollary~\ref{cor:adding:rows:3}}
\label{sec:proof:cor:adding:rows:3}

Let $\matr{A}$ have $L$ rows, let $\vect{a}_{\ell}$, $\ell \in [L]$, be the
vector containing the $\ell$-th row of $\matr{A}$, and let
$\overline{\matr{H}}_{\ell}$, $\ell \in L$, be the matrix that contains the
rows of $\matr{H}$ that are not included in $\matr{H}_{\ell}$. We have
\begin{align*}
  \fph{P}{H}
     &= \bigcap_{\ell \in [L]}
          \big(
            \fp{P}\left( \matr{H}_{\ell} \right)
            \cap
            \fp{P}\left( \overline{\matr{H}}_{\ell} \right)
          \big), \\
  \fp{P}(\matr{H}')
     &= \bigcap_{\ell \in [L]}
          \left(
            \fp{P}
          \left(
            \begin{pmatrix}
              \matr{H}_{\ell} \\
              \vect{a}_{\ell} \cdot \matr{H}
             \end{pmatrix}
          \right)
        \cap
        \fp{P}\left( \overline{\matr{H}}_{\ell} \right)
        \right).
\end{align*}
Using Lemma~\ref{lemma:adding:one:row:1} we conclude that $\fp{P} \left(
\begin{pmatrix} \matr{H}_{\ell} \\ \vect{a}_{\ell} \cdot \matr{H} \end{pmatrix}
\right)$ equals $\fp{P}(\matr{H}_{\ell})$ for all $\ell \in [L]$ and that
therefore $\fp{P}(\matr{H}')$ equals $\fph{P}{H}$.

{
\bibliographystyle{ieeetr}
\bibliography{/home/vontobel/references/references}

\begin{thebibliography}{10}

\bibitem{Gallager:62}
R.~G. Gallager, ``Low-density parity-check codes,'' {\em IRE Trans.\ Inform.\
  Theory}, vol.~8, pp.~21--28, Jan. 1962.

\bibitem{Gallager:63}
R.~G. Gallager, {\em Low-Density Parity-Check Codes}.
\newblock M.I.T. Press, Cambridge, MA, 1963.
\newblock Available online under
  \verb+http://web.mit.edu/gallager/www/pages/ldpc.pdf+.

\bibitem{Zyablov:71:1}
V.~V. Zyablov, ``An estimate of the complexity of constructing binary linear
  cascade codes,'' {\em Probl.~Inform.~Transm.}, vol.~7, no.~1, pp.~3--10,
  1971.

\bibitem{Zyablov:Pinsker:76:1}
V.~V. Zyablov and M.~S. Pinsker, ``Estimation of error-correction complexity of
  {G}allager low-density codes,'' {\em Probl.~Inform.~Transm.}, vol.~11, no.~1,
  pp.~18--28, 1976.

\bibitem{Tanner:81}
R.~M. Tanner, ``A recursive approach to low-complexity codes,'' {\em IEEE
  Trans.\ on Inform.\ Theory}, vol.~IT--27, pp.~533--547, Sept. 1981.

\bibitem{Margulis:82}
G.~A. Margulis, ``Explicit constructions of graphs without short cycles and low
  density codes,'' {\em Combinatorica}, vol.~2, no.~1, pp.~71--78, 1982.

\bibitem{Berrou:Glavieux:Thitimajshima:93}
C.~Berrou, A.~Glavieux, and P.~Thitimajshima, ``{N}ear {S}hannon {L}imit
  {E}rror-{C}orrecting {C}oding and {D}ecoding: {T}urbo-{C}odes (1),'' in {\em
  Proc.\ IEEE Int.\ Conf.\ Communications}, (Geneva, Switzerland),
  pp.~1064--1070, May 1993.

\bibitem{MacKay:Neal:96:1}
D.~J.~C. MacKay and R.~M. Neal, ``Near {S}hannon limit performance of low
  density parity check codes,'' {\em Electronics Letters}, vol.~32, p.~1645, 29
  Aug. 1996.

\bibitem{MacKay:Neal:97:1}
D.~J.~C. MacKay and R.~M. Neal, ``Near {S}hannon limit performance of low
  density parity check codes,'' {\em Electronics Letters}, vol.~33,
  pp.~457--458, 13 Mar. 1997.

\bibitem{MacKay:99:1}
D.~J.~C. MacKay, ``Good error-correcting codes based on very sparse matrices,''
  {\em IEEE Trans.\ on Inform.\ Theory}, vol.~IT--45, no.~2, pp.~399--431,
  1999.

\bibitem{Wiberg:Loeliger:Koetter:95}
N.~Wiberg, H.-A. Loeliger, and R.~K{\"o}tter, ``Codes and iterative decoding on
  general graphs,'' {\em Europ.\ Trans.\ on Telecomm.}, vol.~6, pp.~513--525,
  Sept./Oct. 1995.

\bibitem{Wiberg:96}
N.~Wiberg, {\em Codes and Decoding on General Graphs}.
\newblock PhD thesis, Link\"oping University, Sweden, 1996.

\bibitem{Aji:McEliece:00:1}
S.~M. Aji and R.~J. McEliece, ``The generalized distributive law,'' {\em IEEE
  Trans.\ on Inform.\ Theory}, vol.~IT--46, no.~2, pp.~325--343, 2000.

\bibitem{Kschischang:Frey:Loeliger:01}
F.~R. Kschischang, B.~J. Frey, and H.-A. Loeliger, ``Factor graphs and the
  sum-product algorithm,'' {\em IEEE Trans.\ on Inform.\ Theory}, vol.~IT--47,
  no.~2, pp.~498--519, 2001.

\bibitem{Loeliger:04:1}
H.-A. Loeliger, ``An introduction to factor graphs,'' {\em IEEE Sig.\ Proc.\
  Mag.}, vol.~21, no.~1, pp.~28--41, 2004.

\bibitem{Luby:Mitzenmacher:Shokrollahi:Spielman:98}
M.~G. Luby, M.~Mitzenmacher, M.~A. Shokrollahi, and D.~A. Spielman, ``Improved
  low-density parity-check codes using irregular graphs and belief
  propagation,'' in {\em Proc.\ IEEE Intern.\ Symp.\ on Inform.\ Theory}, (MIT,
  Cambridge, MA, USA), p.~117, Aug.~16-21 1998.

\bibitem{Richardson:Urbanke:00:1}
T.~Richardson and R.~Urbanke, ``Thresholds for turbo codes,'' in {\em Proc.\
  IEEE Intern.\ Symp.\ on Inform.\ Theory}, (Sorrento, Italy), p.~317, June
  25--30 2000.

\bibitem{Richardson:Shokrollahi:Urbanke:01:1}
T.~J. Richardson, M.~A. Shokrollahi, and R.~L. Urbanke, ``Design of
  capacity-approaching irregular low-density parity-check codes,'' {\em IEEE
  Trans.\ on Inform.\ Theory}, vol.~IT--47, no.~2, pp.~619--637, 2001.

\bibitem{Anderson:Hladik:98:1}
J.~B. Anderson and S.~M. Hladik, ``Tailbiting {MAP} decoders,'' {\em IEEE
  J.~Sel.\ Areas Comm.}, vol.~JSAC--16, no.~2, pp.~297--302, 1998.

\bibitem{Aji:Horn:McEliece:98}
S.~M. Aji, G.~B. Horn, and R.~J. McEliece, ``Iterative decoding on graphs with
  a single cycle,'' in {\em Proc.\ IEEE Intern.\ Symp.\ on Inform.\ Theory},
  (MIT, Cambridge, MA, USA), p.~276, Aug.~16-21 1998.

\bibitem{Forney:Kschischang:Marcus:Tuncel:01:1}
G.~D. {Forney, Jr.}, F.~R. Kschischang, B.~Marcus, and S.~Tuncel, ``Iterative
  decoding of tail-biting trellises and connections with symbolic dynamics,''
  in {\em Codes, Systems, and Graphical Models (Minneapolis, MN, 1999)}
  (B.~Marcus and J.~Rosenthal, eds.), pp.~239--264, Springer Verlag, New York,
  Inc., 2001.

\bibitem{Di:Proietti:Telatar:Richardson:Urbanke:02:1}
C.~Di, D.~Proietti, {\d I}.~E. Telatar, T.~J. Richardson, and R.~L. Urbanke,
  ``Finite-length analysis of low-density parity-check codes on the binary
  erasure channel,'' {\em IEEE Trans.\ on Inform.\ Theory}, vol.~IT--48, no.~6,
  pp.~1570--1579, 2002.

\bibitem{MacKay:Postol:03:1}
D.~J.~C. MacKay and M.~S. Postol, ``Weaknesses of {M}argulis and
  {R}amanujan-{M}argulis low-density parity-check codes,'' {\em Electronic
  Notes in Theoretical Computer Science}, vol.~74, 2003.

\bibitem{Richardson:03:1}
T.~Richardson, ``Error floors of {LDPC} codes,'' in {\em Proc.\ 41st Allerton
  Conf.~on Communications, Control, and Computing}, (Allerton House,
  Monticello, Illinois, USA), October 1--3 2003.

\bibitem{Tian:Jones:Villasenor:Wesel:04:1}
T.~Tian, C.~R. Jones, J.~D. Villasenor, and R.~D. Wesel, ``Selective avoidance
  of cycles in irregular {LDPC} code construction,'' {\em IEEE Trans.\ on
  Comm.}, vol.~COM--52, no.~8, pp.~1242--1247, 2004.

\bibitem{Ramamoorthy:Wesel:04:1}
A.~Ramamoorthy and R.~D. Wesel, ``Analysis of an algorithm for irregular {LDPC}
  code construction,'' in {\em Proc.\ IEEE Intern.\ Symp.\ on Inform.\ Theory},
  (Chicago, IL, USA), p.~69, June 27--July 2 2004.

\bibitem{Chernyak:Chertkov:Stepanov:Vasic:04:1}
V.~Chernyak, M.~Chertkov, M.~Stepanov, and B.~Vasic, ``Instanton method of
  post-error-correction analytical evaluation,'' in {\em Proc.\ IEEE Inform.\
  Theory Workshop}, (San Antonio, TX, USA), pp.~220--224, Oct.~24--29 2004.

\bibitem{Stepanov:Chernyak:Chertkov:Vasic:05:1}
M.~Stepanov, V.~Chernyak, M.~Chertkov, and B.~Vasic, ``Diagnosis of weaknesses
  in modern error correction codes: a physics approach,'' {\em available online
  under \verb+http://www.arxiv.org/cond-mat/0506037+}, June 2005.

\bibitem{Frey:Koetter:Vardy:01:1}
B.~J. Frey, R.~Koetter, and A.~Vardy, ``Signal-space characterization of
  iterative decoding,'' {\em IEEE Trans.\ on Inform.\ Theory}, vol.~IT--47,
  no.~2, pp.~766--781, 2001.

\bibitem{Forney:Koetter:Kschischang:Reznik:01:1}
G.~D. {Forney, Jr.}, R.~Koetter, F.~R. Kschischang, and A.~Reznik, ``On the
  effective weights of pseudocodewords for codes defined on graphs with
  cycles,'' in {\em Codes, Systems, and Graphical Models (Minneapolis, MN,
  1999)} (B.~Marcus and J.~Rosenthal, eds.), vol.~123 of {\em IMA Vol. Math.
  Appl.}, pp.~101--112, Springer Verlag, New York, Inc., 2001.

\bibitem{Feldman:03:1}
J.~Feldman, {\em Decoding Error-Correcting Codes via Linear Programming}.
\newblock PhD thesis, Massachusetts Institute of Technology, Cambridge, MA,
  2003.
\newblock Available online under
  \verb+http://www.columbia.edu/~jf2189/pubs.html+.

\bibitem{Feldman:Wainwright:Karger:05:1}
J.~Feldman, M.~J. Wainwright, and D.~R. Karger, ``Using linear programming to
  decode binary linear codes,'' {\em IEEE Trans.\ on Inform.\ Theory},
  vol.~IT--51, no.~3, pp.~954--972, 2005.

\bibitem{Vontobel:Koetter:04:2}
P.~O. Vontobel and R.~Koetter, ``On the relationship between linear programming
  decoding and min-sum algorithm decoding,'' in {\em Proc.\ Intern.\ Symp.\ on
  Inform.\ Theory and its Applications (ISITA)}, (Parma, Italy), pp.~991--996,
  Oct.~10--13 2004.

\bibitem{Pseudocodewords:Website:05:1}
\verb+http://www.pseudocodewords.info+.

\bibitem{Tanner:99:2}
R.~M. Tanner, ``On quasi-cyclic repeat-accumulate codes,'' in {\em Proc.\ of
  the 37th Allerton Conference on Communication, Control, and Computing},
  (Allerton House, Monticello, Illinois, USA), pp.~249--259, Sep.~22-24 1999.

\bibitem{Tanner:Sridhara:Fuja:01:1}
R.~M. Tanner, D.~Sridhara, and T.~Fuja, ``A class of group-structured {LDPC}
  codes,'' in {\em Proc.\ of ICSTA 2001}, (Ambleside, England), 2001.

\bibitem{Laendner:Milenkovic:05:1}
S.~Laendner and O.~Milenkovic, ``Algorithmic and combinatorial analysis of
  trapping sets in structured {LDPC} codes,'' in {\em Proc.~2005 International
  Conference on Wireless Networks, Communications, and Mobile Computing
  (Wirelesscom 2005)}, (Maui, HI, USA), Jun.~13-16 2005.

\bibitem{MacWilliams:Sloane:98}
F.~J. MacWilliams and N.~J.~A. Sloane, {\em The Theory of Error-Correcting
  Codes}.
\newblock New York: North-Holland, 1998.

\bibitem{Boyd:Vandenberghe:04:1}
S.~Boyd and L.~Vandenberghe, {\em Convex Optimization}.
\newblock Cambridge, UK: Cambridge University Press, 2004.

\bibitem{Barvinok:02:1}
A.~Barvinok, {\em A Course in Convexity}, vol.~54 of {\em Graduate Studies in
  Mathematics}.
\newblock Providence, RI: American Mathematical Society, 2002.

\bibitem{Massey:77:1}
W.~S. Massey, {\em Algebraic Topology: an Introduction}.
\newblock New York: Springer-Verlag, 1977.
\newblock Reprint of the 1967 edition, Graduate Texts in Mathematics, Vol. 56.

\bibitem{Stark:Terras:96:1}
H.~M. Stark and A.~A. Terras, ``Zeta functions of finite graphs and
  coverings,'' {\em Adv. Math.}, vol.~121, no.~1, pp.~124--165, 1996.

\bibitem{Polderman:Willems:98}
J.~Polderman and J.~Willems, {\em Introduction to Mathematical Systems Theory}.
\newblock Springer-Verlag New York, Inc., 1998.

\bibitem{Feldman:Karger:Wainwright:03:2}
J.~Feldman, D.~R. Karger, and M.~J. Wainwright, ``{LP} decoding,'' in {\em
  Proc.\ 41st Allerton Conf.~on Communications, Control, and Computing},
  (Allerton House, Monticello, Illinois, USA), October 1--3 2003.
\newblock Available online under
  \verb+http://www.columbia.edu/~jf2189/pubs.html+.

\bibitem{Hwang:79:1}
T.~Y. Hwang, ``Decoding linear block codes for minimizing word error rate,''
  {\em IEEE Trans.\ on Inform.\ Theory}, vol.~IT--25, no.~6, pp.~733--737,
  1979.

\bibitem{Agrell:96:1}
E.~Agrell, ``Vorono\u\i\ regions for binary linear block codes,'' {\em IEEE
  Trans.\ on Inform.\ Theory}, vol.~IT--42, no.~1, pp.~310--316, 1996.

\bibitem{Ashikhmin:Barg:98:1}
A.~Ashikhmin and A.~Barg, ``Minimal vectors in linear codes,'' {\em IEEE
  Trans.\ on Inform.\ Theory}, vol.~IT--44, no.~5, pp.~2010--2017, 1998.

\bibitem{Borissov:Manev:Nikova:01:1}
Y.~Borissov, N.~Manev, and S.~Nikova, ``On the non-minimal codewords in the
  binary {R}eed-{M}uller code,'' in {\em Proc.\ IEEE Intern.\ Symp.\ on
  Inform.\ Theory}, (Washington, D.C., USA), p.~39, June 24-29 2001.

\bibitem{Vontobel:Smarandache:Kiyavash:Teutsch:Vukobratovic:05:1}
P.~O. Vontobel, R.~Smarandache, N.~Kiyavash, J.~Teutsch, and D.~Vukobratovic,
  ``On the minimal pseudo-codewords of codes from finite geometries,'' in {\em
  Proc.\ IEEE Intern.\ Symp.\ on Inform.\ Theory}, (Adelaide, Australia),
  pp.~980--984, Sep.~4--9 2005.
\newblock Available online under \verb+http://www.arxiv.org/abs/cs.IT/0508019+.

\bibitem{Vontobel:Smarandache:05:1}
P.~O. Vontobel and R.~Smarandache, ``On minimal pseudo-codewords of {T}anner
  graphs from projective planes,'' in {\em Proc.\ 43rd Allerton Conf.~on
  Communications, Control, and Computing}, (Allerton House, Monticello,
  Illinois, USA), Sep.~28--30 2005.
\newblock Available online under \verb+http://www.arxiv.org/abs/cs.IT/0510043+.

\bibitem{Smarandache:Wauer:05:1:subm}
R.~Smarandache and M.~Wauer, ``Bounds on the pseudo-weight of minimal
  pseudo-codewords of projective geometry codes,'' {\em submitted, available
  online under \verb+http://www.arxiv.org/abs/cs.IT/0510049+}, Oct. 2005.

\bibitem{Koetter:Li:Vontobel:Walker:05:1:subm}
R.~Koetter, W.-C.~W. Li, P.~O. Vontobel, and J.~L. Walker, ``Characterizations
  of pseudo-codewords of {LDPC} codes,'' {\em submitted, available online under
  \verb+http://www.arxiv.org/abs/cs.IT/0508049+}, Aug. 2005.

\bibitem{Etzion:Trachtenberg:Vardy:99}
T.~Etzion, A.~Trachtenberg, and A.~Vardy, ``Which codes have cycle-free
  {T}anner graphs?,'' {\em IEEE Trans.\ on Inform.\ Theory}, vol.~IT--45,
  pp.~2173--2183, Sept. 1999.

\bibitem{Koetter:Li:Vontobel:Walker:04:1}
R.~Koetter, W.-C.~W. Li, P.~O. Vontobel, and J.~L. Walker, ``Pseudo-codewords
  of cycle codes via zeta functions,'' in {\em Proc.\ IEEE Inform.\ Theory
  Workshop}, (San Antonio, TX, USA), pp.~7--12, Oct.~24--29 2004.
\newblock Available online under \verb+http://www.arxiv.org/abs/cs.IT/0502033+.

\bibitem{Wainwright:Jordan:03:1}
M.~J. Wainwright and M.~I. Jordan, ``Variational inference in graphical models:
  the view from the marginal polytope,'' in {\em Proc.\ 41st Allerton Conf.~on
  Communications, Control, and Computing}, (Allerton House, Monticello,
  Illinois, USA), October 1--3 2003.

\bibitem{Yedidia:Freeman:Weiss:05:1}
J.~S. Yedidia, W.~T. Freeman, and Y.~Weiss, ``Constructing free-energy
  approximations and generalized belief propagation algorithms,'' {\em IEEE
  Trans.\ on Inform.\ Theory}, vol.~IT--51, no.~7, pp.~2282--2312, 2005.

\bibitem{Deza:Laurent:97:1}
M.~M. Deza and M.~Laurent, {\em Geometry of cuts and metrics}, vol.~15 of {\em
  Algorithms and Combinatorics}.
\newblock Berlin: Springer-Verlag, 1997.

\bibitem{Feldman:Malkin:Stein:Servedio:Wainwright:04:1}
J.~Feldman, T.~Malkin, C.~Stein, R.~A. Servedio, and M.~J. Wainwright, ``{LP}
  decoding corrects a constant fraction of errors,'' in {\em Proc.\ IEEE
  Intern.\ Symp.\ on Inform.\ Theory}, (Chicago, IL, USA), p.~68, June 27--July
  2 2004.

\bibitem{Feldman:Koetter:Vontobel:05:1}
J.~Feldman, R.~Koetter, and P.~O. Vontobel, ``The benefit of thresholding in
  {LP} decoding of {LDPC} codes,'' in {\em Proc.\ IEEE Intern.\ Symp.\ on
  Inform.\ Theory}, (Adelaide, Australia), pp.~307--311, Sep.~4--9 2005.
\newblock Available online under \verb+http://www.arxiv.org/abs/cs.IT/0508014+.

\bibitem{Haley:Grant:05:1}
D.~Haley and A.~Grant, ``Improved reversible {LDPC} codes,'' in {\em Proc.\
  IEEE Intern.\ Symp.\ on Inform.\ Theory}, (Adelaide, Australia),
  pp.~1367--1371, Sep.~4--9 2005.

\bibitem{Kou:Lin:Fossorier:01:1}
Y.~Kou, S.~Lin, and M.~P.~C. Fossorier, ``Low-density parity-check codes based
  on finite geometries: a rediscovery and new results,'' {\em IEEE Trans.\ on
  Inform.\ Theory}, vol.~IT--47, pp.~2711--2736, Nov. 2001.

\bibitem{Lucas:Bossert:Breitbach:98:1}
R.~Lucas, M.~Bossert, and M.~Breitbach, ``On iterative soft-decision decoding
  of linear binary block codes and product codes,'' {\em IEEE J.~Sel.\ Areas
  Comm.}, vol.~JSAC--16, no.~2, pp.~276--296, 1998.

\bibitem{Xu:Tang:Kou:Lin:AbdelGhaffar:02:1}
J.~Xu, H.~Tang, Y.~Kou, S.~Lin, and K.~Abdel-Ghaffar, ``A general class of
  {LDPC} finite geometry codes and their performance,'' in {\em Proc.\ IEEE
  Intern.\ Symp.\ on Inform.\ Theory}, (Lausanne, Switzerland), p.~309, June
  30--July 5 2002.

\bibitem{Fossorier:Palanki:Yedidia:03:1}
M.~Fossorier, R.~Palanki, and J.~Yedidia, ``Iterative decoding of multi-step
  majority logic decodable codes,'' in {\em Proc.\ 3rd Intern.~Symp.~on Turbo
  Codes and Related Topics}, (Brest, France), Sept.~1--5 2003.

\bibitem{Wainwright:05:1}
M.~J. Wainwright, ``Codeword polytopes and linear programming relaxations for
  error-control coding.'' Talk at Workshop on "Applications of Statistical
  Physics to Coding Theory", Santa Fe, New Mexico, USA, Jan.~11 2005.
\newblock Available online under
  \verb+http://cnls.lanl.gov/~chertkov/EC_Talks/Wainwright/+.

\bibitem{Hu:Eleftheriou:Arnold:05:1}
X.-Y. Hu, E.~Eleftheriou, and D.~M. Arnold, ``Regular and irregular progressive
  edge-growth {T}anner graphs,'' {\em IEEE Trans.\ on Inform.\ Theory},
  vol.~IT--51, no.~1, pp.~386--398, 2005.

\bibitem{Rosenthal:Vontobel:00}
J.~Rosenthal and P.~O. Vontobel, ``Constructions of {LDPC} codes using
  {R}amanujan graphs and ideas from {M}argulis,'' in {\em Proc. of the 38th
  Allerton Conference on Communication, Control, and Computing}, (Allerton
  House, Monticello, Illinois, USA), pp.~248--257, Oct.~4--6 2000.

\bibitem{Kelley:Sridhara:05:1:subm}
C.~Kelley and D.~Sridhara, ``Pseudocodewords of {T}anner graphs,'' {\em
  submitted to IEEE Trans.\ Inform.\ Theory, available online under
  \verb+http://www.arxiv.org/abs/cs.IT/0504013+}, Apr. 2005.

\bibitem{Vontobel:Koetter:04:1}
P.~O. Vontobel and R.~Koetter, ``Lower bounds on the minimum pseudo-weight of
  linear codes,'' in {\em Proc.\ IEEE Intern.\ Symp.\ on Inform.\ Theory},
  (Chicago, IL, USA), p.~70, June 27--July 2 2004.

\bibitem{Chaichanavong:Siegel:05:1}
P.~Chaichanavong and P.~H. Siegel, ``Relaxation bounds on the minimum
  pseudo-weight of linear block codes,'' in {\em Proc.\ IEEE Intern.\ Symp.\ on
  Inform.\ Theory}, (Adelaide, Australia), pp.~805--809, Sep.~4--9 2005.
\newblock Available online under \verb+http://www.arxiv.org/abs/cs.IT/0508046+.

\bibitem{Horn:Johnson:90}
R.~A. Horn and C.~R. Johnson, {\em Matrix Analysis}.
\newblock Cambridge: Cambridge University Press, 1990.
\newblock Corrected reprint of the 1985 original.

\end{thebibliography}
}

\end{document}